\documentclass[acmsmall]{acmart}
\AtBeginDocument{%
  \providecommand\BibTeX{{%
    \normalfont B\kern-0.5em{\scshape i\kern-0.25em b}\kern-0.8em\TeX}}}

\setcopyright{acmcopyright}
\copyrightyear{}
\acmYear{2020}
\acmDOI{}

\acmJournal{JACM}
\acmVolume{XXX}
\acmNumber{XXX}
\acmArticle{}
\acmMonth{0}

\usepackage{color}
\usepackage{hyperref}
\hypersetup{hypertex=true,
	colorlinks=true,
	linkcolor=blue,
	anchorcolor=blue,
	citecolor=green,
	urlcolor=red}
\usepackage{graphicx}
\usepackage{multirow}
\usepackage{array}
\usepackage{times}
\usepackage{float}
\usepackage{subcaption}
\usepackage{lscape}
\usepackage{bbding}
\usepackage{mathrsfs}
\usepackage{makecell}
\usepackage{verbatim}
\usepackage{longtable}

\usepackage{titlesec}
\titlespacing{\subsubsection}{0pt}{*1}{*1}
\titlespacing{\paragraph}{0pt}{*1}{*1}
\begin{document}

\title{A Comprehensive Survey on Deep Music Generation: Multi-level Representations, Algorithms, Evaluations, and Future Directions}
%\lhead[LO]{A Comprehensive Survey on Deep Music Generation}
%\lhead[O]{A Comprehensive Survey on Deep Music Generation}
\lhead[\small \thepage]{\small A Comprehensive Survey on Deep Music Generation}%{\thepage}
\author{Shulei Ji}
\email{taylorji@stu.xjtu.edu.cn}
%\author{Xinyu Yang}
%\email{yxyphd@mail.xjtu.edu.cn}
\author{Jing Luo}
\email{luojingl@stu.xjtu.edu.cn}
\author{Xinyu Yang}
\email{yxyphd@mal.xjtu.edu.cn}
\affiliation{%
	\institution{School of Computer Science and Technology, Xi’an Jiaotong University}
	\streetaddress{No.28, Xianning West Road}
	\city{Xi'an}
%	\province{Shaanxi}
	\country{China}
	\postcode{710049}
}

\renewcommand{\shortauthors}{Shulei Ji and Jing Luo, et al.}

\begin{abstract}
The utilization of deep learning techniques in generating various contents (such as image, text, etc.) has become a trend. Especially music, the topic of this paper, has attracted widespread attention of countless researchers.The whole process of producing music can be divided into three stages, corresponding to the three levels of music generation: score generation produces scores, performance generation adds performance characteristics to the scores, and audio generation converts scores with performance characteristics into audio by assigning timbre or generates music in audio format directly. Previous surveys have explored the network models employed in the field of automatic music generation. However, the development history, the model evolution, as well as the pros and cons of same music generation task have not been clearly illustrated. This paper attempts to provide an overview of various composition tasks under different music generation levels, covering most of the currently popular music generation tasks using deep learning. In addition, we summarize the datasets suitable for diverse tasks, discuss the music representations, the evaluation methods as well as the challenges under different levels, and finally point out several future directions.
\end{abstract}

\maketitle

\section{Introduction}
\label{1.intro}
\begin{figure}
	%	\centering
	\normalsize
	\setlength{\belowcaptionskip}{0pt}
	\setlength{\abovecaptionskip}{10pt}
	\includegraphics[width=0.9\textwidth]{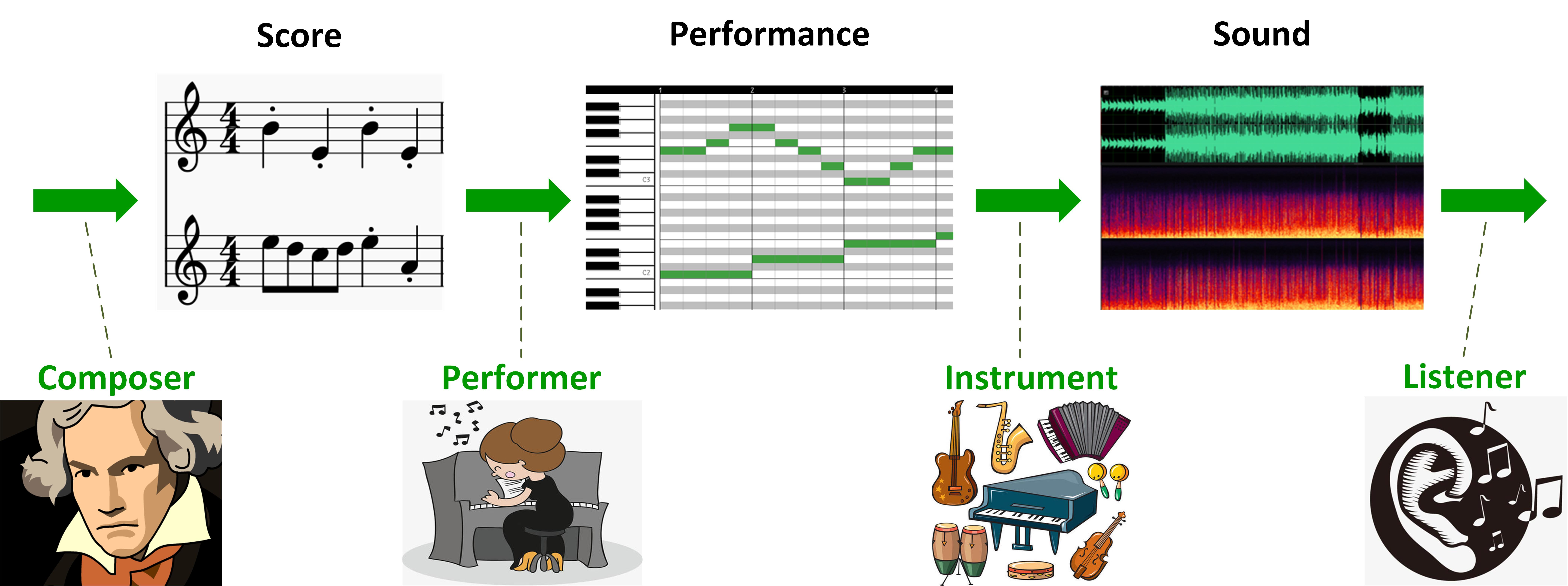}
	\caption{Music generation process}
	\label{fig:111}       % Give a unique label
\end{figure}

About 200 years ago, American poet Henry Wadsworth Longfellow once said, ``music is the universal language of mankind," which was confirmed by researchers at Harvard University in 2019. The relationship between score and real sound is similar to that between text and speech. Music score is a highly symbolic and abstract visual expression that can effectively record and convey music thoughts. While sound is a set of continuous and concrete signal form encoding all the details we can hear. We can describe these two kinds of forms at different levels, with the score at the top and the sound at the bottom. The semantics and expression of music depend largely on performance control, for example, slowing down the rhythm of a cheerful song twice can make it sound sad. Therefore, an intermediate layer can be inserted between the top and the bottom layer to depict the performance. Thus the music generation process is usually divided into three stages \cite{139}, as shown in Figure \ref{fig:111}: in the first stage, composers produce music scores; in the second stage, performers perform scores and generate performance; in the last stage, the performance is rendered into sound by adding different timbres (instruments) and perceived by human beings (listeners). From the above, we divide automatic music generation into three levels: the top level corresponds to score generation, the middle level corresponds to performance generation, and the bottom level corresponds to audio generation. The lower-level music generation can be conditioned on the results of higher-level generation, e.g. predicting performance characteristics through score features.

Researches on generating music of different levels emerge in an endless stream, and the diversification of generation tasks have been increasing. Among them, the study on score generation lasts the longest and pertinent research is the most abundant. From the initial melody generation to now, score generation covers polyphonic music generation, accompaniment generation, style transfer, interactive generation, music inpainting and so on. Score considers musical features such as pitch, duration and chord progression, and its representation is discrete, while performance involves more abundant dynamics and timing information, determining various musical expressions \cite{83}. Performance generation includes rendering performance and composing performance. The former is vested with performance characteristics without changing score features, while the latter models both score and performance features. Audio generation pays more attention to pure continuous and rich acoustic information, such as timbre. The representation of audio is continuous, and two common representations are waveform and spectrum. Adding lyrics to the score then endowing it with timbre can realize the singing synthesis. In addition, style transfer can be performed on music audio and singing voice.

There have been some papers reviewing automatic music generation, \cite{1} reviewed AI methods for computer analysis and composition; book \cite{8} made in-depth and all-round comments on assorted algorithmic composition methods; \cite{76} comprehensively introduced algorithmic composition using AI method, focusing on the origin of algorithmic composition and different AI methods; \cite{78} provided an overview of computational intelligence technology used in generating music, centering on evolutionary algorithm and traditional neural network; \cite{81,82} presented deep learning methods of generating music; \cite{79} put forward the classification of AI methods currently applied to algorithmic composition, methods mentioned cover a wide range, deep learning is only a small branch; \cite{80} analyzed the early work of generating music automatically in the late 1980s, which is a reduced version of \cite{81} with a few additional supplements.

\begin{figure}[H]
	%	\centering
	\normalsize
	\setlength{\belowcaptionskip}{0pt}
	\setlength{\abovecaptionskip}{5pt}
	\includegraphics[width=0.7\textwidth]{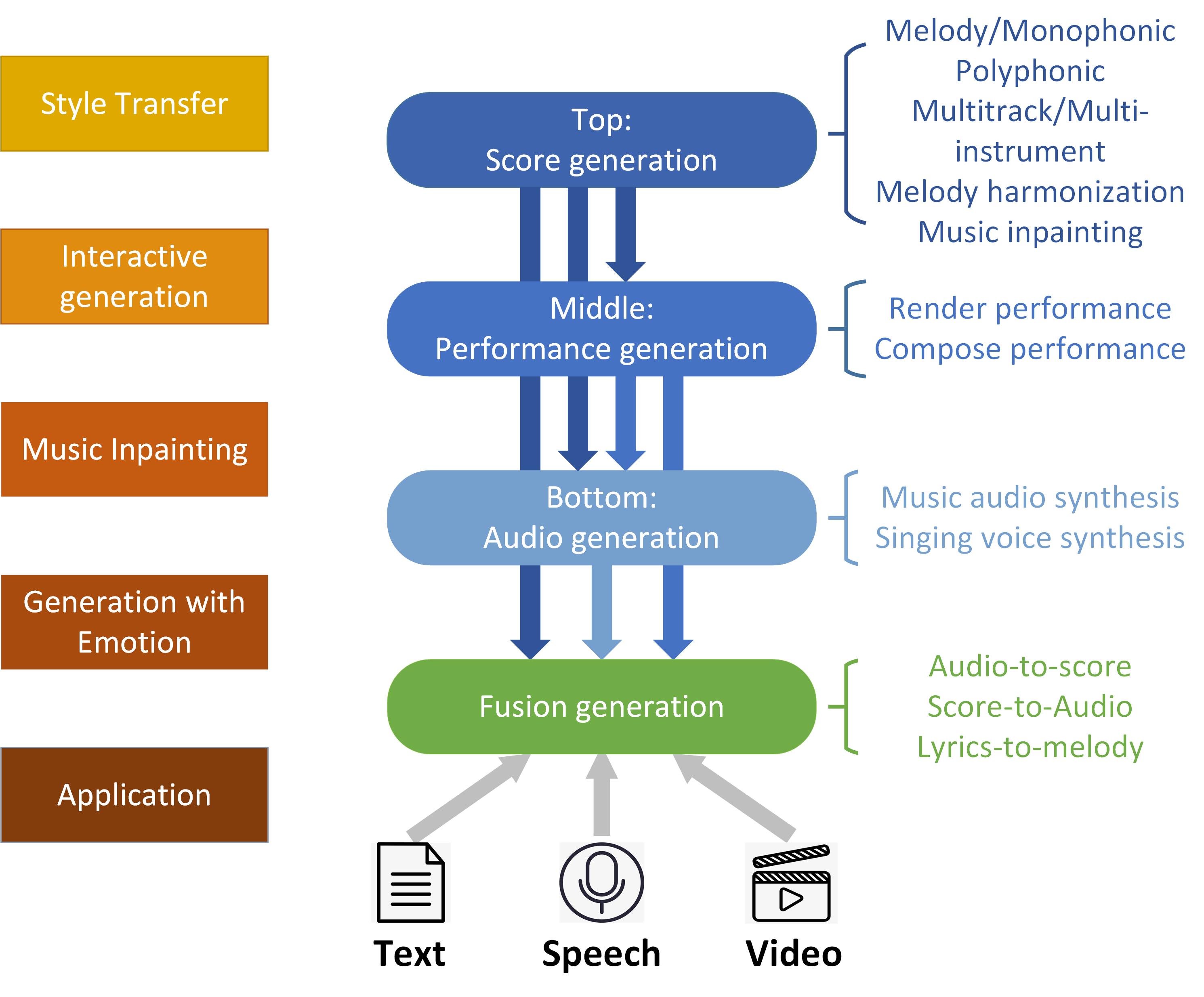}
	\caption{Paper organization}
	\label{fig:222}       % Give a unique label
\end{figure}

However, most of the reviews classified music generation research from the perspective of algorithms, summarizing different music generation tasks using the same algorithm, without classifying and comparing composition tasks from the level of music representation. These works may be comprehensive enough in introducing methods, but for the researchers of a certain subtask of music generation, they may be more inclined to know the work similar to their research. From this point of view, this paper first divides music generation into three levels according to the representations of music, namely score generation, performance generation and audio generation. Then, the research under each level is classified in more detail according to different generation tasks, summarizing various methods as well as network architectures under the same composition task. In this way, researchers can easily find research work related to a specific task, so as to learn from and improve the methods of others, which brings great convenience for the follow-up research. The paper organization is shown in Figure \ref{fig:222}. The blue part in the middle represents the three levels of music generation, and the arrows imply that the lower-level generation can be conditioned on the higher-level results. The green module indicates that the three levels can be fused with each other, and can also be combined with other modal musical/non-musical content (such as lyrics, speech, video, etc.) to produce cross-modal application research, e.g. generating melody from lyrics, synthesizing music for silent videos, etc. The left side displays some special generation tasks that are not limited to a certain level (modality), such as style transfer, including not only symbolic music style transfer, but also audio style transfer. The right side lists several specific generation tasks under each level. One thing that needs to be mentioned is that considering the rise of deep learning and its growing popularity, this paper focuses on the research of music generation using deep learning.

The remaining chapters of this paper are organized as follows. The second section briefly introduces the development history of automatic music generation and summarizes some related review work. In the third section, the music representations based on different domains are introduced in detail. Section \ref{sec:4} divides the music generation into three levels, and the composition tasks under each level are classified more precisely. Section 5 sorts out the available music datasets in different storage formats, so that more researchers can devote themselves to the research of music generation. Section 6 presents the current common music evaluation methods from the objective and subjective aspects. Section 7 points out some challenges and future research directions for reference. The last section draws a general conclusion of this paper.
\section{Related Work}
\label{2.related work}
Automatic music generation has a long history, and it has been developed for more than 60 years. During this period, a large number of researchers utilized different methods, including but not limited to grammar rules, probability model, evolutionary computation, neural network, and carried out numerous researches on various datasets aiming at different generation tasks. With the advent of deep learning, using deep learning technologies to automatically generate various contents (such as images, text, etc.) has become a hot issue. As one of the most universal and infectious art forms in human society, music has attracted plenty of researchers. So far, dozens of subtasks have been derived from automatic music generation, such as melody generation, chord generation, style transfer, performance generation, audio synthesis, singing voice synthesis, etc. There have been quantities of research works for distinct subtasks, and we will elaborate on them in section 3.

With the development in the field of automatic music generation, there have been quite a few overview articles, \cite{1} reviewed the AI methods of computer analysis and composition; \cite{2,3,4} recalled the early history of algorithmic composition; \cite{5} provided a broad perspective for this field, but the content was relatively shallow; \cite{6} analyzed the motivations and methods of algorithmic composition; \cite{7} discussed algorithmic composition from the perspective of music theory and artistic consideration; \cite{8} gave an in-depth and all-round review on various methods; \cite{76} comprehensively introduced the research on algorithmic composition using AI methods, focusing on the origin of algorithmic composition and varied AI methods; \cite{78} provided an overview of computational intelligence technologies utilized in music composition, focusing on evolutionary algorithms and traditional neural networks, such as ART, RNN, LSTM, etc.; \cite{77} presented a functional taxonomy of music generation systems and revealed the relationship between systems. There are similarities with the classification ideas of ours, but it did not clearly divide the levels of music generation, let alone the subdivision under each level; \cite{81,82} mainly classified the existing researches from the perspective of deep network architectures, and a kind of network architecture can be applied to various generation tasks; \cite{79} proposed the classification of AI methods currently applied to algorithmic composition, methods mentioned covered a wide range, deep learning is only a small branch; \cite{80} analyzed the early work of music generation in the late 1980s, then introduced some deep learning conceptual frameworks to analyze assorted concepts and dimensions involved in music generation, and showed many current systems to illustrate the concerns and technologies of distinct researches. It is a reduced version of \cite{81} with some additional supplements. \cite{121} and \cite{126} have conducted extensive and detailed investigations into the field of computational expressive music performance.
\subsection{History}
\label{2.1}
A rapid progress of artificial neural networks is gradually erasing the border between the arts and the sciences. Indeed, there was a number of attempts to automate the process of music composition long before the artificial neural networks era \cite{60}. A popular early example is Musical Dice Game, whereby small fragments of music are randomly re-ordered by rolling a dice to create a musical piece \cite{77}. According to Hedges \cite{10}, there were at least 20 music dice games between 1757 and 1812, enabling novice musicians to compose polonaises, minuets, walzes, etc. John Cage, Iannis Xenakis and other avant-garde composers have continued the idea of chance-inspired composition. John Cage's Atlas Eclipticalis was created by randomly placing translucent paper on a star chart and tracing the stars as notes \cite{11}. With further development, Pinkerton's work \cite{12} may be the first attempt to generate melody by computer. Brooks et al. \cite{13} specifically followed this work and established a Markov transformation model based on a small music corpus. The first computer-generated music appeared in 1957 by a sound synthesis software developed by Bell Labs. ``The Iliac Suite" was the first music score created by computer \cite{14}, making use of stochastic models (Markov chains) for generation as well as rules to filter generated material according to desired properties. Iannis Xenakis, a renowned avant-garde composer, profusely used stochastic algorithms to generate raw material for his compositions. Koenig, another composer, implemented the PROJECT1 algorithm in 1964, using serial composition and other techniques (as Markov chain) to automate the generation of music \cite{3}.

Since then, researches on algorithmic composition has emerged in an endless stream. Traditional music generation algorithms are mainly divided into the following three categories: one is rule-based music generation, the main idea is to use specific rules or grammar for algorithmic composition. Salas et al. \cite{261} use linguistics and grammar to create music. The disadvantage of this method is that different rules need to be created according to different types of music. The composition process is often accompanied by consistency intervention and fine-tuning in the post-processing stage. The other is probability model, including Markov model and hidden Markov model \cite{15} and so on. David cope \cite{262} combines Markov chains and other technologies (as music grammar) into a semi-automatic system for algorithmic composition. However, this kind of model can only produce subsequences existing in the original data, and the model has no memory, inducing that the conditional probability obtained may fluctuate greatly due to the difference of input data each time. Another extremely effective method is to use neural networks to learn music features, and then exploit the features automatically learned by the neural networks to continuously generate new music fragments. Todd et al. \cite{263} used RNNs to generate monophonic music note by note via predicting note pitches and durations. However, the early RNNs have problems such as gradient disappearance, which makes it difficult to generate music structure with long-term consistency. 

In addition, there are some studies using evolutionary algorithm (such as genetic algorithm), intelligent agent \cite{16}, random field and other methods to automatically compose music, and achieving promising results. For example, GenJam \cite{17} is a genetic algorithm-based model that can produce Jazz solos on a given chord progression. Its fitness function needs to interact with people, thus greatly limits its efficiency. Using searching agent, Cope’s Experiments in Musical Intelligence (EMI) program successfully composes music in the styles of many famous composers such as Chopin, Bach, and Rachmanino \cite{18}; Victor et al. \cite{19} used random fields to model polyphonic music. The process of music generation using evolutionary algorithm is too abstract to follow. And it is difficult to extract valuable music ideas and set the fitness function \cite{221}. In addition, these algorithms try to minimize the search space by using simple music representation, resulting in the loss of music quality \cite{240}.

Similarly, previous methods of modeling expressive performance include rule-based methods, such as the early performance model ``Director Musices", which contains rules inferred from theoretical and experimental knowledge \cite{20}; the KTH model consists of a top-down approach for predicting performance characteristics from rules based on local musical context \cite{122}. The other performance modeling method is to use probability models, such as hierarchical HMMs \cite{123}, dynamic Bayesian networks (DBNs) \cite{124}, conditional random fields (CRFs) \cite{125}, and switched Kalman filters \cite{120}. Another kind of method is neural network, such as using feedforward neural networks (FFNNs) to predict expression parameters as functions of music score features \cite{129}, and utilizing RNN to model temporal dependencies between score features and expressive parameters \cite{132}. In addition, Grachten et al. \cite{128} use assorted unsupervised learning techniques to learn features with which they then predict expressive dynamics. On that basis, Herwaarden et al. \cite{127} use an interesting combination of an RBM based architecture, a note-centered input representation, and multiple datasets to predict expressive dynamics. Xia and Dannenberg \cite{130} and Xia et al. \cite{131} show how to use a linear dynamic system trained by spectral learning to predict expressive dynamics and timing of the next score event.

According to whether music audio contains lyrics, we divide music audio into audio and singing voice. Early work on audio synthesis did not focus on music, but on sound itself or speech. Until the last five years, music audio synthesis has gradually developed. There are a variety of traditional methods for sound modeling. A “physical modeling” approach mathematically models the physics that generate a target waveform \cite{267}. An “acoustic modeling” approach uses signal generators and processors for manipulating waveforms such as oscillators, modulators, filters designed to generate acoustic features regardless of physical processes. Concatenative synthesis, more common for speech, draws on a large database of very short snippets of sound to assemble the target audio. Statistical models first came to prominence in the 1980s with the HMM that eventually dominated the fields of automatic speech recognition and generation. Statistical models learn model parameters from data as opposed to expert design in the case of physical and acoustic models \cite{268}. Multiple previous successful singing synthesizers are based on concatenative methods \cite{269,270}, that is, converting and connecting the short waveform units selected from the singer's recording list. Although such systems are the most advanced in terms of sound quality and naturalness, they are limited in flexibility and difficult to expand or significantly improve. On the other hand, machine learning-based approaches, such as statistical parametric methods \cite{271}, are much less rigid and do allow for things such as combining data from multiple speakers, model adaptation using small amounts of training data, joint modeling of timbre and expression, etc. Unfortunately, so far these systems have been unable to match the sound quality of concatenative methods, in particular suffering from over-smoothing in frequency and time. Many of the techniques developed for HMM-based TTS are also applicable to singing synthesis, e.g. speaker-adaptive training \cite{272}. The main drawback of HMM-based method is that phonemes are modeled using a small number of discrete states and within each state statistics are constant. This causes excessive averaging, an overly static “buzzy” sound and noticeable state transitions in long sustained vowels in the case of singing \cite{200}.
\subsection{Deep Music Generation}
\label{2.2}
Deep music generation is to use computers utilizing deep learning network architectures to automatically generate music. In 2012, the performance of deep learning architectures in ImageNet tasks were significantly better than that of manual feature extraction methods. Since then, deep learning has become popular and gradually developed into a rapidly growing field. As an active research field for decades, music generation naturally has attracted the attention of countless researchers. Currently, deep learning algorithms have become the mainstream method in the field of music generation research. This section will review the contribution of deep learning network architectures in the field of music generation in recent years, but we do not deny the value of non-deep learning methods.

RNN is a valid model for learning sequence data and it is also the first neural network architecture for music generation. As early as 1989, Todd \cite{263} used RNN to generate monophonic melody for the first time. However, due to the gradient vanishing problem, it is difficult for RNNs to store long historical information about sequences. To solve this problem, Hochreiter et al. \cite{276} designed a special RNN architecture--LSTM to assist network memorize and retrieve information in the sequence. In 2002, Eck et al. \cite{264} used LSTM in music creation for the first time, improvising blues music with good rhythm and reasonable structure based on a short recording. Boulanger et al. \cite{266} proposed the RNN-RBM model in 2012, which is superior to the traditional polyphonic music generation model in various datasets, but it was still tough to capture the music structure with long-term dependence. In 2016, the Magenta team of Google Brain proposed Melody RNN model \cite{265}, further improving the ability of RNN to learn long-term structures. Later, Hadjeres et al. \cite{22} proposed Anticipation-RNN, a novel RNN model that allows to enforce user-defined positional constraints. Johnson et al. \cite{25} proposed TP-LSTM-NADE and BALSTM using a set of parallel, tied-weight recurrent networks for prediction and composition of polyphonic music, preserving translation-invariance of the dataset.

With the continuous development of deep learning technologies, powerful deep generative models such as VAE, GAN, and Tansformer have gradually emerged. MusicVAE \cite{23} proposed by Roberts et al. is a hierarchical VAE model, which can capture the long-term structure of polyphonic music and has eminent interpolation and reconstruction performance. Jia et al. \cite{35} proposed a coupled latent variable model with binary regularizer to realize impromptu accompaniment generation. Although GANs are very powerful, they are notoriously difficult to train and are usually not applied to sequential data. However, Yang et al. \cite{21} and Dong et al. \cite{32} recently demonstrated the ability of CNN-based GANs in music generation. Specifically, Yang et al. \cite{21} proposed a GAN-based MidiNet to generate melody one bar (measure) after another, and proposed a novel conditional mechanism to generate current bar conditioned on chords. The MuseGAN model proposed by Dong et al. \cite{32} is considered to be the first model that can generate multi-track polyphonic music. Yu et al. \cite{225} successfully applied RNN-based GAN to music generation for the first time by combining reinforcement learning technology. Recently, Transformer models have shown its great potential in music generation. Huang et al. \cite{56} successfully applied Transformer for the first time in creating music with long-term structure. Donahue et al. \cite{39} proposed using Transformer to generate multi-instrument music, and put forward a pre-training technology based on transfer learning. Huang et al. \cite{110,111} proposed a new music representation method named REMI and exploited the language model Transformer XL \cite{112} as the sequence model to generate popular piano music.

There are not too many researches on performance generation using deep learning. Compared with various complex models of score generation, most of the performance generation models are simple RNN-based models. For instance, vanilla RNN is used to render note velocity and start deviation \cite{277}, LSTM is used to predict expressive dynamics \cite{142}, conditional variation RNN is used to render music performance with interpretation variations \cite{134} and \cite{135} represents the unique form of musical score using graph neural network and apply it for rendering expressive piano performance from the music score. Some other researches use DNN models to generate polyphonic music with expressive dynamics and timing \cite{56,139}(as performance RNN \cite{141}), and these models are more like music composition models than expressive performance models with score as input. Except for piano performance, a recent work proposed a DNN-based expressive drum modeling system \cite{145,146}. A bottleneck of performance generation using deep learning is the lack of datasets \cite{144}. The datasets should consist of the scores and the corresponding performance of human musicians so as to render the expressive performance from music scores, and the scores and performance pairs need to be aligned at the note level for effective model training.

Audio synthesis has made great breakthroughs in recent years. In 2016, DeepMind released WaveNet \cite{149} for creating original waveforms of speech and music sample by sample. Since WaveNet was proposed, there have been many follow-up studies. In order to speed up the generation process of WaveNet and improve its parallel generation ability, DeepMind successively released parallel WaveNet \cite{153} and WaveRNN \cite{172}. Engel et al. \cite{152} proposed a powerful WaveNet-based autoencoder model that conditions an autoregressive decoder on temporal codes learned from the raw audio waveform. In February 2017, a team from Montreal published SamplerRNN \cite{151} for sample-by-sample generation of audio exploiting a set of recurrent networks in a hierarchical structure. Later, with the great success of GAN in image synthesis, many research works explored the application of GAN in audio synthesis. Donahue et al. \cite{169,170} first tried to apply GANs to raw audio synthesis in an unsupervised environment. Engel et al. \cite{191} proposed GANSynth, which uses GAN to generate high-fidelity and locally-coherent audio by modeling log magnitudes and instantaneous frequencies with sufficient frequency resolution in the spectral domain. In 2016, Nishimura et al. \cite{273} first proposed a DNN-based singing voice synthesis model, and then Hono et al. \cite{205} introduced GAN into the DNN-based singing voice synthesis system. Juntae et al. \cite{177}, Soonbeom et al. \cite{209} and Juheon et al. \cite{179} proposed Korean singing voice synthesis systems based on LSTM-RNN, GAN and encoder-decoder architectures respectively. Recently, Yu et al. \cite{180} put forward a Chinese singing voice synthesis system ByteSing based on duration allocated Tacotron-like acoustic models and WaveRNN neural vocoders. Lu et al. \cite{183} proposed a high-quality singing synthesis system employing an integrated network for spectrum, F0 and duration modeling. Prafulla et al. \cite{211} proposed a model named Jukebox that generates music with singing in the raw audio domain.

Computer composition and computer-aided composition have been very active fields of commercial software development with a large number of applications such as AI-assisted composition systems, improvisation software\footnote[1]{http://jukedeck.com}, etc. Many commercial software systems have emerged in the last few years, such as the Iamus\footnote[2]{http://melomics.com/} system that can create professional musical works. Some of its works have even been performed by some musicians (as the London Symphony Orchestra). GarageBand\footnote[3]{http://www.apple.com/mac/garageband/}, a computer-aided composition software provided by Apple, supports a large number of music clips and synthetic instrument samples, users can easily create music through combination. PG Music's Band-in-a-Box\footnote[4]{https://www.pgmusic.com/} and Technimo's iReal Pro\footnote[5]{https://irealpro.com/} can generate multi-instrument music based on user-specified chord scales. However, these products only support a limited and fixed number of preset styles and rule-based modifications. Google's Magenta project \cite{281} and Sony CSL's Flow Machine project \cite{278} are two excellent research projects focusing on music generation. They have developed multiple AI-based tools to assist musicians to be more creative \cite{238}, e.g. Melody Mixer, Beat Blender, Flow Composer, etc.

Over the past few years, there have also been several startups that use AI to create music. Jukhedeck\footnotemark[1] and Amber\footnote[6]{http://www.ampermusic.com/} music focus on creating royalty-free soundtracks for content creators such as video producers. Hexachords' Orb Composer\footnote[7]{http://www.orb-composer.com/} provides a more complex environment for creating music, not just for short videos, it is more suitable for music professionals. This program does not replace composers, but helps the composers in an intelligent way by providing adjustable musical ideas. Aiva Technologies designed and developed AIVA\footnote[8]{http://www.aiva.ai} that creates music for movies, businesses, games and TV shows. Recently, Aiva acquired the worldwide status of Composer in SACEM music society, a feat that many artists thought impossible to achieve for at least another decade. These are just a few of the many startups venturing into this uncharted territory, but what all of them have in common is to encourage cooperation between human and machines, to make non-musicians creative and make musicians more effective.
\section{Representation}
\label{sec:3}

Music representation is usually divided into two categories: symbol domain and audio domain. Generally, symbol representations are discrete variables, and audio representations are continuous variables. The nuances abstracted from the symbolic representations are quite important in music, and greatly affect people's enjoyment of music \cite{168}. A common example is that the timing and dynamics of notes performed by musicians are not completely consistent with the score. Symbolic representations are usually customized for specific instruments, which reduces their versatility and means that plenty of work is required to adapt existing modeling techniques to new instruments. Although the digital representation of audio waveforms is still lossy to some extent, they retain all music-related information and possess rich acoustic details, such as timbre, articulation, etc. Moreover, the audio models are more universal and can be applied to any group of musical instruments or non-musical audio signals. At this time, all the symbol abstraction and precise performance control information hidden in the audio digital signal are no longer obvious.
\subsection{Symbolic}
\label{sec:3.1}
Symbolic music representations are usually discrete and include musical concepts such as pitch, duration, chords and so on.
\subsubsection{1D representation}~{}\\
\label{sec:3.1.1}
By sorting the time steps and pitch sequences of note events, music score is often processed as 1D sequence data, although the 1D sequence discards many aspects of the relationship between notes, e.g. it is difficult to indicate that the previous note is still continuing at the beginning of the next note \cite{135}.\\

1) Event-based \\
\begin{table}[H]
	\normalsize
	\setlength{\abovecaptionskip}{5pt}
%	\setlength{\belowcaptionskip}{0pt}
	\centering
	\caption{Common events}
	\label{tab:111}       % Give a unique label
	\begin{tabular}{|m{50 pt}|m{60 pt}|m{150 pt}|}
		\hline
		\textbf{Task} & \textbf{Event} & \textbf{Description}  \\ \hline
		\multirow[c]{10}{*}{Score} & Note On & One for each pitch \\ \cline{2-3}
		& Note Off & One for each pitch \\ \cline{2-3}
		& Note Duration & Inferred from the time gap between a Note On and the corresponding Note Off,by accumulating the Time Shift events in between \\ \cline{2-3}
		& Time Shift & Shift the current time forward by the corresponding number of quantized time steps \\ \cline{2-3}
		& Position & Points to different discrete locations in a bar \\ \cline{2-3}        
		& Bar & Marks the bar lines \\ \cline{2-3}                                                                                             
		& Piece Start & Marks the start of a piece                                                                                                     \\ \cline{2-3}
		& Chord & One for each chord  \\ \cline{2-3}                                                                                                        
		& Program Select/Instrument & Set the MIDI program number at the beginning of each track \\ \cline{2-3}                                                                   
		& Track & Used to mark each track. \\ \hline
		\multirow[c]{2}{*}{Performance} & Note Velocity & MIDI velocity quantized into m bins. This event sets the   velocity for subsequent note-on events \\ \cline{2-3}
		& Tempo & Account for local changes in tempo(BPM) \\ \hline
	\end{tabular}
\end{table}
Musical instrument digital interface (MIDI) \cite{279} is an industry standard that describes the interoperability protocol between various electronic instruments, software and devices, which is used to connect products from different companies, including digital instruments, computers, iPad, smart phones and so on. Musicians, DJs and producers all over the world use MIDI every day to create, perform, learn and share music and art works. MIDI carries the performance data and control data information of specific real-time notes. The two most crucial event information are Note-on and Note-off. Therefore, many researches use MIDI-like events to represent music, we call them event-based music representation. By defining several music events, this method represents music with a sequence of events that progress over time. Generally speaking, it is sufficient to use only Note-on, Note-off events and pitch for music score encoding, while performance encoding uses more events such as velocity and tempo. Table \ref{tab:111} lists some common events in a variety of studies \cite{64,110,111,258}.

The numbers of Note-on and Note-off events are usually 128, corresponding to 128 pitches in MIDI. The number of Program Select/Instrument events is 129, corresponding to 129 MIDI instruments. The number of Time Shift is determined by the beat resolution, or we can fix it to an absolute size that is often used in performance generation, such as the 10ms used in \cite{141}. Velocity event is often quantified into a certain number of bins to determine the velocity value of subsequent notes. MIDI-like representation is very effective for capturing pitch values of notes, but it exists inherent limitations in modeling music rhythm structure \cite{110,111}. When human compose music, they tend to organize recurring patterns regularly in the prosodic structure defined by bars, beats and sub-beats \cite{280}. This structure is clearly indicated in music score or MIDI file with time signatures and vertical bar lines. While in event representation, this information is not obvious. Moreover, it is non-trivial to produce music with stable rhythm using Time Shift events. During addition or deletion of notes, often numerous Time Shift tokens must be merged or split with the Note-on or Note-off tokens being changed all-together, which has caused the model being trained inefficient for the potential generation tasks \cite{292}. It is difficult for sequence model to learn that Note-on and Note-off must appear in pairs, thus generating many hanging Note-on events without corresponding Note-off events. In essence, MIDI is designed as a protocol to convey digital performance rather than digital score. Therefore, MIDI cannot explicitly express the concepts of quarter note, eighth notes, or rests, but can only represent a certain duration of note playing or absence.

To solve the above-mentioned problems of MIDI-like representation, Huang et al. \cite{110,111} proposed REMI (REvamped MIDI-derived events). Specifically, they use Note Duration event instead of Note-off event to promote the modeling of note rhythm, use the combination of Bar and Position to replace Time Shift event, where Bar indicates the beginning of a bar, Position reveals certain positions in a bar, and the combination of Bar and Position provides a clear metric grid to simulate music. In order to simulate the expressive rhythm freedom in music, they also introduced a set of Tempo events to allow local rhythm changes of each beat. In the light of REMI's time grid, each Tempo event is preceded by a Position event.
\begin{comment}
\begin{figure}[H]
%	\centering
\includegraphics[width=0.9\textwidth]{img//3.png}
\caption{Tuple representation}
\label{fig:333}       % Give a unique label
\end{figure}
\end{comment}

In some other studies, a note event is represented as a tuple. In BachBot \cite{51}, each tuple only represents a certain moment of a part. For Bach chorales dataset, each time frame contains four tuples <pitch, tie> representing four parts respectively. Pitch represents MIDI pitch, and tie distinguishes whether a note is a continuation of the same pitch note in the previous frame. The notes in a frame are arranged in pitch descending order ignoring the intersecting parts. Consecutive frames are separated by a unique delimiter `$\mid\mid\mid$', and `(.)' represents the fermatas. Each score contains a unique START and END symbol to indicate the beginning and end of the score respectively. BandNet \cite{37} draws on the representation of BachBot \cite{51} and adopts the same way of score scanning, that is, Z-scan from left to right (time dimension) and top to bottom (channel dimension)\iffalse, as shown in Figure \ref{fig:333}\fi. Similarly, Hawthorne et al. \cite{137} represented each note as a tuple containing note attributes for modeling piano performance, and called it NoteTuple, where the note attributes include pitch, velocity, duration, and time offset from the previous note in the sequence. Compared with the performance representation proposed by Oore et al. \cite{139}, which serializes all notes and their attributes into a sequence of events, NoteTuple shortens the length of music representation without interleaving notes. In addition, BachProp \cite{251}, PianoTree VAE \cite{292} and other studies \cite{219} all represents notes as tuples.\\

2) Sequence-based \\

Sequence-based representation exploits several sequences to represent pitch, duration, chord and other music information respectively. Different sequences of the same music segment have the same length which is equal to the number of notes or the product of the beat number and the beat resolution. Each element in the sequence represents the corresponding information of a note. The encoding way adopted by Dong et al. \cite{32} represented music pieces as sequences of pitch, duration, chord and bar position. The pitch sequence contains the pitch range and a rest; the duration sequence contains all types of note duration in music; the root note of a chord is marked with a single note name in the chord sequence, and the type of chord is represented by a 12-dimensional binary vector; the bar position represents the relative position of the note in a bar, and its value is related to the beat resolution\iffalse, as shown in Figure \ref{fig:444}\fi. Similarly, \cite{72} represents a lead sheet as three sequences of pitch, rhythm and chord, and stipulates that only one note is played at each time step; \cite{257} represents a bar as two sequences with equal length, where the pitch sequence contains all the pitches and uses ``·" as padding, and the rhythm sequence replaces all pitches with the symbol ``O", and uses ``\_" to depict the continuation of a note.
\begin{comment}
\begin{figure}[H]
%	\centering
\includegraphics[width=0.5\textwidth]{img//4.png}
\caption{Sequence representation (1)}
\label{fig:444}       % Give a unique label
\end{figure}
\end{comment}

DeepBach \cite{26} uses multiple lists to represent Bach's four parts chorales, including the lists of melody (pitch), rhythm and fermatas. The number of melody lists is consistent with the number of parts in music, and an additional symbol ``-" is introduced into the melody list to indicate whether to maintain the pitch of the previous note. The rhythm list implies the size of the beat resolution and determines the length of all lists. Anticipation-RNN \cite{22} also uses this encoding method, the difference is that it uses the real note name instead of the pitch value to represent the note. In DeepBach, the time is quantized into 16th notes, which makes it impossible to encode triplets. While dividing 16th notes into 3 parts will increase the sequence length by three times, making the task of sequence modeling tougher. Therefore, Ashis et al. \cite{70} proposed to divide each beat into six uneven ticks to realize the coding of triplets while only increasing the sequence length by 1.5 times. Most previous work \cite{25,27} decomposes music scores by discretizing time, that is, setting a fixed beat resolution for a beat. Although discrete decomposition is very popular, it is still a computational challenge for music with complex rhythms. In order to generate polyphonic music with complicated rhythm, John et al. \cite{36} proposed a new method to operationalize scores that each part is represented as a time series. They regard music as a series of instructions: start playing C, start playing E, stop playing C, etc. Operationalized run-length encodings greatly reduce the computational costs of learning and generation, at the expense of segmenting a score into highly non-linear segments. This encoding method also interweaves the events of different parts. \iffalse As shown in Figure \ref{fig:555}, each column on the left represents a part, and the number of elements in each column is the sum of the number of time steps with notes playing. The left side of the colon represents the duration, the right side indicates the pitch value, and ``*" implies the note continuation.\fi
\begin{comment}
\begin{figure}[H]
%	\centering
\includegraphics[width=0.9\textwidth]{img//5.png}
\caption{Sequence representation (2)}
\label{fig:555}       % Give a unique label
\end{figure}
\end{comment}

In particular, different studies have proposed different methods for the encoding of note continuation in the pitch sequences. A common method is to use special symbols, such as the underline used in \cite{26,257}. \cite{100} employs the ``hold" symbol to represent note continuation, and there are two ways for representation, one uses a ``hold" symbol for all pitches, called ``single-hold", and the other uses a separate ``hold" symbol for each pitch called ``multi-hold". For example, when the beat resolution is 4, the quarter note C4 can be encoded as [C4, hold, hold, hold] or [C4, C4\_hold, C4\_hold, C4\_hold]. On the contrary, another method is to indicate the note continuation by marking the beginning position of notes, e.g. each element of the note sequence in \cite{94} is encoded as a combination of MIDI pitch number and note start mark, and the quarter note C4 is represented as [60\_1,60\_0,60\_0,60\_0].
\subsubsection{2D representation}~{}\\
\label{sec:3.1.2}
Another music representation method draws the notes played over time as a 2D matrix by sampling time. However, compared with the event-based note representation, the sampling-based representation requires a higher dimension. If the music rhythm becomes complex, the time dimension will increase as the required time resolution expands, and the high dimension of time may hinder the model from learning the long-term structure of music.\\

1) Pianoroll \\

The pianoroll representation of a music piece is inspired by the automatic piano. There will be a continuous roll of perforated paper in the automatic piano, and each perforation represents a note control information. The position and length of the perforation determine which note of the piano is activated and how long it lasts. All the perforations are combined to produce a complete music performance \cite{68}. Note events in MIDI can be represented as a $h*w$ real-valued matrix, where h indicates the pitch range of notes, usually adding two additional bits to represent rest and note continuation respectively. E.g. if the pitch range of MIDI is 0-127, then h is equal to 130. The number of time steps per bar depends on the beat resolution. Multi-track music can be represented by multiple matrices. With the help of pretty\_midi and Pypianoroll Python toolkits, MIDI files can be easily converted into pianoroll representations.
\begin{figure}[H]
	%	\centering
	\normalsize
	\setlength{\belowcaptionskip}{0pt}
	\setlength{\abovecaptionskip}{5pt}
	\includegraphics[width=0.5\textwidth]{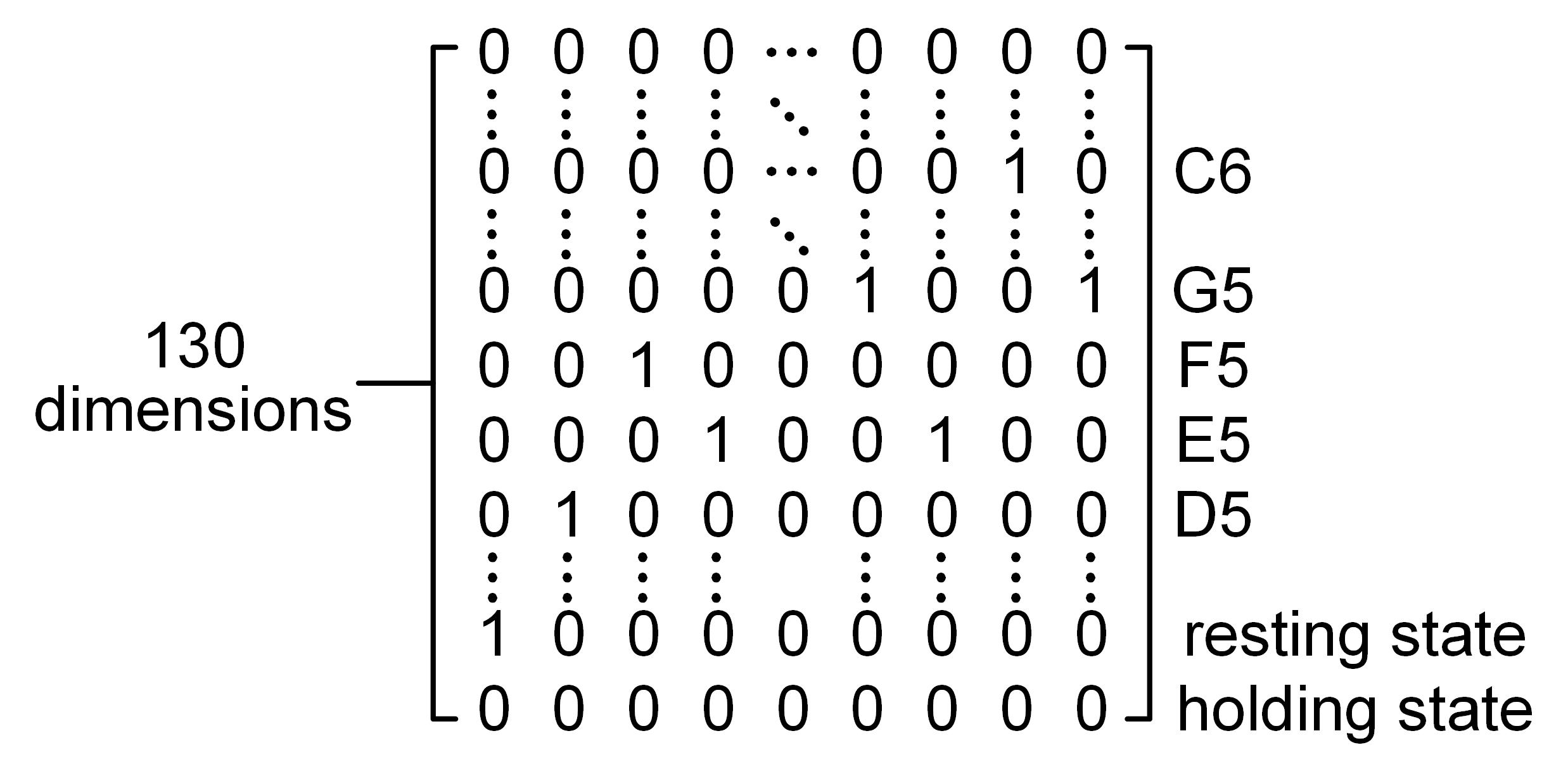}
	\caption{pianoroll representation}
	\label{fig:666}       % Give a unique label
\end{figure}
In practical use, pianoroll is usually binary \cite{65,66,69}, as shown in Figure \ref{fig:666}. Each position in the matrix only indicates whether the note is played or not, thus we call it binary pianoroll(see Figure \ref{fig:666}). But sometimes its value also represents other meanings. For example, in DeepJ \cite{27}, a velocity value is placed at the playing position of a note, and the velocity range is between 0 and 1. In \cite{291}, the playing position records the duration of the note if there is a note onset, and zero otherwise. Even in several representation methods, each position in the 2D matrix stores a non-constant value, e.g. a list that contains information about notes. In \cite{67}, the matrix position $(t,n)$ stores the note of the $nth$ pitch in timestep $t$, each note contains three attributes: play, artificial and dynamic. Play and articulate are binary values that indicate whether the note is playing and articulating; dynamic is a continuous value from 0 to 1\iffalse, as shown in Figure \ref{fig:666}\fi. \cite{142} employs a 2D vectors at each position of the matrix to represent the events of note on, note off and note sustained. Pianoroll expresses music score as a 2D matrix that can be regarded as an image, so some people call it image-like representation \cite{110,111}. Mateusz et al. \cite{68} proposed to translate the MIDI data into graphical images in a pianoroll format suitable for the DCGAN, using the RGB channels as additional information carriers for improved performance.

Drawing on the idea of pianoroll representing notes, chords and drums can also be represented in a similar way, except that the pitch dimension changes to indicate chord types or different components of drums. The time dimension is still quantified by the beat resolution. For example, in \cite{32} and \cite{69}, all chords are mapped to major triads and minor triads with 12 semitones as the root, and then each chord is represented by a 24 dimensional one-hot vector, and a dimension is added to represent the absence of chords (NC); another simpler representation method is to represent chords with 12 dimensional multi-hot vectors directly, where each dimension indicates a semitones pitch class \cite{39}. In \cite{63}, multi-hot vectors are used to represent the nine components of drum: kick, snare, open hi hats, etc. If 100000000 and 01000000 denote kick and snare respectively, the simultaneous playing of kick and snare is 11000000.
\begin{comment}
\begin{figure}[H]
%	\centering
\includegraphics*[width=0.4\textwidth]{img//7-1.png}\\
\includegraphics[width=0.8\textwidth]{img//7-2.png}
\caption{Conlon pianoroll representation}
\label{fig:777}       % Give a unique label
\end{figure}
\end{comment}

Pianoroll is easy to understand and has been used widely. However, compared with MIDI representation, pianoroll representation has no note-off information, so it is unable to distinguish between long notes and continuous repeated short notes (as a quarter note and two eighth notes) \cite{109}. According to \cite{80}, there are several ways to solve this problem: 1) introduce hold/replay as the dual representation of notes \cite{27}; 2) divide the time step into two, one marking the beginning of the note and the other marking the end \cite{84}; 3) use the special hold symbol `\_’ to replace the note continuation \cite{26}, which is only applicable to monophonic melody. In addition, pianoroll representation quantifies time with a specific beat resolution, completely eliminating the possibility of the model learning complicated rhythms and expressive timings \cite{238}. Large temporal resolution results in a long sequence length, leading to the lack of long-term consistency of the generated music. Moreover, aiming at the problem that long notes cannot be distinguished from continuous short notes, Angioloni et al. \cite{352} proposed a new pianoroll representation, Conlon pianololl $({PR}^c)$, which can explicitly represent the duration\iffalse, as shown in Figure \ref{fig:777}\fi. Only two note event types are considered here, $ON(t,n,v)$ and $OFF(t,n)$, where $t$ represents time, $n$ represents pitch, and $v$ represents velocity. They exploited a tensor of size $T\times N\times2$ to represent the speed channel and the duration channel. In the first channel, if the event $ON(t,n,v)$ occurs in $(t,n)$, then $x_{q(t),n,1}=v$ , otherwise $x_{q(t),n,1}=v$; in the second channel, if the event $ON(t,n,v)$ occurs in $(t,n)$, then $x_{q(t),n,2}=d$, otherwise $x_{q(t),n,2}=0$. In addition to distinguishing long notes from continuous short notes, another advantage of the Conlon pianoroll is that all the information about a note is local, so that the convolution network does not need a large receptive field to infer the note duration. However, the $({PR}^c)$ representation still inevitably needs to quantify the time.\\

2) Others \\

Apart from pianoroll, some studies have adopted other 2D representations. These representations have not been used widely, but they may bring us some inspiration for music representations. In order to integrate the existing domain knowledge into the network, Chuan et al. \cite{113} proposed a novel graphical representation incorporating domain knowledge. Specifically, inspired by tonnetz in music theory, they transform music into a sequence of 2D tonnetz matrices, and graphically encode the musical relationship between pitches. Each node in the tonnetz network represents one of the 12 pitch classes. The nodes on the same horizontal line follow the circle-of-fifth ordering: the adjacent right neighbor is the perfect-fifth and the adjacent left is the perfect-fourth. Three nodes connected as a triangle in the network form a triad, and the two triangles connected in the vertical direction by sharing a baseline are the parallel major and minor triads. Note that the size of the network can be expanded boundlessly; therefore, a pitch class can appear in multiple places throughout the network. Compared with the sequences generated by pianoroll models, the music generated by tonnetz has higher pitch stability and more repetitive patterns.

Walder \cite{109} proposed a novel representation to reduces polyphonic music to a univariate categorical sequence, permitting arbitrary rhythm structures and not limited by the unified discretization of time in pianoroll. Specifically, a music fragment is unrolled into an input matrix, a target row and a lower bound row. Each time step a column of the input is sent into the network to predict the value in the target row, which is the MIDI number of the next note to be played. The lower bound is due to the ordering by pitch — notes with simultaneous onsets are ordered such that we can bound the value we are predicting. Apart from MIDI pitch, the matrix has several dimensions to indicate the rhythm structure: $\Delta t_{event}$ is the duration of the note being predicted, $\Delta t_{step}$ is the time since the previous input column, $\epsilon_{onset}$ is 1 if and only if we are predicting at the same time as in the previous column, $\epsilon_{offset}$ is 1 if and only if we are turning notes off at the same time as in the previous column, and t represents the time in the segment corresponding to the current column. This representation allows arbitrary time information and is not limited to a unified discretization of time. A major problem of the uniform discretization time is that in order to represent even moderate complex music, the grid needs to be prohibitively fine grained, which makes model learning difficult. Moreover, unlike the pianoroll representation, this method explicitly represents onsets and offsets, and is able to distinguish two eighth notes with the same pitch and a quarter note.
\subsubsection{Others}~{}\\
\label{sec:3.1.3}
There are some representations that neither belong to the category of 1D representations, nor can they be divided into 2D representations. Here we introduce three additional special representations: text, word2cvec, and graph.\\

1) Text \\

The most common text representation in music is the ABC notation. Encoding music with ABC notation consists of two parts: header and body. The first header is the reference number, when there are two or more pieces on each page, some programs that convert the code into music require each piece to have a separate number, while other programs only allow one reference number. The other headers are title T, time signature M, default note length L, key K, etc. The body mainly includes notes, bar lines and so on, each note is encoded into a token, the pitch class of the note is encoded into its corresponding English letter, with additional octave symbols (e.g. `a’ for two octaves) and duration symbol (e.g. `A2’ for double duration), the rest is represented by `z’, and the bars are separated by `$\mid$’. ABC notation is quite compact, but it can only represent monophonic melody. For a more detailed description of ABC, see \cite{282}. Sturm et al. \cite{86} convert ABC text into token vocabulary text, each token is composed of one or more characters for the following seven types (with examples in parens): meter ``M: 3/4”), key (``K: Cmaj”), measure (``:$\mid$” and ``$\mid$1”), pitch (``C” and ``\^{}C’ ”), grouping (``(3”), duration (``2” and ``/2”), transcription (``$<s>$” and ``$</s>$”). \iffalse Turning the music shown in Figure \ref{fig:888}(a) to take C as the root note produces the token representation in Figure \ref{fig:888}(b).\fi\\
\begin{comment}
\begin{figure}
\subfigure[]{
\includegraphics[width=0.7\textwidth]{img//8-1.png}
}
\subfigure[]{
\includegraphics[width=0.9\textwidth]{img//8-2.png}
}
\caption{ABC notation}
\label{fig:888}       % Give a unique label
\end{figure}
\end{comment}

2) Word2vec \\

Word2vec refers to a group of models developed by Mikolov et al \cite{283}. They are used to create and train semantic vector spaces, often consisting of several hundred dimensions, based on a corpus of text \cite{284}. In this vector space, each word in the corpus is represented as a vector, and the words sharing a context are close to each other geographically. Distributional semantic vector space models have become important modeling tools in computational linguistics and natural language processing (NLP) \cite{285}. Although music is different from language, they share multiple characteristics. Therefore, some researchers have explored the use of word2vec model to learn music vector embedding, and the learned embedding can be employed as the potential input representation of deep learning model.

Chuan et al. \cite{106} implemented a skip-gram model with negative sampling to construct semantic vector space for complex polyphonic music fragments. In this newly learned vector space, a metric based on cosine distance is able to distinguish between functional chord relationships, as well as harmonic associations in the music. Evidence based on the cosine distance between the chord-pair vector shows that there are implicit circle-of-fifths exists in the vector space. In addition, a comparison between pieces in different keys reveals that key relationships are represented in word2vec space. These results show that the newly learned embedded vector representation captures the tonal and harmonic features of music even without receiving clear information about the music content of the segment. What’s more, Madjiheurem et al. \cite{107} proposed three NLP inspired models called chord2vec to learn vector representation of chords. Huang et al. \cite{108} also used word2vec models to learn chord embedding from chord sequence corpus.\\

3) Graph \\

Dasaem et al. \cite{135} represented music score as a unique form of graph for the first time, that is, each input score is represented as a graph $G=(V,E)$, $V$ and $E$ represent nodes and edges respectively. Each note is regarded as a node in the graph, and adjacent notes are connected to different types of edges according to their musical relationship in the score. They defined six types of edges in the score: next, rest, set, sustain, voice, and slur\iffalse, as shown in Figure \ref{fig:999}\fi. The next edge connects a note to its following note; the rest edge links a note with the rest following to other notes that begin when the rest ends. If there are consecutive pauses, they are combined as a single rest; the onset edge is to connect notes beginning together; the notes appearing between the start and the end of a note are connected by sustain edge; the voice edge is a subset of next edge which connects notes in the same voice only. Among voice edges, they add slur edges between notes of the same slur. All edges are directed except onset edges. They consider forward and backward directions as different types of edges, and add a self-connection to every note. Therefore, the total number of edge types is 12.
\begin{comment}
\begin{figure}[H]
%	\centering
\includegraphics[width=0.6\textwidth]{img//9.png}
\caption{Graph representation}
\label{fig:999}       % Give a unique label
\end{figure}
\end{comment}
\subsection{Audio}
\label{sec:3.2}
Audio representations are continuous, which pay more attention to purely continuous and rich acoustic information, such as timbre. Similarly, music representation in audio domain can also be divided into 1D and 2D representation, where 1D representation mainly corresponds to time-varying waveform and 2D representation corresponds to various spectrograms.
\subsubsection{1D representation}
\begin{figure}[H]
	%	\centering
	\normalsize
	\setlength{\belowcaptionskip}{0pt}
	\setlength{\abovecaptionskip}{5pt}
	\includegraphics[width=0.5\textwidth]{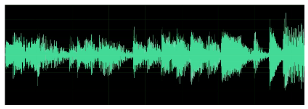}
	\caption{Waveform}
	\label{fig:1010}       % Give a unique label
\end{figure}

~{}\\
\label{sec:3.2.1}
Time domain exists in the real world, its horizontal axis is time and vertical axis is changing signals. Waveform is the most direct representation of original audio signals. It is lossy and can be easily converted into actual sound. Audio waveforms are one-dimensional signals changing with time. An audio fragment at a particular timestep has a smooth transition from audio fragments from previous timesteps \cite{148}. As shown in Figure \ref{fig:1010}, the x-axis represents time and the y-axis represents the amplitude of the signal. WaveNet \cite{149}, SampleRNN \cite{151}, NSynth \cite{152} and other research work all employ the original waveforms as model inputs.
\subsubsection{2D representation}~{}\\
Frequency domain does not exist actually, but is a mathematical structure. Its horizontal axis is the frequency and vertical axis is the amplitude of frequency signals. The time domain signals can be transformed into spectra by Fourier transform (FT) or Short-Time Fourier transform (STFT). The spectrum is composed of amplitude and phase that is a lossy representation of their corresponding time domain signals.\\

1) Spectrogram\\

Spectrogram \cite{155} is a 2D image of a spectrum sequence, in which time is along one axis, frequency is along another axis, and brightness or color represents the intensity of frequency components of each frame. Compared with most of the traditional manual features used in audio analysis, spectrogram retains more information and has lower dimension than the original audio. Spectrogram representation implies that some CNNs for images can be directly applied to sound.
\begin{figure}[H]
	%	\centering
	\normalsize
	\setlength{\belowcaptionskip}{0pt}
	\setlength{\abovecaptionskip}{5pt}
	\includegraphics[width=0.5\textwidth]{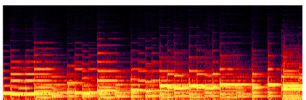}
	\caption{Spectrogram}
	\label{fig:1111}       % Give a unique label
\end{figure}
2) Magnitude spectrogram \\

As for the amplitude spectrum \cite{155}, its horizontal axis is frequency, and the vertical axis is the amplitude of each frequency component. Amplitude spectrum can also be applied to audio generation, but the process of reconstructing the audio signal requires the technique of deriving phases from the characteristics of amplitude spectra. The most common phase reconstruction technology is the algorithm proposed by Griffin-Lim \cite{301}. However, it involves multiple iterations of forward and inverse STFT, and is basically not real-time, that is, the reconstruction of each timestep requires the entire time range of the signal. What’s more, local minimal of the error surface sometimes impedes high-quality reconstruction.\\

3) Mel spectrogram \\

The Mel transform provides a mapping from the actual measured frequency to the perceived Mel center frequency, which is used to transform the log amplitude STFT spectrogram to Mel spectrogram \cite{158}. Mel spectrogram reduces the spectral resolution of STFT in a perceptual consistent way, that is, some information will be lost. It is suitable for neural networks trained on large music audio libraries. The Mel Frequency Cepstrum Coefficient (MFCC) can be obtained by cepstrum analysis (take logarithm, then do DCT transform) on Mel spectrogram.\\

4) Constant-Q Transform spectrogram \\

Constant-Q Transform (CQT) refers to the filter bank whose center frequency is exponentially distributed with different filter bandwidth, but the ratio of center frequency to bandwidth is a constant Q. Different from the Fourier transform, the frequency of its spectrum is not linear, but based on log2, and the length of filter window can be changed according to different spectral line frequency to obtain better performance. Since the distributions of CQT and scale frequency are same, the amplitude value of music signal at each note frequency can be directly obtained via calculating the CQT spectrum of music signal, which directly simplifies the signal processing of music.

Although STFT and Mel spectrograms can represent short-term and long-term music statistics pretty well, neither of them can represent music related frequency patterns (as chords and intervals) that can move linearly along the frequency axis. The CQT spectrogram can express the pitch transposition as a simple shift along the frequency axis. Albeit in some aspects, CQT spectrogram is inferior to Mel spectrogram in terms of perception and reconstruction quality due to lower frequency scaling, it is still the most natural representation for indicating the joint time-frequency spatial features utilizing two-dimensional convolution kernel \cite{158}.
\section{Multi-level Deep Music Generation}
\label{sec:4}
Music representation has the inherent characteristics of multi-level and multi-modal: the high-level is the score representation, including the structure and symbolic abstract features (as pitch, chord); the bottom-level is the sound representation, which contains purely continuous and rich acoustic characteristics (as timbre); the middle-level is the performance representation, consisting of rich and detailed timing and dynamic information, which determines the musical expression of performance. Corresponding to different levels of music representation, music generation can also be divided into three levels: the high level is score generation, the middle level is performance generation, and the bottom level is audio generation. This section reviews the previous work for deep music generation by dividing relevant researches into the above three levels, and further subdividing the generation tasks under each level. For the same kind of generating tasks, such as melody generation and polyphonic music generation, we introduce various deep learning methods used in previous studies in chronological order, and conduct comparative analyses of different methods. With respect to the concrete deep network architecture theory (such as RNN, CNN, etc.), we only give a brief introduction, but no detailed explanation, for it is not this paper’s focus.
\subsection{Top-level: Score Generation}
Music score is a way of recording music with symbols. The algorithmic composition mentioned earlier belongs to the category of score generation. It usually deals with symbolic representation and encodes abstract music features, including tonality, chord, pitch, duration and rich structure information such as phases and repetitions. Music generated by score generation must be converted to MP3, WAV, etc. file formats to be finally heard. In the process of transforming visual score to audio, it is necessary to add expressive information (as dynamics) and acoustic information (as instruments), corresponding to the music characteristics to be encoded in the middle and bottom layers respectively. Previous score generation methods include rule-based method \cite{261}, probability model \cite{262} and neural network \cite{263}. In recent years, there have been numerous researches using deep learning technology to create music automatically, and the generation tasks have gradually become diverse. Here, we divide these studies into two categories. The first category is called creating music from scratch, which is also the first attempt of automatic music generation; the second is called conditional generation, which means that the generation process will be accompanied by some conditional constraints, such as generating melody given chord. This section provides a detailed review of these studies.
\subsubsection{Compose from scratch}~{}
\label{sec:4.1.1}
\paragraph{\large Monophonic/Melody}~{}\\
Melody is a series of notes with same or different pitch, which are organized by specific pitch variation rules and rhythm relations. Melody generation usually uses monophonic data, so at each step, the model only needs to predict the probability of single note to be played in the next time step.
\begin{figure}[]
	%\normalsize
	\setlength{\abovecaptionskip}{10pt}
	\setlength{\belowcaptionskip}{0pt}
	\includegraphics[width=0.65\textwidth]{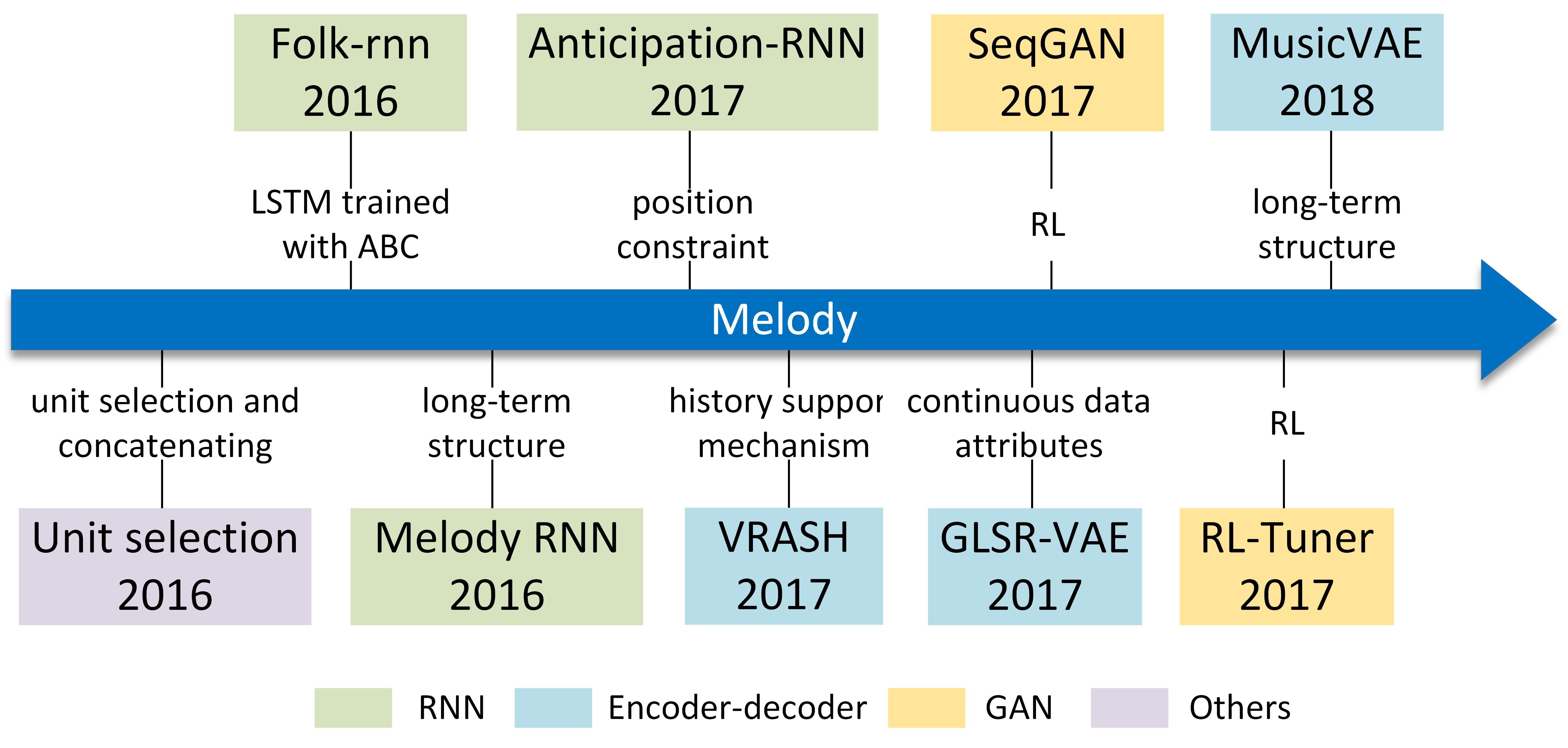}
	\caption{\small Chronology of melody generations. Different colors represent different model architectures or algorithms.}
	\label{fig:1212}       % Give a unique label
\end{figure}

Inspired by the unit selection technology in text-to-speech (TTS) \cite{286}, Bretan et al. \cite{61} proposed a music generation method using unit selection and concatenating. The unit here refers to the variable length music bar. First, develop a deep autoencoder to realize the unit selection and form a finite-size unit library. After that, combine a Deep Structured Semantic Model (DSSM) with an LSTM to form a generative model to predict the next unit. The system only can be utilized to generate monophonic melody, and if there is no good unit, the unit selection may not perform well, but the system is superior to the note-level generation baseline which is based on stacked LSTM.

The most commonly used and simplest model for melody generation is RNNs. Sturm \cite{86} used music transcription represented by ABC to train LSTM to generate music. The training can be character-level (char-RNN) or token-level (which can be more than one character) (folk-RNN). The Melody RNN models \cite{265} put forward by Google Brain’s Magenta project are probably the most famous examples of melody generation in the symbolic domain. It includes a baseline model named basic RNN and two RNN model variants designed to generate longer-term music structure, lookback RNN and attention RNN. Lookback RNN introduces custom input and label, allowing the model to identify patterns that span 1 or 2 bars more easily; attention RNN uses attention mechanism to access previous information without storing it in the RNN unit state. However, RNNs can only generate music sequences from left to right, which makes them far from interactive and creative use. Therefore, Hadjeres et al. \cite{22} proposed a novel RNN model, Anticipation-RNN, which not only has the advantages of RNN-based generation model, but allows the enforcement of user-defined position constraints. Also in order to solve the problem of RNN model sampling from left to right, MCMC-based method has been proposed \cite{26}, but the process of performing user-defined constraints while generating music sequence is almost an order of magnitude longer than the simple left-to-right generation scheme, which is hard to use in real-time settings.

Apart from RNNs, VAE, GAN, etc. generative models have also been used in music composition, and combined with CNN and RNN to produce a variety of variants. In order to solve the problem that the existing recurrent VAE model is difficult to model the sequence with a long-term structure, Roberts et al. \cite{23} proposed the MusicVAE model, which employed a hierarchical decoder to send the latent variables generated by the encoder into the underlying decoder to generate each subsequence. This structure encourages the model to utilize latent variable coding, thus avoiding the ``posterior collapse" problem in VAE, and has better sampling, interpolation and reconstruction performance. Although MusicVAE improved the ability of modeling long-term structure, the model imposed strict constraints on the sequence: non-drum tracks were restricted to monophonic sequences, all tracks were represented by a single velocity and every bar was discretized into 16 timesteps. This is beneficial for modeling long-term structures but at the expense of expressiveness. Although MusicVAE still has many shortcomings, the model provides a powerful foundation for exploring more expressive and complete multi-track latent spaces \cite{64}. Later, Dinculescu et al. \cite{228} trained a smaller VAE on the latent space of MusicVAE called MidiMe to learn the compressed representation of the encoded latent vectors, which allows us to generate samples only from the part of the interested latent space without having to retrain the large model from scratch. The reconstruction or generation quality of MidiMe depends on the pre-trained MusicVAE model. Yamshchikov et al. \cite{60} proposed a new VAE-based monophonic music algorithmic composition architecture, called Variable Recurrent Autoencoder Supported by History (VARSH), which can generate pseudo-live acoustically pleasing and melodically diverse music. Unlike the classic VAE, VRASH took the previous output as an additional input and called it historical input. The historical support mechanism addresses the issue of slow mutual information decline in discrete sequences. Contrary to the MusicVAE \cite{23}, where network generate short loops and then connects them in longer patterns, thus providing a possibility to control melodic variation and regularity, VRASH focuses on the whole-track melody generation. So as to consider the continuous attributes of modeling data during generation, Hadjeres et al. \cite{224} proposed GLSR-VAE architecture to control data embedding in the latent space. First determine the geometric structure of the latent space, and then use the geodesic potential space regularization (GLSR) method to increase the loss of VAE. The variations in the learned latent space reflect the change of data attributes, thus offering the possibility to modulate the attributes of the generated data in a continuous way. GLSR-VAE is the first model specially for continuous data attributes, however, it requires differentiable calculations on data attributes and careful fine-tuning of hyperparameters.

Traditional GAN has limitations in generating discrete tokens. One of the main reasons is that the discrete output of the generator makes it difficult to for the gradient update of the discriminator to be transferred to the generator. The other is that the discriminator can only evaluate an entire sequence. Yu et al. \cite{225} proposed a sequence generation framework SeqGAN to solve these problems, which is the first work to extend GANs to generate discrete token sequences. SeqGAN models the data generator as a stochastic policy in reinforcement learning (RL), and bypasses the generator differentiation problem by directly executing the gradient policy update. The discriminator judges the complete sequence to obtain RL reward signal, and transmits RL reward signal back to intermediate state-action steps using Monte Carlo search. Although SeqGAN has proved its performance on several sequence generation tasks (such as poetry and music), it exists the problem of mode collapse \cite{287}. Later, SentiGAN proposed by Wang et al. \cite{288} alleviated the problem of mode collapse by using penalty target instead of reward-based loss function \cite{219}. Jacques et al. \cite{246} also came up with applying RL to music generation tasks. They proposed a novel sequential learning method for combining ML and RL training called RL tuner, using pre-trained recurrent neural network (RNN) to supply part of the reward value and refining a sequence predictor by optimizing for some imposed reward functions. Specifically, cross-entropy reward was used to augmenting deep Q-learning and a novel off-policy method was derived for RNN from KL control, so that KL divergence can be punished directly from the policy defined by reward RNN. This method mainly depends on the information learned from the data. RL is only used as a way to refine the output characteristics by imposing structural rules. Although SeqGAN obtained improved MSE and BLEU scores on the NMD dataset, it is unclear how these scores match the subjective quality of the samples. On the contrary, RL Turner provided samples and quantitative results demonstrating that the method improved the metrics defined by the reward function. In addition, RL tuner can also explicitly correct the undesirable behavior of RNN, which could be useful in a broad range of applications. After that, Jacques et al. \cite{247} further proposed a general version of the above model called Sequence Tutor for sequence generation tasks, not just music generation.
\paragraph{\large Polyphony}~{}\\
\begin{figure}[]
	%	\centering
	\setlength{\abovecaptionskip}{10pt}
	\setlength{\belowcaptionskip}{0pt}
	\includegraphics[width=0.75\textwidth]{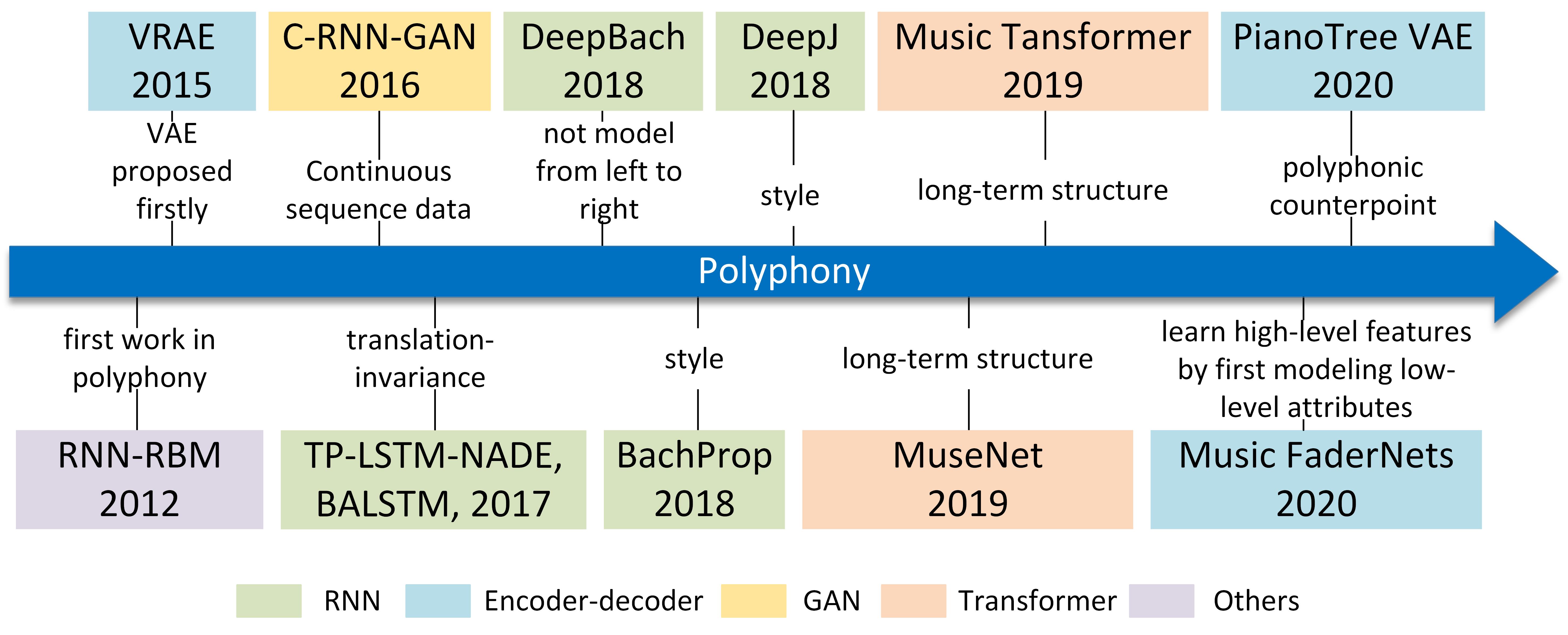}
	\caption{Chronology of polyphonic music generations. Different colors represent different model architectures or algorithms.}
	\label{fig:1313}       % Give a unique label
\end{figure}
Polyphony music is composed of two or more independent melodies, which are combined organically to flow and unfold in a coordinated manner. Polyphonic music has complex patterns along multiple axes: there are both sequential patterns between timesteps and the harmonic intervals between simultaneous notes \cite{67}. Therefore, the generation of polyphony music is more complicated than monophonic music. In each time step, the model needs to predict the probability of any combination of notes played in the next time step.

It may be certain that the first comprehensive consideration of polyphony work is the study of Boulanger-Lewandowski et al. \cite{266}. They proposed RNN-RBM model, which can learn the probability rules of harmony and rhythm from polyphonic music scores with various complexity, and is superior to the traditional polyphonic music model in assorted datasets. Since then, much work on polyphonic modeling has focused on the dataset and coding introduced in \cite{266}. Qi et al. \cite{230,231} proposed LSTM-RTRBM model to generate accurate and flexible polyphonic music. The model combines the ability of LSTM in memorizing and retrieving useful historical information and the advantages of RBM in high-dimensional data modeling. The model embeds long-term memory into the Recurrent Temporal RBM (RTRBM) \cite{296} by increasing a bypassing channel from data source filtered by a recurrent LSTM layer. LATTNER et al. \cite{252} imposed high-level structure on the generated polyphonic music, combining Convolution RBM (C-RBM) as the generation model with multi-objective constrained optimization to further control the generation process. They enforced the attributes such as tonal and music structure as constraints during the sampling process. A randomly initialized sample is alternately updated with Gibbs sampling (GS) and gradient descent (GD) to find a solution that satisfies the constraints and is relatively stable for C-RBM. The results demonstrated that this method can control the high-level self-similar structure, beat and tonal characteristics of the generated music on the premise of maintaining local music coherence.

RNN models is able to generate music in a chronological order, but it does not have transposition-invariance, that is, what the network learns is the absolute relationship between notes, not the relative relationship. Therefore, Johnson \cite{25} proposed two network architectures TP-LSTM-NADE and BALSTM with transposition-invariance, using a set of parallel, tied-weight recurrent networks for prediction and composition of polyphonic music. TP-LSTM-NADE divided RNN-NADE into a group of tied parallel networks, each instance of the network would be responsible for a note, and has associated weight with other network instance, so as to ensure the translation-invariance; BALSTM replaced the NADE portion of the network with LSTMs that have recurrent connections along the note axis, which reduced the disadvantage that the TP-LSTM-NADE network must use windowed and binned summaries of note output. Although the BALSTM model proposed by Johnson possesses translation-invariance, it cannot maintain the consistency of output styles when training with multiple styles, leading to a shift to different music styles in one music piece. Hence Mao et al. \cite{27} proposed the DeepJ model based on the BALSTM architecture, which is different from the BALSTM model in that each layer of the model uses style conditioning. DeepJ model can create polyphony music according to a specific music style or a mixture of multiple composer styles, and can learn music dynamics. Colombo \cite{251} proposed an algorithmic composer called BachProp for generating music of any style. BachProp is a LSTM network with three continuous layers, and the LSTM unit blocks are connected in a feedforward structure. In addition, skip connection is added to the network to facilitate gradient propagation. The model first predicts the timing dT of the current note with respect to the previous note according to the information of the previous note, then predicts the duration T of the current note according to the information of the previous note and the dT of the current note, and finally predicts the pitch according to the information of the previous note and the dT and T of the current note. The evaluation results on distinct datasets showed that BachProp can learn many different music structures from heterogeneous corpora and create new scores based on the extracted structures. Moreover, in order to generate polyphonic music with complicated rhythm, John et al. \cite{36} proposed a novel factorization that decomposes a score into a collection of concurrent, coupled time series: parts. They proposed two network structures to generate polyphonic music: hierarchical architecture and distributed architecture. Considering that polyphonic music has abundant temporal and spatial structure, it may be suitable for weight-sharing scheme. They also explored various weight-sharing ideas. However, this method can only capture the short-term structure of the dataset, and the Markov window in the model makes the model unable to capture the long-term structure in music.

Generative models such as VAE, GAN, etc. have been gradually applied to polyphonic music generation. In 2015, Fabius et al. \cite{254} first proposed the Variational Auto-Encoder (VAE) and applied it to the generation of game songs. Zachary et al. \cite{344} studied the normalized flow in the discrete environment within the VAE framework, in which the flow modeled the continuous representation of discrete data through a priori model, and realized the generation of polyphonic music without autoregressive likelihood. Inspired by models such as MusicVAE and BachProp, Lousseief \cite{242} proposed VAE-based MahlerNet to generate polyphonic music. Tan et al. \cite{322} proposed a novel semi-supervised learning framework named Music FaderNets, by first modeling the corresponding quantifiable low-level attributes, then learning the high-level feature representations with a limited number of data. Through feature disentanglement and latent regularization techniques, low-level attributes can be continuously manipulated by separate ``sliding faders". Each ``fader" independently controls a low-level music feature without affecting other features, and changes linearly along with the controlled attributes of the generated output. Then, Gaussian Mixture Variational Autoencoders (GM-VAEs) is utilized to infer high-level features from low-level representation though semi-supervised clustering. In addition, they applied the framework to style transfer tasks across different arousal states with the help of learnt high-level feature representations. Wang et al. \cite{292} proposed a novel tree-structure extension model called PianoTree VAE, which is suitable for learning polyphonic music. It is first attempt to generate polyphonic counterpoint in the context of music representation learning. The tree structure of music syntax is employed to reflect the hierarchy of music concepts. The whole network architecture can be regarded as a tree, each node represents the embeddings of either a score, simu\_note or note, and the edge is bidirectional. The recurrent module in the edge can either encode the children into the parent or decode the parent to generate its children. Mogren \cite{24} proposed a generative adversarial model named C-RNN-GAN based on continuous sequence data to generate polyphonic music. Different from the symbolic music representation commonly adopted in RNN models, this method trained a highly flexible and expressive model with fully continuous sequence data for tone lengths, frequencies, intensities, and timing. It has been proved that generative adversarial training is a feasible network training way for modeling the distribution of continuous data sequence.

There are also some attempts to apply Transformer-based models for polyphonic music generation. Huang et al. \cite{56} combined relative attention mechanism into Transformer and proposed Music Transformer, which can generate long-term music structures of 2,000 tokens scale, generate continuations that coherently elaborate on a given motif, and generate accompaniments according in a seq2seq setting. It is the first successful application of Transformer in generating music with long-term structure. Payne \cite{220} created MuseNet based on GPT-2, which can generate 4-minute music with 10 different instruments combining various styles. MuseNet has full attention in the context of 4096 tokens, which may be one of the reasons why it can remember the medium and long-term structure of segments. Both Music Transformer and MuseNet are constructed using decoder-only transformers. They are trained to predict the next token at each time step, and are optimized with cross-entropy loss as the objective function. However, this loss function has no direct relationship with the quality of music generated. It is only an indicator for the training process rather than the generated results. Although these models using the teacher forcing training strategy produce some good pieces, generally speaking, the generated music pieces are of poor musicality and the attention learned is messed and of poor structure. Therefore, Zhang \cite{219} proposed a novel adversarial Transformer to generate music pieces with high musicality. By combining generative adversarial learning with self-attention architecture, the generation of long sequences is guided by adversarial objectives to offer a powerful regularization to force transformer to focus on global and local structure learning. In order to accelerate the convergence of training process, adversarial objective is combined with teacher forcing target to guide the generator. Instead of using the time-consuming Monte Carlo (MC) search method commonly used in existing sequence generation models, Zhang proposed an efficient and convenient method to calculate the rewards for each generated step (REGS) for the long sequence. The model can be utilized to generate single-track and multi-track music. Experiments show that the model can generate higher quality long music pieces compared with the original music transformer. Wu et al. \cite{256} also tried for the first time to create jazz with Jazz Transformer based on the Transformer-XL model.

In addition, a dataset frequently used in polyphonic music generation is J.S.Bach four parts chorales dataset (JSB Dataset), which was introduced by Allan and Williams \cite{297}, and has since become a standard benchmark for evaluating the performance of generation models in polyphonic music modeling. Many researches have tried to generate Bach four-part chorales music, including various network variants based on LSTM \cite{231}, GRU \cite{232}. However, these pianoroll-based models are too general to reflect the characteristics of Bach’s chorales. Moreover, they lack flexibility, always generate music from left to right, and cannot interact with users. Hence Hadjeres et al. \cite{26} proposed a LSTM-based model called DeepBach, which does not sample and model each part separately from left to right, and permits users to impose unary constraints, such as rhythms, notes, parts, chords and cadences. Liang et al. \cite{51} also put forward a LSTM-based model named BachBot. This model can generate high-quality chorales almost without music knowledge. But compared with DeepBach, the model is not universal and is of poor flexibility. Additionally, the model generates chorales which are all in C major and can only fix the soprano part. However, BachBot's ancestor sampling method only needs to pass forward once without knowing the number of timestamps in the sample in advance. When DeepBach generates samples, it needs to initialize a predetermined number of timestamps randomly, and then perform MCMC iterations for many times. In order to make the process of machine composition closer to the human way of composing music rather than generating music in a certain chronological order, Huang et al. \cite{52} trained a convolutional neural network Coconet to complete partial music score and explored the use of blocked Gibbs sampling as an analogue to rewriting. The network is an instance of orderless NADE. Peracha \cite{351} proposed a GRU-based model TonicNet, which can predict the chords and the notes of each voice at a given time step. The model can be conditioned on the inherent salient features extracted from the training dataset, or the model can be trained to predict the features as extra components of the modeled sequence. Both DeepBach and Coconet need to set the length of the sample in time-steps in advance, so as to use the orderless sampling method similar to Gibbs sampling. DeepBach further requires fermata information. On the contrary, TonicNet employs ancestor sampling to generate music scores, and predicts continuous markers in a purely autoregressive manner. It does not need any preset information related to length or phrase, which has achieved the most state-of-the-art results on JSB dataset. Yan et al. \cite{47} also proposed a neural language model to model multi-part symbolic music. The model is part-invariant, that is, the structure of the model can explicitly capture the relationship between notes in each part, and all parts of the score share this structure. A simple training model can be exploited to process/generate any part of the score composed of any number of parts. Hadjeres et al. \cite{233} also proposed a representation learning method for generating discrete sequence variants. Given a template sequence, they aim at generating a new sequence with perceptual similarity to the original template without any annotation. They first use Vector Quantized Contrastive Predictive Coding (VQ-CPC) to learn the meaningful allocation of basic units over a discrete set of codes and the mechanism to control the information content of these discrete representations. Then, they utilize these discrete representations in the Transformer architecture to generate variants of the template sequence. They have verified the effectiveness of the above method in symbolic music generation on JSB datasets. In addition, the above-mentioned two models with translation-invariance \cite{25} are also evaluated on JSB dataset.

Orchestral music refers to other types of works performed by orchestra other than concertos and symphonies. An orchestra is usually composed of strings, woodwinds, brass, percussion and other instruments. By learning the inherent regularities existing between piano scores and their orchestrations by well-known composers, Crestel et al. \cite{241} proposed the first system of automatic orchestral arrangement from real-time piano input based on conditional RBM (cRBM) which is called Live Orchestral Piano (LOP), and referred to this operation of extending a piano draft to an orchestral score as projective orchestration. This process is not just about assigning the notes in the piano score to different instruments, but means harmony enhancement and timbre manipulation to emphasize the existing harmonic and rhythmic structure. Specifically, they represented the piano score and orchestral score as two time-aligned state sequences, and used cRBM to predict the current orchestral state given current piano state and the past orchestral states. The evaluation results demonstrated that Factored-Gate cRBM (FGcRBM) had better performance.
\paragraph{\large Multi-track/Multi-instrument music}~{}\\
\newline
Multi-track/multi-instrument music belongs to multi-track polyphony music, which usually consist of multiple tracks/instruments with their own temporal dynamics, and these tracks/instruments are interdependent in terms of time.
\begin{figure}[]
	%	\centering
	\includegraphics[width=0.8\textwidth]{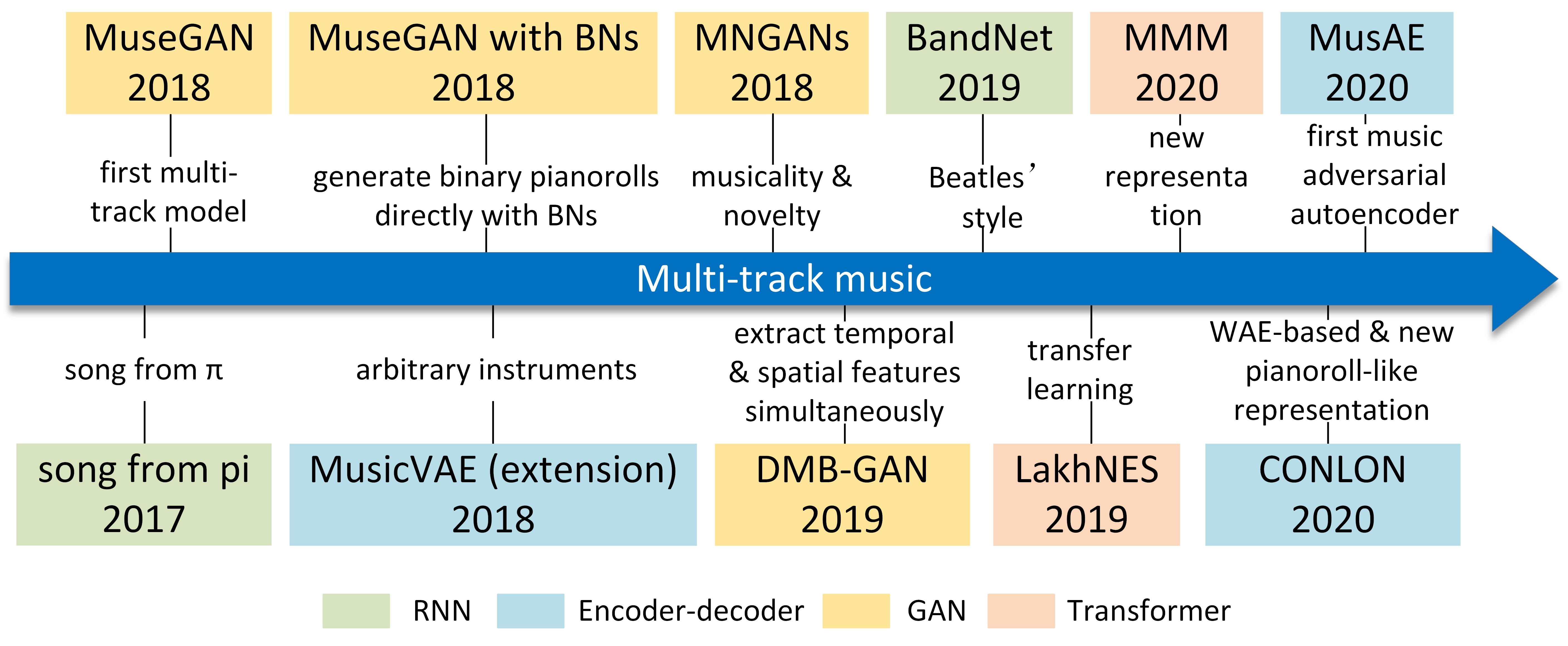}
	\caption{Chronology of multi-track music generations. Different colors represent different model architectures or algorithms.}
	\label{fig:1414}       % Give a unique label
\end{figure}

Chu et al. \cite{58} proposed to use hierarchical RNN to generate pop music, in which the network layer and hierarchical structure encode prior knowledge about how pop music is composed. Each layer of the model produces a key aspect of the song, the bottom layer generates melody and the higher level generates drums and chords. Inspired by song from $\pi$\footnote[9]{https://youtu.be/OMq9he-5HUU}, they condition the model on the scale type, allowing the melody generator to learn the notes played on a particular scale type. Similarly, Kang et al. \cite{59} also tried to accompany random melody with drums, but the scale type was mandatory in their model.

In terms of multi-track music generation, the most well-known model is Muse-GAN proposed by Dong et al. \cite{32} which is considered to be the first model that can generate multi-track music. They first put forward three multi-track music generation models based on WGAN-GP which differ in basic assumptions and corresponding network structure, and then proposed temporal models that can generate multiple bars. MuseGAN is an integration and extension of multi-track model and temporal models, which aims at generating multi-track polyphonic music with harmonic and rhythmic structure, multi-track interdependence and temporal structure. However, the music generated by MuseGAN has multiple shortcomings, such as the number of qualified notes generated is much less than that of the training data, and the generated music contains a large number of fragmented notes. Moreover, MuseGAN’s pre-trained work is too cumbersome, and requires some music knowledge and early labeling work. SeqGAN \cite{225} proposed by Yu et al. can solve the above problem \cite{221}. Since MuseGAN proposed, there have been many follow-up studies. In view of the problem that existing models can only generate real-value pianorolls, further post-processing such as HT (hard threshing) or BS (Bernoulli sampling) is necessary to obtain the final binary results, Dong et al. \cite{33} utilized binary neurons to generate binary pianorolls directly based on the MuseGAN model. Compared with MuseGAN, the use of binary neurons indeed improved the ratio of qualified note and reduced the number of fragmented notes in the generated music. However, above GAN-based multi-track music generation methods are always unstable, and the harmony and coherence of the generated music are not good enough. Part of the reason is that the above methods only use convolution to extract features, which cannot effectively extract temporal features. Therefore, Guan et al. \cite{34} proposed a Dual Multi-branches GAN architecture (DBM-GAN) with self-attention mechanism. The self-attention mechanism can help the model extract temporal and spatial features at the same time, and the multi-branches architecture can coordinate various instruments across time, each branch for one track. Compared with MuseGAN, the number of effective notes generated by the model is about twice that of MuseGAN, and the training time of each batch is only 1/5 of that of MuseGAN. Valenti et al. \cite{234} introduced the first music adversarial autoencoder (MusAE), which can reconstruct phrases with high precision, interpolate between latent representations in a pleasant way, and change specific attributes of songs by modifying their respective latent variables. The above methods all employ the same dataset as MuseGAN, that is, the LMD/LPD dataset commonly used for multi-track music generation. Most of the objective evaluation metrics also use the same metrics used in MuseGAN. We will introduce these metrics in detail in section 6.1.2. Although all of these models can arrange instruments in multi-track music, they cannot specify which track plays the melody.

In addition to the work related to MuseGAN, there are also studies focusing on various aspects of multi-track music, such as musical instruments, novelty, style and so on. MuseGAN \cite{32} exploited the cross-track latent space to deal with the interdependence among instruments. However, the instrument set in MuseGAN is a fixed quintet composed of bass, drum, guitar, piano and string. Therefore, Simon et al. \cite{64} proposed an extension of MusicVAE model which can generate multi-track music with arbitrary instruments. Unlike MusicVAE, both the encoder and decoder of this model adopt hierarchical architecture. However, the model can only represent an individual bar, it is difficult to generate long-term structure. Adding chord condition can improve this shortcoming to some extent. Moreover, MuseGAN focuses more on accompaniment and generation, which is unable to manipulate the existing music. While the system proposed by Simon et al. \cite{64} can not only generate from scratch, but operate existing music through latent space. Different from most of the existing work focusing on the imitation of musicality, i.e. the characteristics that sound like music, Chen et al. \cite{218} emphasized the imitation of artistic creation, i.e. the novelty of generated samples. They proposed the Musicality-Novelty Generative Adversarial Nets for algorithmic composition. Using the same generator, two adversarial networks alternately optimize the musicality and novelty of music. For the reason that novelty is hard to define and quantify, they put forward a new model called novelty game to maximize the minimal distance between machine-created music samples and any human-created music samples in the novelty space, while all well-known human composed music samples are far away from each other. This is the first GAN aiming at music novelty. Zhou et al. \cite{37} collected a new Beatles dataset from the Internet, and proposed BandNet, a music composition model based on RNN, to generate multi-instrument music with Beatles' style. In order to improve the quality of generated music, they integrated expert knowledge in the model learning process. They use a template-based method to generate structured music for modeling the repetitive structure in music. BandNet generated variable length music clips for each part of a pre-defined song structure template (such as AABA), and then combined the generated fragments into a complete song. Since BandNet can only generate music with Beatles' style, Jin et al. \cite{221} explored the utilization of reinforcement learning to automatically generate music with specific styles, including classical, popular, jazz, etc. They also designed a fusion network to deal with the feature extraction and control of music style \cite{222}, but the results show that only 30$\%$ of the samples can meet the actual needs of people for music with a specific style. Hence further research is needed to generate music with specific styles. Angioloni et al. \cite{352} proposed a pattern-based pseudo-song generation method called CONLON. CONLON takes Wasserstein Autoencoder (WAE) as the underlying generative model, and employs a new lossless pianoroll-like data description called $PR^C$, in which the velocity and duration are stored in seperate channels. MIDI pseudo-songs are obtained by concatenating patterns decoded from smooth trajectories in the embedding space. This method aims to generate smooth results in the pattern space by calculating optimal trajectory. Compared with MuseGAN based on pianoroll, songs generated by CONLON are more useful in music production.

Transformer architecture has recently shown promising results in piano score generation. Thus Donahue et al. \cite{39} proposed to use transformer to generate multi-instrument music, and called the proposed generative model LakhNES. They mainly proposed a pre-training technology based on transfer learning, i.e. pre-train the model on one dataset (Lakh MIDI), and then fine-tune the model to another dataset (NES-MDB) to improve the model performance by using the information from another dataset. MuseGAN \cite{32} or MuseGAN with BNs \cite{33} can only generate fixed-length music, but this method can generate arbitrary-length sequences. This method is similar to MuseNet \cite{220}, and these two studies are carried out simultaneously. Ens et al. \cite{289} proposed the Multi-Track Music Machine (MMM) based on transformer. In previous work, the multi-track music was represented as a single time-ordered sequence, which led to the interleaving of music events corresponding to different tracks. Ens et al. created a time-ordered sequence for each track and concatenated multiple tracks into a single sequence. This non-time-ordered representation is directly motivated the nature of transformer attention mechanism. What’s more, they also provided an interactive demonstration to achieve track-level and bar-level inpainting and offer control over track instrumentation and note density.
\subsubsection{Conditioning}~{}\\
\label{sec:4.1.2}
The music generation architectures proposed in section 4.1.1 focus on the directly generating sequence without considering constraints. Apart from generating music from scratch, another way of music generation is to adjust the music generation process by conditioning models on additional information (such as melody, chord, music style, etc.).
\paragraph{\large Given chord, compose melody}~{}\\
\begin{figure}[]
	%	\centering
	\includegraphics[width=0.4\textwidth]{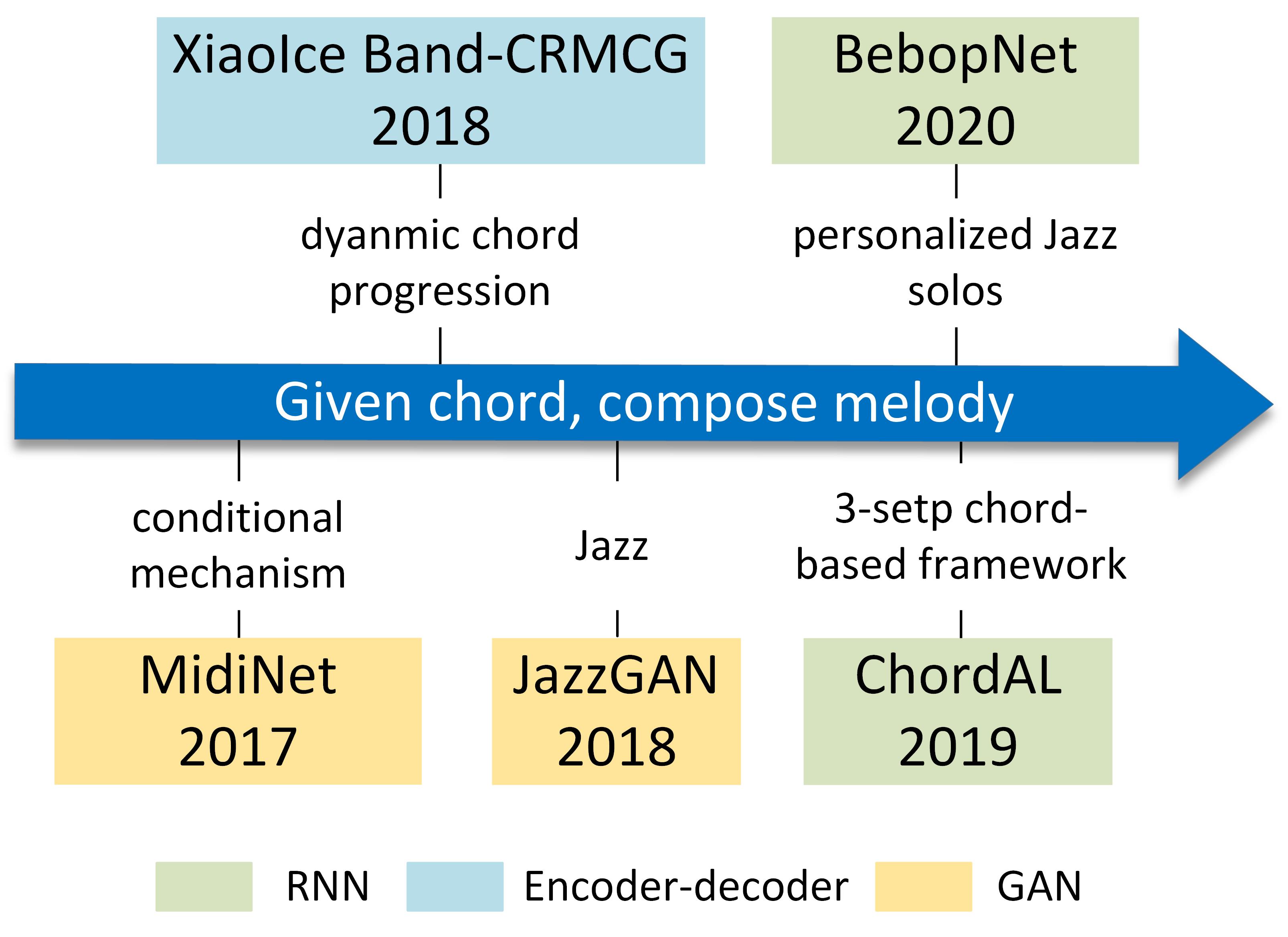}
	\caption{Chronology of melody generations given chords. Different colors represent different model architectures or algorithms.}
	\label{fig:1515}       % Give a unique label
\end{figure}
A chord is a harmonic collection of arbitrary pitches composed of two or more notes that sound as if they are playing at the same time. A series of ordered chords is called chord progression. The monophonic melody generation systems usually only learn melody ignoring any form of explicit harmonic background, which is a crucial guide for note selection \cite{31}. Therefore, lots of researches try to generate melody given chord information, hoping to bring better music structure for the generated melody.

Companies providing music services, such as Jukedeck1 and Amper6, usually just perform arpeggios on basic chords to generate melody. Teng et al. \cite{54} proposed a hybrid neural network and rule-based system to generate pop music. Firstly, temporal production grammar was used to augment machine learning to restore the temporal hierarchy and generate the overall music structure and chord progressions, and then the conditional variational recurrent autoencoder was employed to generate compatible melody. Inspired by WaveNet \cite{149} using convolution to generate audio music, Yang et al. \cite{21} proposed a CNN-based GAN model named MidiNet to generate melodies one bar after another. However, GAN alone could not consider the temporal dependence among different bars, so they proposed a conditional mechanism to utilize chords or melody of previous bars to adjust the generation of current bar. In view of the role of chord progression in guiding melody in pop songs, Zhu et al. \cite{55} proposed XiaoIce Band, an end-to-end melody and arrangement generation framework for song generation, which includes a Chord based Rhythm and Melody Cross-Generation Model (CRMCG) to generate melody with chord progression. Compared with other methods that use single chord as input feature to generate melody \cite{54}, this method also considers the dynamic progression of chord when generating melody, and the CRMCG model is better than Magenta \cite{265} and MidiNet \cite{21} in term of generation result. Tan \cite{69} proposed a chord-based melody generation system called ChordAL. Firstly, chord generator was exploited to generate chords, and then the generated chord sequence was sent into chord-to-note generator to generate melody. Finally, the generated chord and melody were sent into music styler for postprocessing and styling, and the two parts were combined into a complete piece. ChordAL is able to generate coherent harmony, but the generated rhythm and structure need to be improved. The automatic generation of jazz has plenty of challenges, such as frequent and diverse key changes, unconventional and off-beat rhythms, flexibility with off-chord notes and so on. In order to solve these problems, Trieu et al. \cite{104} proposed a jazz melody improvised over chord progression, compared the effect of three rhythm representations encoding on generated outputs, and introduced the use of harmonic bricks for phrase segmentation. The results show that the music generated by JazzGAN is more popular than that generated by Magenta's ImprovRNN \cite{105}. Hakimi et al. \cite{353} studied the task of monophonic, harmony-constrained jazz improvisations, which was the first work to generate personalized Jazz solos using deep learning techniques. They adopted BebopNet, a music language model trained on a corpus of jazz improvisations by Bebop giants. It can generate improvisation based on any given chord, and can use BebopNet's personalized variant of beam-search to generate personalized jazz improvisation for a specific user based on the user-specific metrics trained on a personalized dataset labeled by a specific user.
\paragraph{\large Melody harmonization}~{}\\
Melody harmonization refers to the generation of harmonic accompaniment for a given melody \cite{74}. Here, the term harmony or harmonization can refer to both chord accompaniment and polyphony/multi-track accompaniment, also known as arrangement.
\begin{figure}[H]
	%	\centering
	\includegraphics[width=0.6\textwidth]{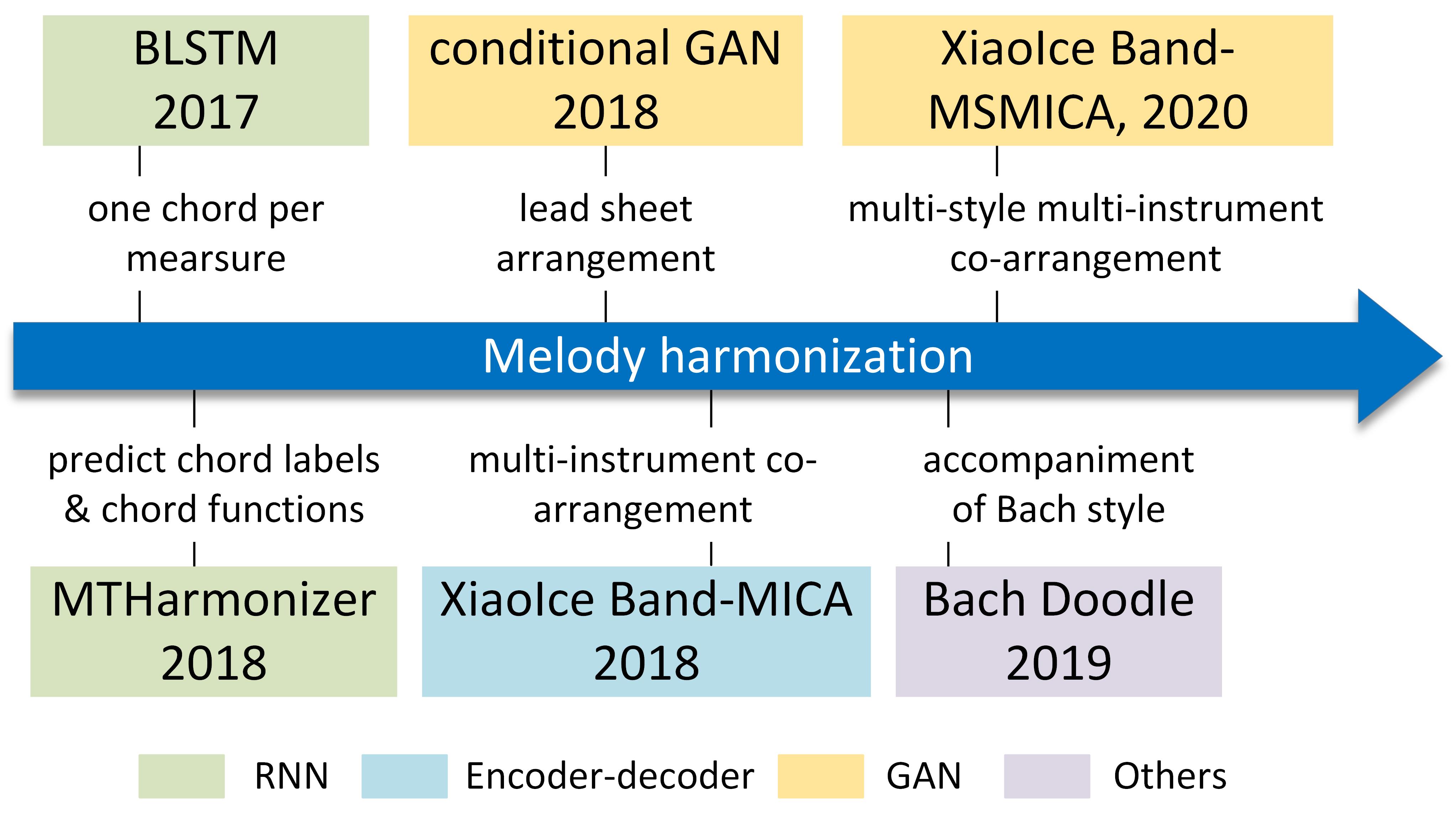}
	\caption{Chronology of melody harmonization. Different colors represent different model architectures or algorithms.}
	\label{fig:1616}       % Give a unique label
\end{figure}
Chords play a key role in music creation and arrangement. However, it is difficult for people who have no music composition experience or professional knowledge to create chords conditioned on melody. There have been many researches on chord generation based on traditional machine learning methods. One of the most popular methods is probabilistic modeling. The commonly used models are Markov model, hidden Markov model (HMM), etc. However, the method based on HMM model usually cannot retain the long-term consistency of chord progression, and cannot explicitly describe the rhythm and hierarchical structure of chords. Lim et al. \cite{28} proposed to use two BLSTM network layers and a full connection layer to learn the correspondence between melody and chord sequence pairs, which achieved better performance than HMM model in both quantitative and qualitative evaluation. But the model enforces that only one chord can be generated per measure, and chord type must be triad. Yeh et al. \cite{74} proposed that for longer phrases, the BLSTM model proposed by Lim et al. possessed two main defects: one is the excessive use of common chords, which makes the chord progressions monotonous; the other is the non-congruent phrasing between the melody and chords, indicating a musical cadence in an unfit location, which may bring unnecessary sense of ending in the middle of a chord sequence. Therefore, they proposed an extended BLSTM model to improve these two defects, called MTHarmonizer. Its core idea is to not only predict chord labels, but also predict chord functions when training models \cite{75}. Compared with the distribution of chord labels, the distribution of chord functions is relatively balanced, which is easier for model learning. Secondly, since chord labels and functions are interdependent, adding a chord function as a target will inform the model which chord labels share the same function and may thus be interchangeable. Yang et al. \cite{237} proposed a novel LSTM-based model CLSTM to generate matching chord accompaniment for a given melody. CLSTM consists of two LSTM models, one of which focuses on the relationship between bar note information and corresponding chords, and the other explores chord conversion rules. However, this method can also generate only one chord for each bar.

Arranging is an art of giving an existing melody musical variety. Liu et al. \cite{29} defined lead sheet arrangement as the process that takes as input a lead sheet and generates as output piano-rolls of a number of instruments to accompany the melody of the given lead sheet. They proposed a novel conditional generation model based on convolutional GAN to implement lead sheet arrangement, which is the first research work of lead sheet arrangement. After that, they \cite{57} also adopted a new hybrid generation model to re-implement the task, in which the lead sheet generation adopts the recurrent VAE model, and the arrangement generation still adopts convolutional GAN. The new hybrid model takes advantage of two generative models to obtain better results. Considering the harmony of music arrangement, Zhu et al. \cite{55} proposed XiaoIce Band, an end-to-end melody and arrangement generation framework for pop music, which includes a Multi-Instrument Co-Arrangement model (MICA) to generate multi-track accompaniment for melody. The results demonstrated that the quality of music generated by MICA model is better than song from PI \cite{58} in both monophonic and multi-track, and the harmony among multiple tracks was improved significantly. Later, in order to generate music with a specific style, they \cite{248} extended MICA to a Multi-Style Multi-Instrument Co-Arrangement Model (MSMICA). MSMICA employed adversarial training to learn music style, the model can not only maintain the harmony of the generated music, but also control the music style for better utilization. Inspired by SeqGAN \cite{225}, they train the model by regarding rewards as policy gradients , and developed a SeqGAN variant named Multi-SeqGAN with new sampling and reward method to process multi-track sequences. Melody harmonization encourages human-computer interaction, especially suitable for novices \cite{43}. Therefore, many human-computer interaction interfaces and tools have been developed for this task, such as MySong \cite{44}, Hyperscope \cite{45}, etc. Huang et al. \cite{43} also designed a Google Doodle, the Bach Doodle, based on artificial intelligence. Users can create their own melody, or generate accompaniment for melody with Bach style under the help of Coconet model \cite{52}.
\paragraph{\large Given conditional tracks, compose remaining tracks}~{}\\
The task of generating other tracks given the condition track is similar to melody harmonization. They both generate accompaniment for a given music. The difference is that the conditional track in this type of task is not necessarily a melody, but can be any one/multiple track(s) in multi-track music.

To solve this task, a track-conditional generation method is proposed in Muse-GAN \cite{32}. For improving the performance of impromptu accompaniment and solving the non-differentiable optimization problem of discrete and binary pianoroll, Jia et al. \cite{35} proposed a coupled latent variable model with a binary regularizer. On the one hand, the model used a coupled mechanism to learn a latent variable that simultaneously captures the internal distribution of each track and the joint distribution of multiple tracks. On the other hand, they propose to reformulate the discrete and binary properties of pianoroll representation into a convex constraint, thus obtaining a differentiable optimization problem that can be solved by neural networks. They compared the music generated by MuseGAN \cite{32} and MuseGAN with BNs \cite{33} from quantitative and qualitative evaluation, and the results showed that the quality of music generated by the MuseGAN with BNs was better. Previous work usually produced multiple tracks seperatele, and there was no clear interdependence between notes in different tracks, which would damage harmony modeling. In order to improve the harmony, Ren et al. \cite{295} proposed a pop music accompaniment generation system called PopMAG, which generated five instrumental accompaniment tracks under the condition of chord and melody. Particularly, they proposed a novel MUlti-track MIDI representation (MuMIDI), which can generate multiple tracks in a single sequence simultaneously and explicitly model the dependence of the notes from different tracks. Although MuMIDI greatly improved harmony, it expanded the sequence length and thus brought new challenges for long-term music modeling. Therefore, they adopted the transformer-based seq2seq model to solve this challenge from two aspects: 1) modeling multiple attributes of a note (such as pitch, duration, velocity) with one rather than multiple steps, thereby shortening the MuMIDI sequence; 2) introducing extra long-context as memory in the encoder and decoder of seq2seq model to capture the long-term dependency in music. Experimental results demonstrated that MuMIDI had great advantages in modeling harmony compared with previous representation methods such as MIDI and REMI, and PopMAG is superior to MuseGAN in both subjective and objective evaluation \cite{32}.

Additionally, in order to achieve rhythm coherence, drum patterns tend to correlate with other instruments (e.g., rhythm guitar and bass guitar), so the structure and rhythm information from other instruments is crucial for designing reasonable drum patterns \cite{98}. Some researches guide the generation of drum patterns by adding additional conditional information, such as metrical structure, melody, and other instrument tracks. Makris et al. \cite{62} proposed a neural network architecture combining LSTM and feedforward neural network, which can learn long drum sequences under the constraints imposed by metrical rhythm information and a given bass sequence. The LSTM module learns sequences of consecutive drum events, while the feedforward layer receives the information about metrical structure and bass movement as the condition layer. LSTM with feedforward condition layer can generate drum rhythm pattern closer to ground truth. Recently, they improved the FF module by using two FF modules to process past and future conditional inputs respectively \cite{99}.
\subsection{Middle-level: Performance Generation}
\label{sec:4.2}
Music performance is one of the most basic activities in music. Good performance not only needs to convert the notes in the score into physical actions with precise timing and correct pitch on the instrument, but also needs to transfer emotion and information through subtle control of tempo, dynamics, articulations and other performance elements \cite{133}. Expression is an intuitive aspect of music performance, which is regarded as the strategies and changes which are not marked in the score, but which performers apply to the music. It is also one of the reasons why music sounds interesting and vivid \cite{259}. A complete definition is given in \cite{300}, relating the liveliness of a score to ``the artist's understanding of the structure and `meaning’ of a piece of music, and his/her (conscious or unconscious) expression of this understanding via expressive performance."

Score generation has generated music works equivalent to human composition, but these works are rendered strictly accordance with the metrical grid written on the score, and most of them lack expression. Therefore, many researchers have begun to explore expressive performance generation. The result of performance generation is the music performed by musicians in accordance with the score (not necessarily strictly obey the score information). It adds music dynamics and timing information on the basis of score generation so that the generated music has a variety of musical expressions. Some score generation models mentioned in Section 4.1 encode music dynamics as well, such as DeepJ \cite{27}, MIDI-VAE \cite{48}, etc., but their focus is still on score generation. The encoding characteristics of performance generation include timing, tempo, dynamics, etc. The generated result is not the final musical sound, but is still regarded as middle-level abstraction. To transform the result into audio, it is necessary to customize the acoustic features of music (as instruments), that is, the underlying music features, and convert them into audible audio file format. Similar to the early algorithm for score generation, previous performance generation methods also include rule-based method \cite{122}, probability model \cite{123,124} and neural network \cite{129,132}. Recent studies have begun to try to use deep learning to model performance. Here, we divide these studies into two categories: one is rendering expressive performance, which refers to adding performance characteristics to a given score; the other is composing expressive performance, which refers to modeling score features and performance features simultaneously to create novel expressive works. This section will review these studies in detail, but compared with the traditional methods, deep learning researches on expressive performance modeling is far from enough.
\subsubsection{Performance features}~{}\\
Before introducing the research work of performance generation, it is necessary to introduce several crucial performance characteristics, including tempo, timing, dynamics, articulation, and the method of extracting performance characteristics from music score and performance data.

Broadly speaking, tempo is the approximate speed of musical events. It may refer to the global tempo of a performance (usually roughly specified by time signature in the music score), or refer to the local tempo, that is, the occurrence rate of events within a smaller time window, which can be regarded as local deviation from the global tempo. Expressive timing refers to the deviation of the individual events from local tempo \cite{144}. When the score is performed, most of the timing information in the score will be deliberately ignored. For example, rubato and swing developed in classical music beginning in the 19th century is a defining quality of many African American music traditions \cite{139}. Articulation refers to the direction or playing skill that affects the conversion or continuity of single note or multiple notes/sounds. On the music score, articulation is the cohesive mark of musical expressiveness and appeal, such as legato, staccato, stress, etc. Dynamics (also known as velocity and volume) refers to how music becomes louder and quieter. Although information about dynamics is sometimes provided in the score, the effectiveness of dynamics depends largely on conventions that are not written into the score. For example, the ‘p’ written on the score means to play quietly, but it doesn't tell performers how quietly, nor will all the notes be equally quiet. In addition, when playing polyphonic piano music, the notes played at the same time are usually played at different dynamics levels and articulated differently from one another in order to bring out some voices over others \cite{139}.

Jeong et al. \cite{136} gave a detailed process of extracting corresponding features from music score and performance data. They extracted more detailed note-level score features, including the duration of following rest, articulation markings (e.g. trill, staccato, tenuto), distance from the closest preceding tempo and dynamic directions, slur and beam status and some note-level performance features, including tempo, onset deviation, velocity and pedal. Here tempo represents the number of beats played in a minute. The beat is defined by the time signature of the bar. For each beat in a piece of music, the time interval between two beats in the performance is calculated. Onset deviation is to explain a micro-timing or asynchronous of performance, or tempo rubato within a single beat. For each note, the shift of the note onset in its ``in tempo" position is calculated based on the pre-defined rhythm. The dynamics of performance can be represented with MIDI velocities of the corresponding notes. The modeling of piano pedal is one of the most challenging problems in performance rendering. By observing the pedal states in four different positions of notes, the pedal information in MIDI is encoded as note-level features. The four positions are: start of the note, offset of the note, minimum pedal value between note, on set and offset, minimum pedal value between note offset and the following new onset. They have released Python-based MusicXML feature extraction as a publicly available software library\footnote[10]{https://github.com/jdasam/pyScoreParser}.
\subsubsection{Render expressive performance}~{}\\
\label{sec:4.2.2}
\begin{figure}[H]
	%	\centering
	\includegraphics[width=0.4\textwidth]{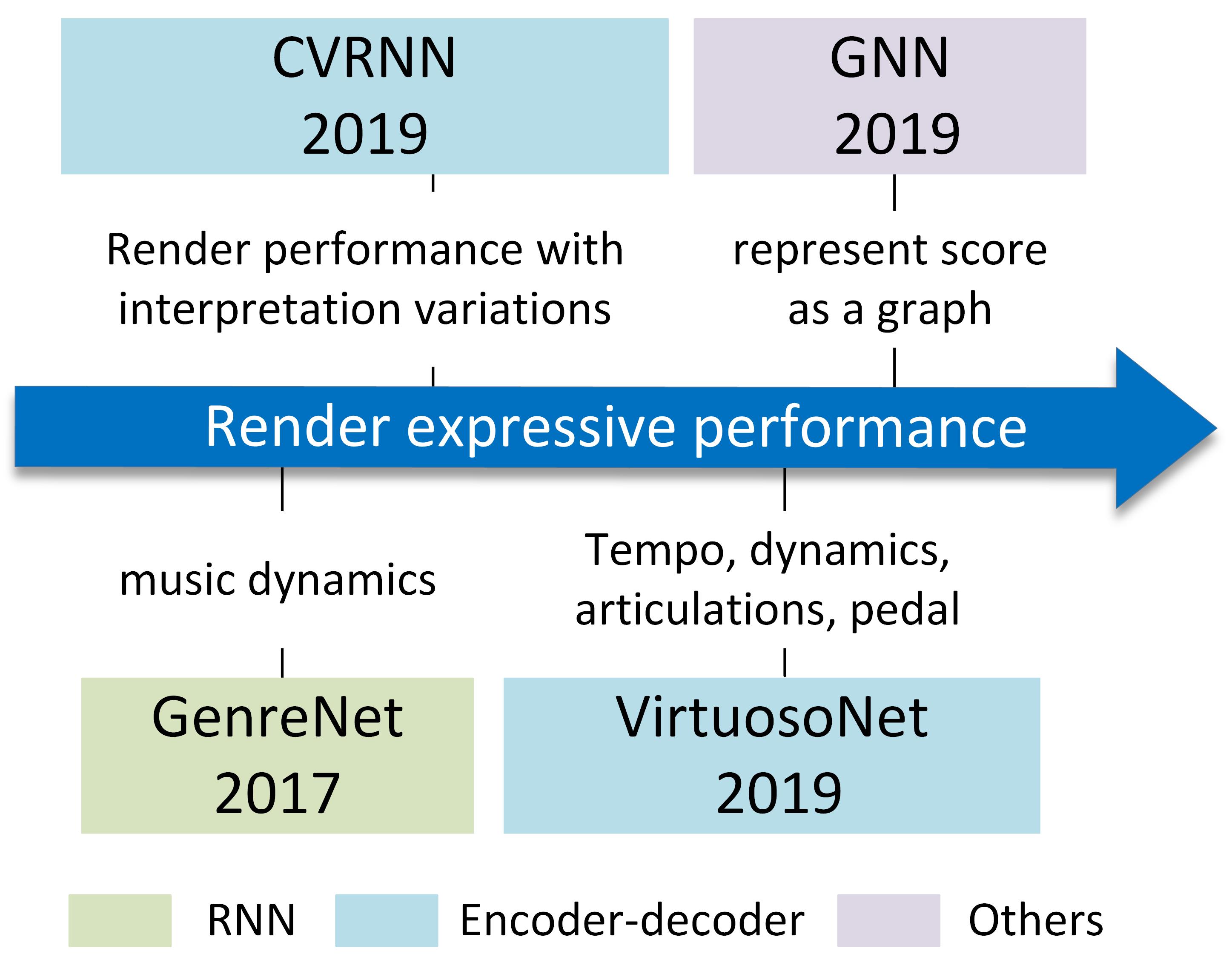}
	\caption{Chronology of rendering expressive performance. Different colors represent different model architectures or algorithms.}
	\label{fig:1717}       % Give a unique label
\end{figure}
Rendering expressive performance refers to endow a given score with expression to make it easier to perform. In this case, there is no need to re-model the pitch, duration and other information that have already provided in the score, but to generate a set of expressive performance features for the notes in the score, such as timing, dynamics, etc.

The mapping between score and performed music is not a bijection. Musicians will inject their own style which can be rendered by dynamics when performing music. Malik et al. \cite{142} explored how to automatically generate music performances with specific styles that are indistinguishable from human performances. They proposed a GenreNet architecture composed of BLSTM layer and linear layer to generate dynamics information for score. However, GenreNet is limited to learning the dynamics of a specific genre, and finally employs the rendition model StyleNet to learn various styles. The experimental results showed that the model can play music in a more humanized way, but it is not enough to distinguish the characteristics of different music styles. Most methods allow the control of specific musical concepts, such as tempo and velocity, but cannot manipulate abstract musical concepts that cannot be labeled, such as liveliness within a performance. The root of this limitation lies in the attempt to establish a direct mapping between score and expressive performance. Therefore, Maezawa et al. \cite{134} proposed a music performance rendering method which can explicitly model the performance with interpretation variations of a given music segment. They first explored the use of LSTM-based VAE to generate music performance conditioned on the music score \cite{138}, and then proposed a new model CVRNN, which extended the variational recurrent neural network (VRNN) to accept position-dependent conditional inputs. Finally, the trained model can automatically generate expressive piano performance based on music score and interpretation sequence. Jeong et al. \cite{133} used a deep neural network based on hierarchical RNN to model piano performance, mainly imitating the pianist's expressive control of tempo, dynamics, articulations and pedaling. They used CVAE to model different performance styles with the same input. CVAE was first proposed in \cite{138}, but previous work encoded the latent vector in note-level, while Jeong et al. \cite{133} encoded performance styles in a longer-level (such as an entire piece). In order to keep the music expressions consistent over long-term sections, the model used multi-scale approach to predict the performance characteristics, that is, to predict the tempo and dynamics in bar-level first, and then fine-tunes them in note level according to the results. Based on the above work, Jeong et al. \cite{135} proposed to use graph neural network (GNN) model to represent music score as a unique form of graph, and apply it to render expressive piano performance from music score, which is the first attempt to adopt GNN to learn the note representation in western music score.

In addition, not only is the existing composition work ignoring the expressive characteristics required for reasonable music interpretation, but most of the datasets only contain the information necessary to model the semantics of music works and lack of details on how to transform these works into subtle performance, resulting in rigid and boring music generated by using these datasets. Therefore, Donahue et al. \cite{115} created a new dataset NES-MDB. Each song in the dataset contains scores of four kinds of instrument sounds, as well as the dynamic and timbre performance properties of each sound. For a detailed introduction to the dataset, please refer to section 5.6. Later, they proposed to exploit LSTM to learn a mapping between score and actual expressive attributes to model performance.
\subsubsection{Compose expressive performance}~{}\\
\label{sec:4.2.3}
\begin{figure}[]
	%	\centering
	\includegraphics[width=0.6\textwidth]{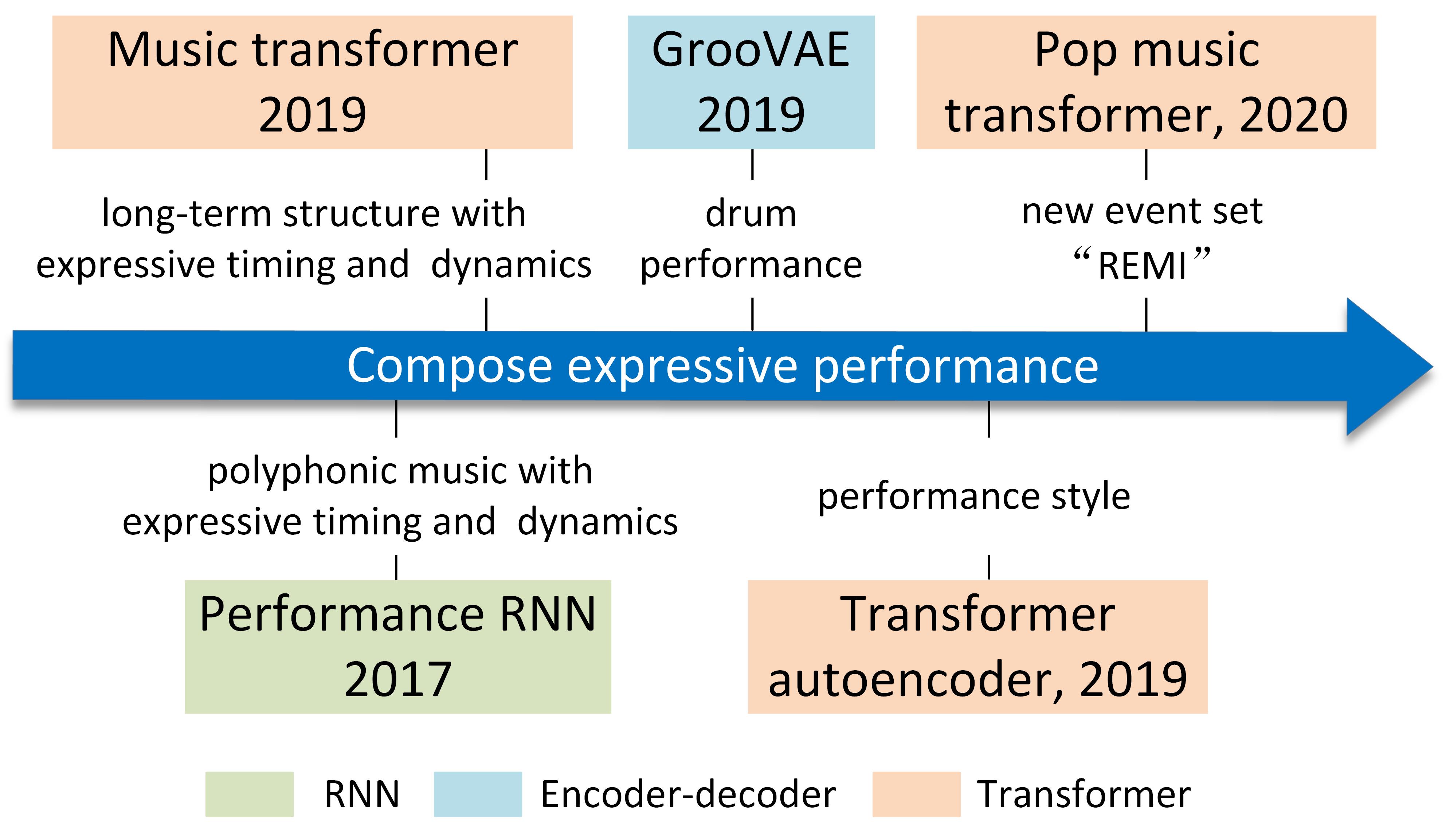}
	\caption{Chronology of composing expressive performance. Different colors represent different model architectures or algorithms.}
	\label{fig:1818}       % Give a unique label
\end{figure}
Composing expressive performance refers to the direct generation of works with expressive performance characteristics, that is, the model needs not only to learn the score features of notes, but also to model the corresponding performance characteristics. Compared with performing specific works, its advantages are more embodied in improvisation. At present, there are not many researches on composing performance using deep learning techniques. Compared with various complicated models of score generation, most models of performance generation are simple LSTM-based models, so there is still a large space for future research. Because the piano performance is easier to quantify, and the piano datasets are easier to obtain, current performance generation is mainly limited to the generation of piano performance. Moreover, a few researches have attempted to compose drum performance automatically.

Simon et al. \cite{141} proposed a LSTM-based model named Performance RNN to compose polyphonic music with expressive timing and dynamic. The performance generated by this model lacks overall consistency, but local features (phrasing within a 1 or 2 second time window) are quite expressive. Music Transformer proposed by Huang et al. \cite{56} can also achieve expressive piano performance modeling, and is superior to performance RNN \cite{141}. Oore et al. \cite{139,143} also proposed to use LSTM-based model to jointly predict notes and their expressive timing and dynamics. In order to capture expressiveness, they put forward a performance representation that serialized all notes and their attributes into a series of events. Compared with the previous model, Oore et al. produced more human-like performances with this new representation. But this representation caused already lengthy musical pieces to be on average 3-4x longer. In addition, the attributes of a note, such as note play and note off, may be far apart in the event list. Therefore, Hawthorne et al. \cite{137} proposed a novel representation called NoteTuple, which groups the attributes of a note as one event, thus reducing the length of the sequence. Choi et al. \cite{217} performed advanced control over the global structure of sequence generation based on the Music Transformer architecture, and proposed a transformer autoencoder to aggregate encodings of the input data across time to obtain a global representation of style from a given performance. They showed that combining this global embedding with other temporally distributed embeddings can better control the separate aspects of performance style and melody. The experimental results demonstrated that above conditional model enable generate performance similar to the input style, and generate accompaniment that conforms to the given performance style for the melody. Based on Transformer-XL \cite{112}, Huang et al. \cite{110} proposed Pop Music Transformer for modeling popular piano music performance. They focused on improving the way of music score transforming into events. The new event set is called ``REMI" (REvamped MIDIderived events). The event set provided a metrical context for modeling the music rhythm patterns of music, so that the model could be aware of the beat-bar-phrase hierarchy in music more easily, while allowing for local tempo changes. In addition, the event set explicitly establishes a harmonic structure to make chord progression controllable. It also facilitates coordinating the different tracks of a music piece, such as piano, bass and drum. The results implied that music generated by Pop Music Transformer has more reasonable rhythm structure than music transformer \cite{56}. Later, they also embedded music information retrieval (MIR) techniques such as automatic transcription, downbeat estimation and chord recognition into the training process, and proposed beat-based music modeling and generation framework \cite{111}.

Apart from the above research on the piano performance generation, Gillick et al. \cite{145,146} proposed to compose drum performance automatically, using seq2seq and recurrent variable information bottleneck (VIB) model to transform abstract music ideas (as score) into expressive performance. They proposed a class of models called GrooVAE, which are used to generate and control the expressive drum performance, and proposed three specific applications using these models: Humanization, generated performance for a given drum pattern, and realized the style transfer of drum performance (groove transfer) by disentangling performance characteristics from the score to perform same drum pattern in different ways; Tap2drum, came up with drums that match the sense of groove on another instrument (or just tapping on a table); Infilling, completed or modified the drum beat by generating or replacing the part for the desired instrument. In addition, they also created a new drum dataset named Groove MIDI Dataset with paired scores and performance for training above models.
\subsection{Bottom-level: Audio Generation}
\label{sec:4.3}
Compared with the higher-level representations mentioned earlier such as score and performance, audio refers to the sound that can be heard directly, which is the most intuitive and easily perceived music representation form for human beings. Audio generation is to generates audible music audio files directly, also known as music audio/sound synthesis. The previous sound modeling methods mainly include physical modeling \cite{267}, concatenative synthesis \cite{269,270}, and statistical model-based methods \cite{268,271}. The deep learning methods of audio synthesis has shown their strong ability in the past five years. Here, we divide music audio synthesis into two categories in the light of whether the synthesized audio contains lyrics or not, namely audio synthesis and singing voice synthesis.
\subsubsection{Audio synthesis}~{}\\
\label{sec:4.3.1}
\begin{figure}[]
	%	\centering
	\includegraphics[width=1.0\textwidth]{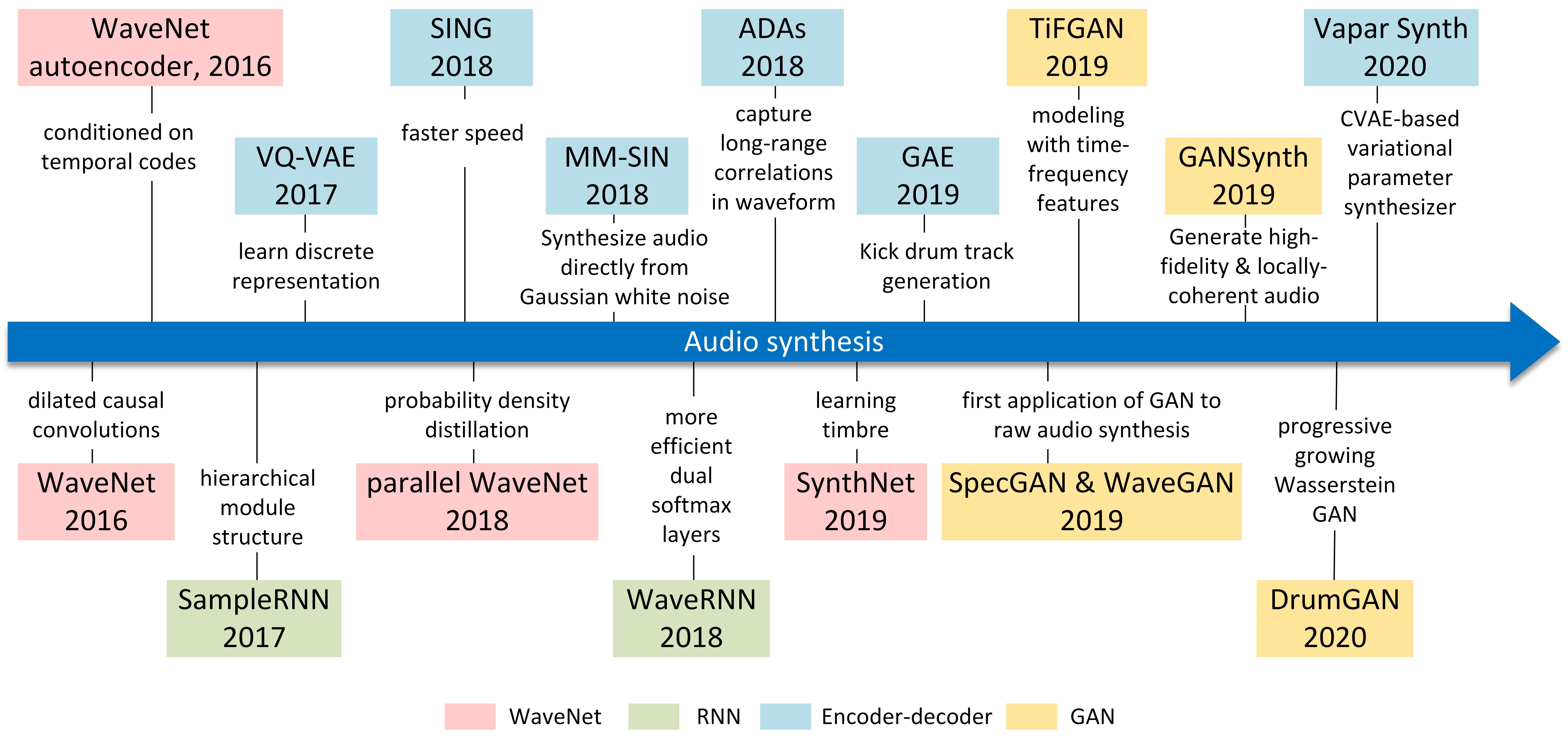}
	\caption{Chronology of music audio synthesis. Different colors represent different model architectures or algorithms.}
	\label{fig:1919}       % Give a unique label
\end{figure}
All the sounds that human beings are able to hear can be called audio, including speech, music, noise, etc. In the field of speech research, there are many audio synthesis tasks, such as speech synthesis, speech conversion, text to speech. Here we focus on music audio synthesis. Some models for music audio synthesis may also be widely used in the field of speech research, which we will not elaborate on. In addition, most of the audio generation work is based on audio synthesis technology, so we do not make a clear distinction between the two, and both are regarded as audio synthesis.

The emergence of WaveNet \cite{149} model made a breakthrough in audio synthesis task. WaveNet is a PixelCNN-based \cite{150} generative model proposed by Google DeepMind. It utilized dilated causal convolutions to create original waveforms of speech and music, and guided music generation by introducing conditional features. WaveNet can generate novel and highly realistic music clips, but the generated music still lacks long-term consistency, resulting in second-to-second variations in genre, instrumentation, volume and sound quality. Since WaveNet was proposed, many follow-up studies have appeared. Engel et al. \cite{152} proposed a WaveNet-based autoencoder model, which enable the autoregressive decoder to be conditioned on the temporal codes learned from the original audio waveforms. The model learned a manifold of embeddings, allowing morphing between instruments, and meaningfully interpolating in timbre to create new, realistic and expressive sounds. In addition, the model can be conditioned on global attributes by utilizing a one-hot pitch embedding. Compared with WaveNet \cite{149}, WaveNet autoencoder aims at obtaining long-term consistency structure without relying on external conditions. WaveNet is able to predict its conditional probability in parallel during training, while sample generation remains inherently sequential and therefore slow, so it is not suitable for today's large-scale parallel computers and is hard to deploy in real-time production environment. IAFs represent a dual formulation of deep autoregressive modeling, in which sampling can be performed in parallel, while the inference process required for likelihood estimation is sequential and slow. Therefore, Oord et al. \cite{153} combined the parallel training of WaveNet and the parallel sampling of IAFs, and proposed a new method Probability Density Distillation that trained parallel WaveNet (students) from learned WaveNet (teacher). In comparison to the original WaveNet, parallel WaveNet achieves several orders of magnitude acceleration without affecting the quality of generation, and has been deployed online in Google Assistant to offer real-time queries to millions of users. The rapid generation of parallel WaveNet comes at the cost of increasing the training time. Therefore, Défossez et al. \cite{274} studied a more computationally efficient model of generating waveform frame by frame with large strides. Specifically, they proposed a lightweight audio synthesizer named SING, that could generate the waveform for entire audio frames of 1024 samples at a time. The training speed is 32 times faster and the generation speed is 2500 times faster than the WaveNet-based autoencoder model \cite{152}. Schimbinschi et al. \cite{196} proposed a network architecture called SynthNet based on WaveNet, which can directly learn the mapping between notes and instrument timbres from the original audio waveform and the score in the binary pianoroll format. Their research focused on learning timbre, while controlling the given melody content and avoiding any rhythm variations. Compared with WaveNet \cite{149}, SynthNet owned faster training speed and generated almost the same high-quality audio as real samples. The main difference from \cite{152} is that SynthNet can learn timbre from the entire song, while WaveNet autoencoder requires separately marked notes. High quality speech synthesis conditioned on detailed language features, but the instrumental neural synthesis has not yet reached a comparable control level.

Since the audio waveform is a one-dimensional continuous signal that changes over time, it is easy to think of using RNN to model the time-varying audio signals. However, it is difficult to model audio waveform with high temporal resolution (such as 16kHz) using RNN directly. To solve this problem, Mehri et al. \cite{151} proposed a novel model named SampleRNN for unconditional audio generation. SampleRNN adopted hierarchical module structure, and each module ran at different temporal resolution. The lowest module processes individual samples, which is called sample-level module. Each higher module runs with an increasingly longer timescale and a lower time resolution, which is called frame-level module. Carr et al. \cite{161} employed unconditional SampleRNN to generate an album in the time domain. which imitated bands belonging to assorted music genres, such as metal, rock, punk, etc. Similar to WaveNet \cite{149}, SampleRNN generated one acoustic sample at a time based on previously generated samples. The difference is that there are different modules running at different clock rates in SampleRNN. In Comparing to WaveNet, SampleRNN alleviates the problem of long-term dependence through hierarchical structure and stateful RNN. In view of the low efficiency of serial sampling in sequence generation model, kalchbrenner et al. \cite{172} proposed a single-layer RNN called WaveRNN with a dual softmax layers, which matched the quality of the state-of-the-art WaveNet model. Hantrakul et al. \cite{193} exploited domain specific conditioning features to drive a simpler and more efficient WaveRNN \cite{172} model to generate instrumental audio. They explored various conditioning features and structures, and proved that the WaveRNN-based model can synthesize real audio faster than real-time through fine-grain time control over several music features (such as loudness and fundamental frequency).

Models such as WaveNet \cite{149}, SampleRNN \cite{151}, NSynth \cite{152} are still very complex, requiring a large number of parameters, long training time and plenty of examples, and lack of interpretability. Therefore, some studies consider using other models to synthesize audio signals. Esling et al. \cite{162} explored the relationship between timbre space and latent space. They bridged analysis, synthesis and perceptual audio research by regularizing the learning of latent spaces, which were matched with the perceptual distance from timbre studies. Latent spaces can be directly used to synthesize sounds with continuous evolution of timbre perception, which can be turned into generative timbre synthesizers. Andreux et al. \cite{163} utilized a scattering autoencoder to synthesize and modify music signals from Gaussian white noise. The statistical properties of the synthesized signals are limited by ensuring that the synthesized signals have the same moment in the scattering space as the input signal. The network thereby is called the Moment Matching-Scattering Inverse Network (MM-SIN). It has a causal structure similar to WaveNet \cite{149}, which allows for synthesizing audio step by step and provides a simpler mathematical model related to time-frequency decomposition. Although its synthesis quality is lower than the state-of-the-art methods, it is still a promising new method to synthesize audio signal directly from Gaussian white noise without learning encoder or discriminator. Oord et al. \cite{167} proposed to employ the Vector Quantised-Variational AutoEncoder (VQ-VAE) to learn useful discrete representations in a completely unsupervised way. The model optimized the maximum likelihood while preserving important features of the data in the latent space. Audio with high-quality can be synthesized by pairing learnt representations with an autoregressive prior. In order to capture the high-level structure and model the long-term consistency of music, Dieleman et al. \cite{168} put forward to capture the long-range correlations in the waveform using the autoregressive discrete autoencoders (ADAs). The model can unconditionally generate tens of seconds piano music with style consistency directly in the original audio domain, and can further capture long-range correlations in music exploiting the separately trained autoregressive model at different levels of abstraction. There is still some work on drum audio generation. Lattner et al. \cite{101} proposed a GAE-based conditional kick drum track generation model with existing music materials as input. The model learned from the music dataset a low-dimensional code which can encode the rhythmic interactions of the kick drum vs. bass and snare patterns, which is called mapping code and has rhythm and time shift invariance. After that, a variety of musically plausible kick drum tracks music can be generated by sampling mapping codes. In addition, the model can transfer kick drum patterns from one song to another as well.

GAN has achieved widespread success in synthesizing real images. It is common to apply image processing algorithms to audio tasks in discriminative setting. Donahue et al. \cite{169,170} studied the time-domain and frequency-domain strategies of using GANs to synthesize audio. They proposed the frequency-domain method SpecGAN and the time-domain method WaveGAN based on the image synthesis architecture DCGAN, a first attempt at applying GANs to raw audio synthesis in an unsupervised setting. The results showed that although SpecGAN model achieved higher inception score, humans prefer sound quality synthesized by WaveGAN. The synthesis speed of fully parallelized WaveGAN is several orders of magnitude faster than autoregressive models such as WaveNet \cite{149}. Marafioti et al. \cite{174} demonstrated the potential of modeling using time-frequency features by training GAN on short-time Fourier features. They employed DCGAN-based TiFGAN, which can directly generate reversible STFT representation, for unconditional and unsupervised audio synthesis. The results showed that the model is superior to the most advanced GAN for waveform generation. GANs possess global latent conditioning and efficient parallel sampling, but struggle to generate locally-coherent audio waveforms. Therefore, Engel et al. \cite{191} proposed GANSynth, which uses GAN to generate high-fidelity and locally-coherent audio by modeling log-magnitudes and instantaneous frequency with sufficient frequency resolution in the spectrum domain. They conducted global conditioning on latent and pitch vector to adjust GAN to generate perceptually smooth interpolation in timbre and consistent timbral identity across pitch. Moreover, Experiments showed that GANSynth is superior to the powerful WaveNet baseline in automated and manual evaluation metrics, and can efficiently generate audio several orders of magnitude faster than autoregressive models. However, the model cannot deal with fine-scale adjustment or variable length sequences. Due to the lossy characteristics of spectrograms and the oversmoothing artifacts caused by insufficiently expressive models, it is always a challenge to generate high-fidelity audio with the help of existing spectrogram model. To solve this problem, Vasquez et al. \cite{192} designed a model MelNet which can generate high-fidelity audio samples by employing time-frequency representation, combining a highly expressive fine-grained autoregressive model with multi-scale generation process, the model captures structure at timescales which has not been implemented in the time domain model. \cite{167} and \cite{168} utilized the hierarchical structure of the autoencoder to capture the long-term correlation in the waveform, this method requires multi-stage model which must be trained in sequence, while the multi-scale method in MelNet can be parallelized at multiple levels. In the case of MIDI conditional music generation, WaveNet relies on symbolic music representation, while MelNet needs no intermediate supervision. MelNet is complementary to WaveNet in many aspects, MelNet is better at capturing high-level structures, while WaveNet can generate higher-fidelity audio. Moreover, Nistal et al. \cite{349} proposed DrumGAN, which synthesizes drum audio via progressive growing Wasserstein GAN (PGAN), and conditoned the model on continuous perceptual features computed with a publicly available feature extractor.

Most of the audio generation models mentioned above directly generate samples in time or frequency domain. Although these representations are sufficient to express any signal, they are inefficient for utilizing no existing knowledge of sound generation and perception. Another method (vocoder/synthesizer) successfully combines the powerful domain knowledge of signal processing and perception, but due to the limited representativeness and the difficulty of integrating with modern automatic differentiation-based machine learning methods, there are few studies. Subramani et al. \cite{350} proposed a CVAE-based variational parameter synthesizer, VarPar Synth, which employed a conditional VAE trained on a suitable parameter representation to synthesize audio, and reconstructed and generated instrument tones through flexible pitch control. The parameter representation adopted by them involves the application of source filter decomposition to the harmonic components of the spectrum extracted from the harmonic model. The filter is estimated as the envelope of the harmonic spectrum and represented by low dimensional cepstral coefficients. Compared with the raw waveform or spectrogram, an advantage of this powerful parameter representation is that it can synthesize high-quality audio with less training data. In addition, Engel et al. \cite{198} introduced the Differentiable Digital Signal Processing (DDSP) library, which can directly integrate the classical signal processing elements and deep learning methods, that is, the interpretable signal processing elements can be integrated into the modern automatic differential software such as Tensorflow. In terms of audio synthesis, they achieved high-fidelity generation without the need for large autoregressive models or adversarial losses. DDSP also offered an interpretable and modular approach for generative modeling, allowing each module to be manipulated separately without sacrificing the benefits of deep learning.
\subsubsection{Singing Voice Synthesis(SVS)}~{}\\
\label{sec:4.3.2}
Singing voice synthesis (SVS) refers to the generation of singing voice from a music score containing linguistic (lyrics) and musical features (note, rhythm, etc.). The SVS system consists of two parts: first, the linguistic and musical features are synthesized, and then the singing voices are synthesized with the help of the model learned by using the composed input features \cite{177}. SVS differs from traditional text-to-speech synthesis (TTS) because singing voice is distinct from speech. It contains more local dynamic movements of acoustic features. For example, the F0 contour of singing voice contains many dynamic F0 patterns related to singing, such as vibratos. The spectral characteristics of singing voice are also affected by F0 movements \cite{178}. Previous successful singing synthesizers are based on concatenative methods \cite{269,270}. Although such systems are state-of-the-art in terms of sound quality and naturalness, their flexibility is limited and it is difficult to expand or significantly improve. Methods based on machine learning, such as statistical parameter method \cite{271}, are not so strict, and allow the combination of data from multiple speakers, the use of a small amount of training data for model adaptation, etc. These systems cannot match the sound quality of the concatenative methods, especially existing over-smoothing in frequency and time. Many TTS techniques based on HMM are also suitable for singing voice synthesis, such as speaker adaptive training \cite{272}. The main drawback of HMM-based method is that a small number of discrete states are used to model phonemes, and the statistics within each state are constant. This leads to excessive averaging, an overly static ``buzz" sound, and significant state transition in long sustained vowels during singing \cite{200}.
\begin{figure}[]
	\includegraphics[width=0.6\textwidth]{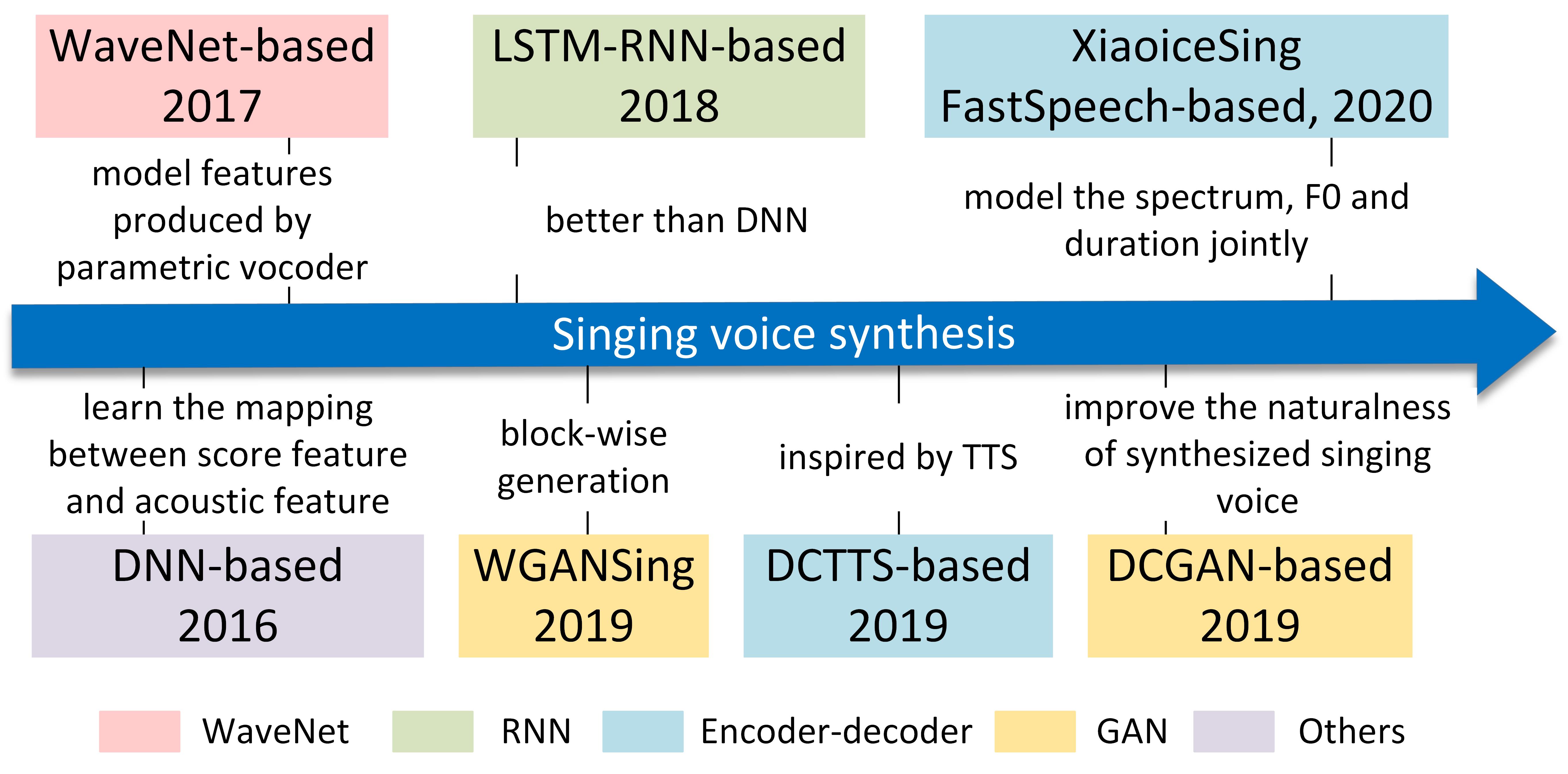}
	\caption{Chronology of singing voice synthesis. Different colors represent different model architectures or algorithms.}
	\label{fig:2020}    
\end{figure}

The initial attempt to apply deep learning method to the task of SVS is to use deep neural network (DNN). The steps for SVS are usually as follows: first use well-trained HMM to align score feature sequence with acoustic feature sequence, next exploit DNN to learn the mapping relationship between score feature and acoustic feature, obtaining acoustic feature sequence at frame-level, and then employ parameter generation algorithm to obtain a natural a singing voice trajectory, finally transformed the parameters into audio waveform with the help of vocoder \cite{175}. Nishimura et al. \cite{273} used the above-mentioned DNN-based method for singing voice synthesis. Hono et al. \cite{176} introduced a DNN-based SVS system named Sinsy which provided on-line services for SVS. Meanwhile, trajectory training, a vibrato model, and a time-lag model were introduced into the system to synthesize high quality singing voices. Experimental results showed that the DNN-based methods are better than the HMM-based methods which have been used widely before. In addition, Blaauw et al. \cite{200} proposed a new singing synthesis model based on a modified version of WaveNet, which modeled the features produced by the parametric vocoder. Parametric vocoder separated the influence of pitch and timbre, so that the pitch can be modified easily to match any target melody. This work focuses on the timbre generation without considering pitch and timing. Experimental results demonstrated that this method is superior to the existing parameter statistics and the state-of-the-art concatenative method. The system proposed in \cite{200} only produced timbre features, but does not produce features related to music expression, such as F0 and phonetic timing. Therefore, Blaauw et al. \cite{201} extended \cite{200} to include additional parts to predict F0 and phonetic timing from the score with lyrics. Compared with the concatenative method, the proposed model is more flexible and robust to small deviations between phonetic and acoustic features in training data. Kim et al. \cite{177} proposed a Korean singing voice synthesis system based on LSTM-RNN, and focused on a novel feature synthesis method based on Korean syllable structure, including linguistic features and musical features. At the same time, some post-processing methods were proposed to synchronize the duration of the synthesized singing voice with the musical note duration. Experimental results showed that LSTM-RNN is better than DNN for SVS. In the above-mentioned systems, the duration, F0 and spectrum models are trained independently, which usually leads to the neglect of their consistency in singing voice and the dependence of music features. For this reason, Lu et al. \cite{183} proposed a high-quality SVS system called XiaoiceSing, which employed an integrated network to jointly model the spectrum, F0 and duration. The system is based on the TTS model FastSpeech \cite{184} and adds singing-specific design to suit SVS tasks. Traditional DNN-based SVS lacks inter-utterance variation due to the determination of the synthesis process. Therefore, Tamaru et al. \cite{260} proposed a random modulation post-filter based on Generative Moment Matching Network (GMMN) to provide inter-utterance pitch variation for DNN-based singing voice synthesis.

The acoustic models based on DNN are usually trained by minimum Mean Square Error (MSE) criterion or maximum likelihood criterion. However, the prediction of acoustic characteristics is problematic. It is well known that the distribution of acoustic features is multi-modal, as humans can sing the same lyrics in many different ways. Traditional neural network training methods cannot learn to model acoustic feature distribution which is more complex than unimodal Gaussian distribution, leading to that the generated acoustic parameters are often over-smoothed, resulting in deterioration of the naturalness of synthesized audio. Research shows that adversarial training is able to effectively alleviate the over-smoothing effect of acoustic parameters and significantly improve the naturalness of synthesized audio \cite{205}. Therefore, like other music tasks mentioned before, GAN is also widely used in singing voice synthesis. Hono et al. \cite{205} introduced the GAN into the DNN-based singing synthesis system. Different from the vanilla GAN, the generator input here is not random noise, but the music score feature sequence. After training, the generator predicts the acoustic characteristics, and then generates smooth parameter trajectory through the maximum likelihood parameter generation (MLPG) algorithm. Later, they proposed another SVS approach based on conditional GAN (CGAN) \cite{206}. The training method based on GAN and CGAN alleviated the over-smoothing, and improved the naturalness of synthesized singing voice compared with the DNN-based method. Inspired by the Deep Convolutions Generative Adversarial Networks (DCGAN) architecture, Chandna et al. \cite{208} proposed a novel block-wise generation network for SVS, and optimized it with Wasserstein-GAN algorithm. The network takes a block of consecutive frame-wise linguistic and fundamental frequency features, along with global singer identity as input and outputs vocoder features. This block-wise training approach allows time-dependent feature modeling. In addition, Choi et al. \cite{209} also proposed a Korean SVS system based on GAN, which employed an auto-regressive algorithm with the boundary equilibrium GAN (BEGAN) objective to generate spectrogram, and finally used Griffin-Lim algorithm to reconstruct phase information and generate waveform. The above-mentioned GAN-based SVS systems only use a discriminator which directly operates on the entire sampling sequence, resulting in a limitation, i.e. the lack of diversity in sample distribution assessment. Therefore, Wu et al. \cite{181} introduced multiple random window discriminators (MRWDs) \cite{182} into the multi-singer singing voice model to make the network be a GAN. MRWDs is a set of discriminators that operated on random subsample segments of samples and allowed for the evaluation of samples in different complementary ways. Similar to \cite{202}, Wu et al. only focused on spectrum modeling and assumed that F0 is known. At the same time, in order to avoid the influence of different duration models, the real phoneme duration is employed in both training and inference stages.

There are also some researches using encoder-decoder architecture for singing voice synthesis. Aiming at the limitation that the previous network predicting the characteristics of the vocoder cannot exceed the upper limit of the performance of the vocoder, Lee et al. \cite{179} explored the direct generation of linear-spectrogram utilizing an end-to-end architecture. Similar to the goal of \cite{177}, they proposed to use the encoder-decoder architecture to synthesize Korean songs. In view of the similarity between TTS and SVS in synthesizing natural human speech, they modified a TTS model DCTTS \cite{357} to adapt to SVS tasks. After that, Lee et al. \cite{204} also extended the above-mentioned single-singer song synthesis system into a multi-singer system. Their latest development divided singer identity into two independent factors: timbre and singing style, and used two independent decoders to model the two concepts separately under the condition of encoded singer identity, two independent decoders are used to model the two concepts respectively \cite{345}. Blaauw et al. \cite{202} also exploited encoder-decoder architecture for SVS, and both of them initially aligned the input states to the output timesteps. Considering that the phonetic timings in singing voice are highly constrained by the music score, they first derived an approximate initial alignment by using an external duration model, and then refined the initial alignment results by using a decoder based on the feedforward variant of Transformer model to obtain the target acoustic characteristics. This method has faster inference speed and avoids the inductive bias problem of autoregressive model \cite{178,201} trained via teacher forcing. However, in order to achieve high performance, the seq2seq singing voice synthesizer requires a large amount of training data from a singer, which is difficult and expensive to collect in specific application scenarios. For reducing the training data of target singer, Wu et al. \cite{181} designed a multi-singer seq2seq singing voice model based on adversarial training strategy to utilize all the existing singing voice data of different singers. To reduce the imbalance of scores among singers, they employed a singer classifier incorporating an adversarial loss to encourage the encoder to learn the singer-independent representation from the score, so as to reduce the dependence of the encoder output on the singer.

The following studies are not strictly SVS, they are also discussed here for the reason that the lyrics are involved in the generation process. Previous SVS work required pre-given score and lyrics information. Liu et al. \cite{203} explored the singing voice generation without pre-assigned scores and lyrics. They proposed three unconditional or weakly conditional singing voice generation schemes. The three schemes are: 1) a free singer only uses random noise as input; 2) an accompanied singer learns over a piece of instrumental music; 3) a solo singer only uses noise as input, but it first uses noise to generate some ``inner idea" of what to sing. The accompanist singer adopts the way of bypassing the music score to generate the spectrogram directly, which is the first attempt to produce singing voice given an accompaniment. Since multi-track music is difficult to obtain, and the existing multi-track datasets containing clear voice are usually different in music genre and singing timbre, it is hard for the model to learn the relationship between singing and accompaniment. Therefore, they implemented a vocal source separation model to provide as much training data as possible, but the singing voice generation model may be affected by this model. Although the above models are tougher to train than SVS models, they enjoy more freedom in generating output. Considering the artistic nature of singing, this freedom may be expected. However, the model proposed by Liu et al. \cite{203} cannot generate samples with satisfactory perceptual quality. In addition, whether the model can be applied to non-singing voice audio remains to be explored. Therefore, Liu et al. \cite{210} extended the previous work and proposed the UNAGAN architecture based on BEGAN to generate the mel-spectrogram of singing voice unconditionally. To avoid the mode collapse problem in previous work, they introduced a cycle regularization mechanism to the generator to enhance the cyclic consistency between each input noise vector and the corresponding segment in the output mel-spectrogram. This method can not only generate singing voice but speech and instrument music (as piano and violin). The above-mentioned singing synthesis system almost only synthesizes vocal without background music. Based on VQ-VAE, Dhariwal et al. \cite{211} proposed a model that generated music with singing in the audio domain, including both vocal and background music. Moreover, they conditioned on artist and genre to steer the musical and vocal style, and on unaligned lyrics to make the singing more controllable. The model can generate high-fidelity and diverse songs with a coherence of several minutes. Yusong et al. \cite{346} even explored the composition of Peking Opera singing voice based on the Duration Informed Attention Network (DurIAN) framework.
\subsection{Fusion Generation}
\label{sec:4.4}
There are some researches that are not limited to a certain modal, but combine the multi-modal music representations in one generation framework. For example, the three-stage music generation process is directly regarded as a whole, that is, performing the generated music score and synthesizing audio. Here, we call the music generation including multi-modal music representations as fusion generation. We mainly discuss several kinds of fusion generation, one is to generate audio from music score, or MIDI-to-Audio mentioned in some researches, and the other is the reverse process of Score-to-Audio, i.e. the audio is parsed back to music score. In addition, there are some work on generating melody from lyrics.
\subsubsection{Score-to-Audio}~{}\\
\label{sec:4.4.1}
Symbolic music usually needs intermediate steps to be converted into audio output. Although models such as WaveNet \cite{149} can generate more realistic sounds, the computational cost of directly using synthesis models to generate audio is relatively high and they can't deal with the long-term dependence in music. Although the generated music is expressive and novel, it doesn't sound real due to the lack of music structure. Therefore, Manzelli et al. \cite{46} combined symbolic music model with audio model to create structured, human-like music works. They first used the biaxial LSTM model to learn the melody structure, and then use the output of LSTM as additional conditional input to generate the final audio music with the help of the WaveNet-based audio generator. Using notes as intermediate representation, Hawthorne et al. \cite{116} used musical notes as intermediate representations and trained a set of models that can transcribe, compose and synthesize audio waveforms with coherent musical structures on timescales spanning six orders of magnitude (~0.1 ms to ~100 s), a process called Wave2Midi2Wave, this model can generate about one minute of coherent piano music. The model consists of three parts, the first part is an onsets and frames transcription model which generates symbol representation (MIDI) from raw audio \cite{306}, the second part is music language model based on self-attention \cite{307}; the last part is a WaveNet model which generates performance audio based on MIDI generated by the second part. Synthesizing piano audio from MIDI requires either a large number of recordings of individual notes or a model to simulate the actual piano. However, these methods have two main drawbacks: (I) the ``stitching" of a single note may not optimally capture the various interactions between notes; and (II) the quality of synthetic audio is limited by the recording in the sound library or the piano simulator. In terms of performance style, current research either focuses on direct mapping without additional controllable conditions for performance style \cite{116}, or only focuses on global attributes, such as synthesizing different instruments \cite{255}. Therefore, Tan et al. \cite{249} put forward a GM-VAE-based controllable neural audio synthesizer to map MDI to realistic piano performance in audio domain, conditioning on two basic temporal characteristics of piano performance (articulation and dynamics). The model can achieve fine-grained style morphing by linearly interpolating between Gaussian mixture priors, and conduct  performance style transfer by interpreting each onset note with the style inferred from a given audio performance.

The above methods can only synthesize the audio of piano performance. Although WaveNet \cite{149} adopted by Manzelli et al. is able to generate any sound, it is not aimed at music. Thereby Wang et al. \cite{140} proposed to construct an AI performer to realize the conversion from music score of any instrument to expressive audio music. Specifically, they proposed a deep convolution model named PerformanceNet to learn the mapping between the symbolic representation of pianorolls and the audio representation of spectrogram in an end-to-end manner. The goal of PerformanceNet is to predict the performance-level attributes that performers may apply when performing music, such as tempo changes and pitch changes (as vibrato). Subjective evaluation results showed that this model performed better than WaveNet-based model \cite{46} and two existing synthesizers. PerformanceNet can generate audio for any instrument, but it requires to train a model for each instrument. Most of the audio synthesizers change the timbre by presetting different instrument categories. For a single instrument type, the space of timbre transformation available is very limited. One disadvantage of the previous timbre morphing method \cite{308} is that morphing can only be applied among the range of timbres covered by a specific synthesis model, whose expressiveness or parameter set may be limited. To overcome this, Kim et al. \cite{255} adopted a data-driven music synthesis method, which can be generalizable to all timbres in the dataset. Specifically, they proposed a neural music synthesis model called Mel2Mel with flexible timbre control, which can synthesize polyphonic music represented as pianoroll containing note timings and velocity into audio. The model aims to synthesize real sound while flexibly controlling timbre. The embedding space learned by Mel2Mel successfully captures all kinds of changes of timbre in a large dataset, and realizes timbre control and timbre morphing by interpolation between instruments in the embedded space. Moreover, they have developed an interactive web demo that can convert preloaded or user-provided MIDI files into Mel-spectrograms using user-selected points in the embedding space.
\subsubsection{Audio-to-Score}~{}\\
\label{sec:4.4.2}
The research on generating symbolic score from audio is relatively rare. Hung et al. \cite{195} used the neural network of music transcription based on encoder-decoder architecture to take a polyphonic music audio as input and predict its score as output, not only predicting the note pitch, but also predicting the instrument that plays each note. They can also generate new score rearrangement from audio input by modifying the timbre representations. Haki et al. \cite{173} proposed to use seq2seq learning based on word-RNN to generate bass music score which is consistent with the given drum audio in style and rhythm. Wei et al. \cite{98} put forward a VAE-based conditional drum pattern generation model. The symbolic drum pattern was generated given the accompaniment composed of melody sequence. During this process, self-similarity matrix (SSM) was used to encapsulate the global structure of melody sequence and investigate its potential impacts on generated drum sequence. Fusion generation is embodied in that the CQT spectrogram converted from audio domain data is utilized for melody SSM calculation, and symbolic drum track data is directly employed for drum SSM calculation; the input of drum pattern generator is spectrogram, and the generated drum track is symbolic data.
\subsubsection{Lyrics-to-Melody}~{}\\
\label{sec:4.4.3}
Generating melody from lyrics is to compose the corresponding melody and exact alignment between the generated melody and the given lyrics using a piece of text as input. There have been some non-deep learning researches on the generation of melodies using lyric conditions \cite{326,327}, which usually regard the task of melody composition as a classification or sequence labeling problem. They first determine the number of notes by counting the syllables in the lyrics, and then predict the notes one after another by considering the previously generated notes and the corresponding lyrics. However, these works only consider the ``one-to-one" alignment between melody and lyrics, which may lead to deviation in melody composition. Therefore, Bao et al. \cite{328} proposed a melody composition model, songwriter, which can generate melody from lyrics and handle the ``one to many" alignment between the generated melody and the given lyrics. Songwriter is the first end-to-end neural network model for melody composition from lyrics. They first divided the input lyrics into sentences, then used the seq2seq-based model to generate melodies from the sentences one by one, and finally merge these segments into a complete melody. Yu et al. \cite{329,330} utilized a new deep generative model conditional LSTM-GAN to generate melody from lyrics, in which the generator and discriminator were LSTM networks with lyrics as conditions. Specifically, firstly, the skip-gram model is exploited to encode the syllables/words of lyrics into a series of embedding vectors, and then the LSTM-GAN is employed to generate the melody matching the given lyrics. They also publish an aligned English lyrics-melody paired dataset. In addition, iComposer developed by Lee et al. \cite{323} can not only automatically generate melody accompaniment for a given text, but generate a set of lyrics for arbitrary melody (MIDI file) given by users.
\subsection{Style Transfer}
\label{sec:4.5}
Gatys et al. \cite{298} introduced the term style transfer in the context of neural network, which usually refers to preserving the explicit content features of an image and applying to it explicit style features of another image. Music style transfer is to generate creative and human-like novel music by separating and recombining the music content and music style of different music pieces. Music style can literally refer to any level of musical characteristics. The boundary between content and style is highly dynamic, and different objective functions in timbre, performance style or composition are related to distinct style transfer problems \cite{166}. In general, style transfer can be accomplished in two steps: first, the ``style" code is separated from the latent vector representation which is used to generate music, and then the style code is ``inserted" into an appropriate sequence generation framework that retains all other elements. We will classify and discuss the music style transfer according to distinct separation styles below.
\subsubsection{Score style transfer}~{}\\
\label{sec:4.5.1}
For music score, music style refers to the composition characteristics of tone, rhythm and chord sequence. Here we mainly introduce the style transfer in the popular sense, i.e. genre transfer. Disentanglement learning is usually employed to separate style and content. Theoretically, disentanglement learning can disentangle any two interpretable variables, such as the disentanglement of rhythm-content and chord-texture introduced below. In addition, there are also several few researches on style fusion, which we will briefly introduce here.
\paragraph{\large Genre style transfer}~{}\\
In such tasks, ``style" stands for gene, which is the most intuitive understanding of style for human beings. The main difficulty of the current genre style transfer task is the lack of alignment data, so most of the researches related to this task adopt unsupervised learning frameworks, such as MDI-VAE \cite{48}, and apply them to datasets with genre labels. Brunner et al. \cite{48} proposed a VAE-based model called MIDI-VAE, which can realize the style transfer of symbolic music by automatically changing pitch, dynamics and instrument. It is the first successful attempt to apply unaligned style transfer to complete music works. MIDI-VAE contains a style classifier that forces the model to encode the style information in the shared latent space, so as to manipulate the existing songs and change their styles effectively. The model can generate multi-instrument polyphonic music, and learn the dynamics of music by incorporating note durations and velocities. Lu et al. \cite{50} studied the style transfer of homophonic music, which only considered the style transfer of the accompaniment part, and took the melody as the condition of the network. They employed the DeepBach-based model \cite{26} to model the temporal information combined with the autoregressive model WaveNet \cite{149} to model the pitch information. They especially improved the DeepBach \cite{26} model so that it is no longer limited to the number of parts and can be applied to various polyphonic music. Exploiting the Gibbs sampling on the above generative model to modify the accompaniment part, the model can transfer arbitrary homophonic music scores into the styles of Bach’s 4-part chorales and Jazz. The GAN model with multiple generators and cycle consistency loss has achieved great success in image style transfer. Therefore, Brunner et al. \cite{65} adopted the previous image style transfer model named CycleGAN to realize symbol music gene transfer. Since no alignment data is required, the model can be easily retrained and utilize music data from other styles, which is also the first application of GAN in symbolic music transfer. However, this gene transfer only changes the pitch of the note and does not include variations in instrument, rhythm and dynamics information. Compared with MIDI-VAE \cite{48}, the CycleGAN-based model does not limit the number of notes to be played simultaneously, thus making the music sound richer. The work of Wang et al. \cite{66} continues the work of Brunner et al. \cite{65} based on CycleGAN. Specifically, they proposed a two-GAN architecture, in which one GAN was used to transfer information from style a to style B, and the other was to transfer information in the opposite direction. Similarly, the style transfer proposed by them only involves the change of pitch, and does not include music dynamics information. In order to solve the problem of the lack of aligned data, Cífka et al. \cite{40} designed a synthetic data generation scheme, using RealBand in Band-in-a-Box software toolkit to generate almost unlimited amount of aligned data. Then, they proposed a symbol music accompaniment style transfer algorithm based on encoder-decoder architecture, which was the first research work that completely employed a supervised algorithm to achieve style transfer. The lack of jazz data in MIDI format hinders the construction of jazz generation model. Therefore, Hung et al. \cite{73} tried to solve this problem by transfer learning. They used a recurrent VAE model as the generative model, a genre-unspecified dataset as the source dataset, and a jazz-only dataset as the target dataset. After that, they evaluated two transfer learning methods. The first method first trained the model on the source dataset, and then fine-tunes the model parameters on the target dataset; the second method trained the model on both the source dataset and the target dataset, but added genre label to the latent vector, and used gene classifier to improve jazz generation. The evaluation results demonstrated that these two methods are better than the baseline method which only uses the jazz data for large-scale training. In addition, Voschezang \cite{103} proposed a lightweight but high-resolution variational recurrent autoencoder based on MusicVAE \cite{23} to realize drum pattern style transfer.
\paragraph{\large Rhythm style transfer}~{}\\
The ``style" in this type of task refers to the rhythm in music. Generally speaking, the rhythm is the duration and strength of notes in music movements. Here we do not consider the strength. Yang et al. \cite{41} proposed a new MusicVAE-based \cite{23} method  to disentangle pitch and rhythm dimensions in latent representation. But the model can only handle two-bar music segments, and employed violent splicing for longer music pieces. This kind of disentanglement is similar to style transfer. The difference is that the disentanglement of content and style in style transfer is carried out automatically by the model, while the disentanglement mentioned in \cite{41} is to manually select the latent space dimensions that can optimally represent pitch information or rhythm information. This work laid the foundation for their later research on rhythm style transfer. Based on the above work, Yang et al. \cite{38} proposed an explicitly-constrained conditional variational autoencoder (EC2-VAE), which is dedicated to disentangling the pitch and rhythm representation of 8-beat music clips conditioned on chords and then concatenating the rhythm features and pitch latent vector to reconstruct the new melody, so as to realize the rhythm transfer between music. However, changing the rhythm characteristics of melody will inevitably affect the pitch contour, which would lead to the complicated behaviors of the latent space difficult to be explained intuitively.
\paragraph{\large Others}~{}\\
Inspired by the idea of content-style disentanglement in style transfer method, Wang et al. \cite{291} proposed to use VAE to learn two interpretable latent factors of polyphonic music (8-beat long piano piece): chord and texture. Chord can be analogized as content, which refers to chord progression; texture can take analogy as style, referring to chord arrangement, rhythmic pattern and melody contour. The disentanglement of chord-texture provides a controllable generation way, which leads to a wide range of applications, including compositional style transfer, texture variation and accompaniment arrangement. The core of the model design lies in the encoder, which contains two inductive biases to successfully disentangle chord-texture. The former employed a rule-based chord recognizer to embed information into the first half of the latent representation. The latter regarded music as a two-dimensional image, extracted texture information by chord-invariant convolution network, then stored it in the second half of latent representation. The decoder adopted the design of PianoTree-VAE \cite{292}, which can reconstruct polyphonic music hierarchically from latent representation.

Style fusion is the mixing of two or more styles to generate a new unknown style. Chen et al. \cite{253} proposed a new gene fusion framework FusionGAN based on three-way GAN, which combines the advantages of generative adversarial network and dual learning to generate music with fusion music genre. In order to effectively quantify the differences between different domains and avoid the problem of gradient vanishing, FusionGAN proposed a Wasserstein based metric to approximate the distance between the target domain and the existing domain, and utilized Wasserstein distance to create a new domain by combining the patterns of the existing domains using adversarial learning. The experimental results showed that the method can effectively merge two music genres.
\subsubsection{Audio Style Transfer}~{}\\
\label{sec:4.5.2}
The style in audio style transfer generally refers to audio features such as timbre and texture. This section mainly discusses two types of style transfer tasks: timbre transfer and non-timbre transfer. Non-timbre characteristics refer to other musical characteristics besides timbre/instrument, such as genre and texture. Since audio can be represented as 2D spectrogram, CNN is often used in audio style transfer.
\paragraph{\large General style transfer}~{}\\
As Ulyanov et al. \cite{156} said, ``Most probably you would say that style transfer for audio is to transfer voice, instruments, intonations. In fact, neural style transfer does none aim to do any of that. So we call it style transfer by analogy with image style transfer because we apply the same method." Styles discussed in this section refer to styles other than timbre/instrument, such as genre, texture, etc.

The spectral representation implies that CNN applied to image can also be applied to audio. Wyse \cite{155} uses the pre-trained VGG-19 network to realize audio style transfer based on spectrogram. Firstly, it is necessary to solve the problem that the image network processes three-channel RGB input, while the spectrogram has single-channel magnitude value. The single-channel magnitude value of the spectrogram must be replicated on 3 channels in order to work with the pre-trained network. Since color channels are processed differently from each other in the neural network, the post-processed synthetic color image must be converted back to a single channel based on luminosity to be meaningful as a spectrogram. This method is not as effective as image style transfer because the frequency axis does not possess translation-invariance. In addition, audio objects are non-locally distributed over a spectrogram, and visual objects are often composed of neighboring pixels in images. Multiple objects in audio can have energy at the same frequency, while a given pixel in visual images is almost always matched to only one object. Ulyanov et al. \cite{156} performed an audio style transfer algorithm using the log-magnitude STFT of audio with the help of a convolution layer with random untrained weights. They represented $F$(Frequency bins)$\times T$(Time bins) spectrogram as $1\times T$ images with $F$ color channels. Therefore, convolution is only carried out in the time dimension, while the patterns and relationship between frequencies are localized within the style loss. The method proposed by Ulyanov et al. \cite{156} produced results as good as any other method, and the sound examples published by them are indeed convincing \cite{155}.

Ulyanov’s method \cite{156} needs to reverse the optimized magnitude value back to the audio signal using the Griffin-Lim phase reconstruction algorithm iteratively. The utilization of phase reconstruction ultimately means that the stylization process does not model the fine temporal characteristics of the audio signal contained in its phase information. And the quality of phase reconstruction will be limited by the Griffin-Lim algorithm. Therefore, Mital \cite{157} explored to directly optimize the time-domain audio signals by eliminating the phase reconstruction process, so as to realize the audio style transfer or synthesis application. They explored the process of style transfer on various neural networks with time-domain signals as input which all adopted Ulyanov's style transfer technology. The results showed that only the network using combinations of real, imaginary, and magnitude can produce meaningful audio. Compared with the Ulyanov’s method, the noise produced by this method is much smaller. Moreover, the method of Ulyanov et al. \cite{156} cannot capture rhythm or chord information, i.e. neither it is able to represent the long-term characteristics of audio signals effectively, nor can it represent music-related frequency patterns along the frequency axis that can be shifted linearly, such as chords and intervals. To solve this problem, Barry et al. \cite{158} expanded the algorithm proposed by Ulyanov et al. to explore long-term and high-quality audio style transfer and texture synthesis in time domain, and capture acoustic features related to music style, such as harmony, rhythm and timbre. Apart from the log-magnitude STFT representation used by Ulyanov et al., they proposed two additional audio representations with accompanying neural structure: Mel spectrogram and constant-Q transformation (CQT) spectrogram. Mel spectrogram is used to capture rhythmic information better, and CQT spectrogram is used to represent harmonic style. Like the STFT representation in Ulyanov's method, the mel-frequency axis is also regarded as channel axis rather than spatial axis. The results implied that the advantages of the three audio representations are fully utilized in the parallel neural network to help capture various desired characteristics of the audio signal, resulting in the most convincing examples of style transfer.

Texture refers to the collective temporal homogeneity of acoustic events. Texture generation can be regarded as a special case of audio style transfer. By optimizing only those terms of the loss function associated with the texture, and not involving any loss terms related to the content (i.e. the content weight is set to zero), the texture generation can be realized, such as \cite{156,158}.
\paragraph{\large Timbre style transfer}~{}\\
Timbre is a set of attributes that distinguish one instrument from another under the same pitch and loudness. Timbre transfer applies part of the auditory characteristics of one instrument to another. The transfer of all timbre characteristics of an instrument will result in domain conversion, and partial modification of these characteristics will lead to style transfer \cite{165}. This paper does not make an obvious distinction between domain conversion and style transfer, which are collectively referred to as timbre style transfer.

Strictly aligned training data is difficult to obtain, and there are several methods to avoid the limitation of training data. Mor et al. \cite{159} proposed an unsupervised method for music translation across musical instruments, genres and styles based on a multi-domain WaveNet autoencoder. They adopted the same autoencoder architecture as \cite{152}, and used a domain confusion network in the latent space to provide adversarial signals to the encoder to ensure that the domain-specific information is not encoded. At the same time, random local pitch modulation was applied to the input audio to make the encoder encode the input signal in a semantic way rather than memorize it. In terms of encoding, the network can successfully handle unseen instruments or other sources, such as whistles; in terms of decoding, new instruments can be added without retraining the entire network. In order to realize multi-modal and non-deterministic mapping between different domains, Lu et al. \cite{166} proposed an unsupervised method without the need for parallel data to realize one-to-many style transfer music generation. They adopted the Multi-modal Unsupervised Image-to-Image Translation (MUNIT) framework to represent the multi-modal distribution of music pieces, which can generate various outputs from the learnt latent distribution of the content and style. Experimental results proved that the proposed method has the advantages of improving the sound quality in music style transfer and allowing users to manipulate the output. Although the universal music translation (UMT) proposed by Mor et al. \cite{159} supported translation across multiple complicated audio domains, it needs to learn a separate decoder for each domain, resulting in long training time. In addition, it does not provide control over audio synthesis, and the learnt representation cannot be directly visualized, nor can it only transfer specific parts of timbre attributes. Therefore, Biton et al. \cite{165} introduced the Modulated Variational auto-Encoder (MoVE) to perform many-to-many timbre transfer. They used Feature-wise Linear Modulation (FiLM) to condition the existing domain translation techniques (UNIT) \cite{160}, and realized timbre transfer by switching the categorical condition between different instruments. By further conditioning the system on several different instruments, they generalized the system to many-to-many transfer within a single variational architecture able to perform multi-domain transfers.

None of the above research on timbre transfer solves the problem of learning pitch and timbre latent variables. Hung et al. \cite{164} proposed two CNN (DuoAE and UnetAE) using encoder/decoder and adversarial training to learn the disentangled representation for timbre and pitch in music audio, and used the learned features to generate new audio signals to evaluate the effectiveness of disentanglement. The experimental results showed that UnetAE had better ability to create new instruments without losing pitch information. However, one drawback of the current model is that the timbre embedding must be selected from another music piece. In addition, the model adopted audio synthesized from MIDI and relied on clean frame-wise labels, which are rarely found. Luo et al. \cite{194} used GM-VAEs to learn the disentangled representation of timbre and pitch for musical instrument sounds. The model allows controllable synthesis of the selected instrument sounds by sampling from the latent space. The model can also realize the timbre transfer between different instruments. Different from the previous research on timbre transfer, the proposed framework realized the timbre transfer between multiple instruments without training a domain-specific decoder for each instrument as in \cite{159}, and inferred pitch and timbre latent variables without requiring categorical conditions of source pitch and instrument as in \cite{165}. In addition, inspired by Tacotron2 processing the time-frequency representation of speech and using WaveNet to output high-quality audio conditioned on the generated Mel spectrogram, Huang et al. \cite{197} introduced a timbre transfer method called TimbreTron, which applied the ``image" domain style transfer to a time-frequency representation of audio signals. Specifically, they employed CycleGAN to achieve timbre transfer in the log-CQT domain. 
\subsubsection{Singing Style Transfer}~{}\\
\label{sec:4.5.3}
Singing style transfer is to change the singer's timbre and the way the song is performed, as known as singing voice conversion (SVC). In addition, there is a special style transfer task named speech-to-sing (STS), which refers to the conversion of speech into singing voice. The ``style" here not only refers to incorporating melody, but also changing the timbre of the speaker into that of singing.
\paragraph{\large Singing Voice Conversion(SVC)}~{}\\
Singing Voice Conversion (SVC) refers to converting a song of a source singer to sound like the voice of a target singer without changing the linguistic content \cite{188}. In the field of SVC, there are relatively few researches on the singing voice transfer using deep learning. At present, it is mainly based on statistical methods, and most methods can only realize the transformation with parallel training datasets, that is, different singers are required to sing the same song. It is hard to collect parallel datasets, even if parallel datasets are collected, data alignment is a time-consuming and laborious task \cite{189}.

Xin Chen et al. \cite{190} proposed a many-to-one SVC method without parallel data, which was the first study to train SVC system using non-parallel data. The SVC they understand is to convert the person-dependent content from the source to the target while preserving the person-independent content. They viewed the phonetic posterior as the person-independent content of SVC, which was first generated by decoding singing voices through a robust Automatic Speech Recognition Engine (ASR). Then, RNN was used to model the mapping from the person-independent phonetic posteriors to the acoustic features of target person including both person-dependent and person-independent content. Finally, F0 and aperiodic information were extracted from the original singing voice together with the target acoustic features to reconstruct the target singing voice through vocoder. However, this method needs to train different models for different target singers. And because the system uses an automatic speech recognition engine, the performance of the system also depends on the quality of speech recognition. Nachmani et al. \cite{187} proposed a deep learning method based on WaveNet autoencoder for unsupervised singing voice conversion. The model neither requires parallel training data between singers, nor does it need to transcribe audio into text or musical notes. Each timestep the conditioning of decoder is on a concatenation of the target singer's embedding vector stored in a look-up table (LUT) and the output of the encoder to generate the probability of the signal in the next timestep. The model also contains a domain confusion network that enforces the latent representation to be singer-agnostic. In addition to the problem of relying on parallel data, most SVC methods can only achieve one-to-one conversion, and the generalization ability is weak. Therefore, Hu et al. \cite{189} proposed a cycle-consistent generative adversarial learning model, MSVC-GAN, which can use non-parallel data to realize many-to-many SVC. Through adversarial loss, MSVC-GAN learns the acoustic feature distribution of different singers, and establishes the forward and reverse mapping among the acoustic features of different singers, that is, the model can learn a unique function for all many-to-many mappings. Moreover, Sisman et al. \cite{188} adopted the GAN-based conversion framework SINGAN for singing voice conversion, which was also the first attempt for GAN in SVC tasks. Blaauw et al. \cite{185} utilized the voice cloning technology in TTS for SVS based on the autoregressive neural network architecture. Firstly, a multi-speaker model is created by using the data from multiple speakers, and then a small amount of target data is employed to adapt the model to unseen sounds. They only focus on the cloning of timbre without considering the expressive aspects related to the interpretation of the given musical score. Nercesian \cite{354} proposed to use speaker embedding network together with the WORLD vocoder to perform zero-shot SVC, and designed two encoder-decoder architectures to realize it, which is also the first work of zero-shot SVC. The speaker embedding network can adapt to the new voice on-the-fly, train the unlabeled data model, and conduct the initial training on a large speech dataset which are more widely available, followed by model adaptation on a smaller singing voice dataset.
\paragraph{\large Speech-to-Sing(STS)}~{}\\
Although the existing SVS algorithms can produce natural singing voices, they usually need a large amount of singing data to train new sounds. Compared with the normal speech data, the collection of singing voice data is more difficult and costly. To alleviate this limitation, someone recently proposed a more effective SVS method \cite{185}, i.e. a multi-speaker SVS model trained from the minimum amount of singing voice data of the target speaker. There are also some unsupervised SVC methods \cite{187} which can also generate new sounds through SVC. Although data-efficient SVS methods and unsupervised SVC methods can effectively generate new sounds, they still need to obtain a small number of singing voice samples from the target speaker, which limits the application of singing voice synthesis to relatively limited scenarios, that is, the singing voice of the target speaker must be available \cite{302}. Normal speech samples are much easier to collect than singing voices, so the speech-to-sing task was born. Speech-to-sing (STS) can be regarded as an example of the general problem of style transfer. Compared with other style transfer work, the ``style" of STS not only refers to the incorporation of the required melody, but also requires maintenance of the speaker's identity while transferring the timbre of speech to that of singing \cite{186}.

The applicability of the techniques previously applied to STS tasks is limited because they require additional inputs, such as high-quality singing templates \cite{305} or phoneme-score synchronization information \cite{304}, which are difficult to deduce for any general speech and melody. Zhang et al. \cite{302,303} proposed an algorithm to directly synthesize high-quality target speaker singing voice by learning speech features from normal speech samples. The key to the algorithm is to integrate speech synthesis and singing voice synthesis into a unified framework and learn the universal speaker embeddings that can be shared between speech synthesis and singing voice synthesis tasks. Specifically, they proposed DurIAN-4S, a speech and singing voice synthesis system based on the previously proposed autoregressive generation model DurIAN. The entire model is trained together with the learnable speaker embedding as the conditional input of the model. By selecting different speaker embedding in the process of singing voice generation, the model can generate different singing voices. The algorithm was evaluated on the task of SVC, and results showed that the algorithm can produce high-quality singing voice which is highly similar to the target speaker's voice given the normal speech samples of the target speaker. Parekh et al. \cite{186} explored a method to achieve STS conversion by employing the minimal additional information over the melody contour. Given the time-frequency representation and the target melody contour of the speech, they adopted an encoder-decoder network based on popular U-net architecture to learn the encoding that can synthesize the singing voice, which is able to preserve the speaker's linguistic content and timbre while maintaining the target melody.
\subsection{Interactive Generation}
\label{sec:4.6}
Automatic music generation can indeed bring inspiration to human creation, but the current music generated automatically by machines is not comparable to human creation. Although some generation may be equivalent to the level of human creation, it is limited to a certain music style like Bach. Therefore, for now it is not hoped that machines will drive the whole music generative process. Human participation can enable the generation to capture subjective and context-dependent preferences, and simplify the modeling problem in complex tasks, so as to shape the results more meaningfully \cite{97}. In order to achieve the above creative purposes, some researchers have begun to devote themselves to the interactive generation music. Interactive music generation refers to the cooperation between human and machine to generate music in a certain way. Though automatic music generation has been a very active research field, there is little research on real-time human-computer interaction music generation. The concept of designing an AI-based interactive interface for music creators is still developing rapidly. According to \cite{91}, interactive music generation can be divided into two categories: call and response and accompaniment. For call and response, machine learns to play a solo alternately with the user in real-time, i.e. the machine and the user play in turn; accompaniment generation requires machines to adapt and support human performance. Pachet \cite{93} also called these two types of music interaction protocols as Continuation mode and Collaboration mode.
\subsubsection{Call and response}~{}\\
\label{sec:4.6.1}
Bretan et al. \cite{95} proposed a system that employs a deep autoencoder to learn the semantic embedding of music input in order to study the music interaction scenario of call and response. The system first transforms these embeddings, and then reconstructs the transformation vectors by employing a combination of generation and unit selection to generate appropriate music responses, where the selection is based on the nearest neighbor search within the embedding space. During the live demonstration, the human plays MIDI keyboard, and the computer generates the response and plays it through the speaker. There are multiple systems (such as GenJam \cite{17}, AI Duet \cite{92}, etc.) that also perform in the configuration of ``call and response" for human-computer cooperation \cite{94}. However, one shortcoming of call and response that it does not allow human and machine to play at the same time. To solve this problem, Benetatos et al. \cite{94} proposed a system called BachDuet which can perform real-time counterpoint improvisation between human and machine. The system utilizes RNN to process the monophonic performance of human musicians on a MIDI keyboard, and generates the monophonic performance of machine in real-time. BachDuet supports two kinds of human-machine music improvisation: one is free counterpoint improvisation; the other is counterpoint improvisation with a given bass part as condition. Most of the deep learning models run in ``offline" mode with almost no limitation on processing time. Integrating these models into real-time structured performances is a challenge. In addition, these models are often agnostic to the style of a performer, making them impractical for live performances. Castro \cite{235} proposed a software system integrating off-the-shelf music generation models into improvisation. The system combines human improvisation with the melody and rhythm generated by deep learning model to solve the real-time compatibility and style personalization problems of previous models. Specifically, he exploited melody generated by Melody RNN \cite{265} and rhythm provided by human performers to improvise, which not only produced novel melody, but maintained the style of human improvisation. The impromptu part of the system adopts call and response mode as well.
\subsubsection{Accompaniment}~{}\\
\label{sec:4.6.2}
Bach Doodle \cite{43} mentioned earlier belongs to interactive music generation in the form of accompaniment. Users create their own melodies and then generate accompaniment of Bach style for melody using improved Coconet model. However, the system cannot generate accompaniment in real-time, but utilizes Gibbs sampling to generate other three parts offline. After that, Jiang et al. \cite{100} proposed a new deep reinforcement learning algorithm RL-duet for online accompaniment generation. Compared with Bach Doodle, RL-duet can realize real-time interaction with human input without any delay. RL-duet is also the first reinforcement learning model that uses a set of reward models to generate music accompaniment online. The key of the algorithm is to propose an effective reward model that considers both the inter-part and intra-part compatibility of the generated notes from horizontal and vertical perspectives. Reward model is learned from training data, not defined by manual music composition rules. SequenceTutor \cite{247} also gives partial rewards through training neural networks, but it mainly relies on more than a dozen handmade music rules, and SequenceTutor does not consider the coordination between polyphonic parts, which is very crucial for human-computer interaction improvisation. Additionally, Lee et al. \cite{323} developed an interactive web-based bidirectional song (Chinese pop music) creation system named iComposer, which can automatically generate melody accompaniment for a given text, or generate a set of lyrics for arbitrary melody (MIDI file) given by users. The iComposer assists human creators by greatly simplifying music production procession.
\subsubsection{Others}~{}\\
\label{sec:4.6.3}
Apart from call and response and accompaniment, there are quite a few music interactive generation systems interacting with users in other ways. For example, Roberts et al. \cite{96} used the latent space in MusicVAE \cite{23} to develop a music sequence browse-based interface for the generation of two-bar melody loop and drum beat. Users can also define 1D trajectories in 2D space for autonomous and continuous morphing during the process of improvisation. Huang et al. \cite{97} proposed a generation framework of mixed-initiative co-creativity. Mixed-initiative means that human and AI systems can ``take the initiative" to make decisions. Co-creativity means that the output of generation system is driven by the creative input from humans and generative technologies. Bazin et al. \cite{90} proposed an interactive music generation interface named NONOTO based on the inpainting model. The proposed system is similar to the FlowComposer system \cite{299}, but FlowComposer does not have real-time audio and MIDI playback, so the level of interactivity is greatly reduced, limiting the tool only to solely studio usage and making the user experience less spontaneous and reactive. Donahue et al. \cite{243} designed Piano Genie, an intelligent instrument, with which users can generate piano music in real-time by playing on eight buttons. The process is realized by training a recurrent autoencoder with discrete bottlenecks. In order to better meet the needs of human and AI co-creation, Louis et al. \cite{244} developed Cococo (collaborative co creation) based on Coconet, a music editor web-interface for novice-AI co-creation. It augments the standard generative music interface through a set of AI-steering tools, including Voice Lanes used to limit the generation of content into specific sounds (such as soprano, bass, etc.); Example-based Sliders used to control the similarity between generated content and existing music samples; Semantic sliders used to promote music generation towards high-level directions (e.g. happier/sadder, more conventional/more surprising); and Multiple Alternatives used audition and select the generated content. These tools not only allow users to express music intention better, but also contribute to a greater sense of self-efficacy and ownership of the composition relative to the AI. Few studies have focused on the multi-track music generation considering emotion and the participation of a large number of people, thus Qi et al. \cite{229} developed an interactive music composition system named Mind Band, which takes expressions, images and humming as input and outputs music. The output music matches their input emotions. Mind band can be roughly divided into three main parts: input, music generation model and output. The input of music generation model can be humming, VAE-GAN music composing and image/emoji. Music generation database is maintained by music conversion, music generation and music indexing. Similarly, Liu et al. \cite{325} introduced a WeChat application, Composer4Everyone, which can generate music for a recording. Users can input audio through humming, singing and even speaking, then Composer4everyone takes the input audio as the motivation to develop into music of 20-30 seconds. It can convert piano music into other styles as well, such as pop, electronic, classical, etc.

There are also some researches on interactive generation of drum patterns. Hutchings et al. \cite{102} proposed a Musically Attentive Neural Drum Improver (MANDI) system based on temporal convolution network (TCN), which can generate drum patterns in real-time and improvise with human performers. Aouameur et al. \cite{347} also proposed a real-time drum sound generation system. The system consists of two parts: a drum sound generation model and a Max4Live plugin providing intuitive control during the generation process. The generation model is comprised of a Conditional Wasserstein autoencoder (CWAE) coupled with a Multi-Head Convolutional Neural Network (MCNN). CWAE learns to generate Mel-scaled spectrograms of short percussion samples, and MCNN estimates the corresponding audio signal from the magnitude spectrogram. The Max4Live plugin allows users to synthesize a variety of sounds by exploring a continuous and well-organized parameters latent space. Alain et al. \cite{348} proposed a drum loop generation tool named DeepDrummer based on deep learning model to learn human users' preferences from a small number of interactions using active learning. The system consists of three parts: interface, critic and generator. Users listen to the generated drum loops by the generator in turn on the interface, and score them by clicking the like or dislike button, so as to update the critic model by giving feedback to the network.

In addition, the interaction with audio generation system has been studied in many aspects: from helping users generate audio to automatically creating audio content according to the target timbre given by users. Sarroff et al. \cite{147} developed an interactive music audio synthesis system by using the stacked autoencoding neural network. Users can directly interact with the model parameters to generate music audio in real-time. However, the performance of this autoencoding architecture is not good. Therefore, based on the work of \cite{147}, Colonel et al. \cite{154} proposed a novel architecture to generate audio music by improving the autoencoder. Roberts et al. \cite{96} developed an instrument timbre browsing interface using the latent space of instrument samples generated by WaveNet autoencoder \cite{152}. The interface provides an intuitive and smooth search space for the learnt audio latent space, which is utilized for the morphing between different instrument timbres. Moreover, the SVS system Sinsy proposed by Hono et al. \cite{176} can offer online singing voice synthesis services.
\subsection{Music Inpainting/Completion}
\label{sec:4.7}
Music inpainting/completion refers to filling the missing information in a piece of music, which is more in line with the human creation process.
\subsubsection{Score Inpainting}~{}\\
Pati et al. \cite{70} defines the music score inpainting problem as: given the past music context $C_p$ and the future music context $C_f$, an inpainted sequence $C_i$ is generated to connect $C_p$ and $C_f$ in a musically meaningful manner. Previous music generation models mostly assume that music is generated continuously, that is, the generated music segment only depends on the previous music segments. This method is inconsistent with the typical human composition practice, which is usually iterative and non-sequential. And the sequential generation paradigm seriously limits the degree of interactivity allowed by these models, once generated, it is impossible to adjust the specific parts of the generation to meet the user's aesthetic sensibilities or compositional requirements. Music inpainting is more in line with the process of human music creation. Users input a part of the music fragments that have been created, and complete the rest with the help of computer, so as to bring some inspiration for human creation.

Compared with other generation tasks, there are not many researches on music score inpainting generation. The previously mentioned Anticipation-RNN \cite{22} can selectively regenerate the notes at a specific position. DeepBach \cite{26} predicts the current note with the help of the surrounding note context, and can also be used to complete Bach's four parts chorales. Huang et al. \cite{52} trained a convolutional neural network Coconet to complete partial music score, which is an instance of orderless NADE, and explored the use of blocked Gibbs sampling as an analogue to rewriting.  Neither the model nor the generating process is related to the specific causal direction of composition. Inspired by Coconet, Ippolito et al. \cite{89} trained a Transformer to infill the missing part of MIDI transcription of performed piano music. Composers can utilize this infilling technique to select contiguous sections of music to be rewritten by neural networks, or gradually morph one piece into another through iterative Gibbs sampling. Recently, Pati et al. \cite{70} proposed a score inpainting model based on latent representation, Music InpaintNet, which considers both the past and the future music context, and allows users to change the musical context to achieve interactive music generation. Some previous work interacts with latent space relying on simple methods such as attribute vectors or linear interpolation, they trained a conditional RNN to use latent embeddings to learn complex trajectories in the latent space, and then sent the output of the RNN to the VAE to generate multiple music bars connecting two music clips. Nakamura et al. \cite{88} proposed a melody completion method based on image completion network. They first represent melody as images considering pitch and rhythm, and then train melody completion network to complete these images conditioned on given chord progression. The results showed that the inpainted music is original, but the algorithm of converting images into actual notes needs to be improved. Inspired by the work of sketching and patching in computer vision, Chen et al. \cite{257} proposed a neural network framework for music completion: Music SketchNet, which focuses on filling missing measures in incomplete monophonic music according to the surrounding context and user-specified music ideas (as pitch and rhythm segments). Unlike previous music inpainting work, in addition to the surrounding context, Music SketchNet also considers user preferences. The experimental results demonstrated that the generated melody follows the constraints from users, and Music SketchNet is superior to the previous Music InpaintNet, both objectively and subjectively. Wayne et al. \cite{53} introduced a new method of music generation using self-correcting, non-chronological, autoregressive model named ES-Net. ES-Net can not only add notes to complete music score, but also delete wrong notes. This method allows the model to correct the previous mistakes, prevent accumulation of errors of the autoregressive model, and allows for finer, note-by-note control in the process of human and AI collaborative composition. The experimental results showed that ES-Net produces better results than orderless NADE and Gibbs sampling. The difference between ES-Net and Coconet is that Coconet uses Gibbs sampling for inference, while ES-Net employs ancestor sampling. And Coconet did not explicitly train the model to delete notes, but during the Gibbs sampling masking process, notes may be removed; ES-Net explicitly models note removal.
\subsubsection{Audio Inpainting}~{}\\
\label{sec:4.7.2}
Audio inpainting refers to the restoration of local lost information, which is also called audio interpolation and extrapolation \cite{336,337}, or waveform substitution \cite{338} in the literature. Audio inpainting techniques can be roughly divided into two categories. For short damage less than tens of milliseconds, the goal of audio inpainting algorithm is to accurately recover the lost information \cite{339}; for long corruptions, this goal becomes unrealistic, so audio inpainting algorithms attempt to reduce the damage by preventing audible artifacts and introducing new coherent information. New information needs to be semantically compatible, which is a challenging task for music, because music usually has a strictly underlying structure with long dependencies. Previous audio inpainting work can only solve the problem of short interval \cite{341}, or rely on copying partial available information samples from other signal \cite{340}. In recent two years, researchers have gradually used deep learning and time-frequency representation to solve audio inpainting problems.

Marafioti et al. \cite{341} used time-frequency representation to restore the musical and instrumental audio in the range of tens of milliseconds. DNN is comprised of convolution layers and full connection layers, and its input is context, i.e. the signals surrounding the gap. Therefore, this algorithm is also called audio-inpainting context encoder. The results showed that the model has better performance in inpainting musical audio, and the ability of inpainting instrumental audio is still inferior to the reference model (LPC). Chang et al. \cite{342} explored the possibility of applying deep learning frameworks in different fields (including audio synthesis and image inpainting) to audio inpainting, designed novel waveform-based and spectrogram-based models for long audio inpainting, and discussed the effects of gap size, receptive field and audio representation on the inpainting performance. It seems the first to use neural networks to evaluate long audio (100 to 400 ms) inpainting method. In addition, in order to achieve the goal of inpainting long-gap audio by generating new content, Marafioti et al. \cite{343} proposed GACELA, a new audio inpainting algorithm based on GAN, which can restore missing musical audio data with a duration ranging between hundreds of milliseconds to a few seconds. Specifically, it utilizes five parallel discriminators to consider various time scales of audio information to improve the resolution of receptive fields. Secondly, it is conditioned not only on the available information (i.e. context) around the gap, but on the latent variables of conditional GAN. This solves the inherent multi-modal problem of audio inpainting at such long gaps and provides user-defined inpainting options.
\subsection{Generation with Emotion}
\label{sec:4.8}
As German philosopher Hagel has said \cite{67}: ``Music is the art of mood. It is straightly directed against mood." The above deep learning model shows promising results in the automatic music creation. However, it is difficult to control the model to guide the composition to the desired goal, especially since the model can hardly generate music with specific emotions. This section will introduce several previous attempts to generate music with emotion. The emotions here are all from simple emotion classification. Therefore, there is still a broad research space on the emotion music generation, which needs further exploration by researchers in the future.

Ferreira et al. \cite{42} proposed a deep generative model based on mLSTM, which is able to control the polyphonic music generation with given emotion. The mLSTM controls the emotion of generated music by optimizing the weight of specific neurons responsible for emotional signals. At the same time, these neurons insert logical regression into mLSTM, and classify the emotions of symbolic music encoded with the hidden state of mLSTM by training logical regression. Therefore, the model can be regarded as both a generator and an emotion classifier. It is the first deep learning model for emotional music generation. Zhao et al. \cite{67} extended the BALSTM network \cite{25} proposed by Johnson to generate polyphonic music with specific emotions. They first divided four basic emotions by Russell's two-dimensional Valence-Arousal (VA) emotion space, and then used emotion vectors as global conditions to train the network and designed controllable parameters to generate music with corresponding emotion. Moreover, Ferreira et al. \cite{290} presented a system called Bardo Composer, which generates background music for tabletop role-playing games. They first used speech recognition system to translate player speech into text, and then utilized the VA emotional model to classify the text into four categories: happy, calm, agitated and suspenseful. The Stochastic Bi-Objective Beam Search (SBBS) and a neural model were used to generate music clips that convey the desired emotions. The experimental results demonstrated that the subjects can accurately recognize the emotions of the  generated music works, just as they can correctly recognize the music created by human beings.
\subsection{Application}
\label{sec:4.9}
Current research on generative music systems is still focused on viewing music modeling as the ultimate goal, rather than studying the availability of such system in actual scenarios. People have been committed to developing novel architectures and proving that the quality of music generated by these architectures is comparable to that of music created by human beings. This is a necessary first step, because the system must be able to generate convincing music before it can function in the actual environment \cite{289}. It is encouraging that some automatic music generation techniques have been gradually applied to the actual environment. Most of these applications are interactive applications, i.e. human and AI work together to create music.

Section 4.6 has mentioned quite a few interactive generation applications, such as the Bach Doodle system for generating Bach style accompaniment, the bidirectional Chinese pop song creation system iComposer, the intelligent instrument Piano Genie for real-time generation of piano performance, the MANDI system for real-time generation of drum pattern, the music editor web interface Cococo for novice-AI collaborative creation, emotional matching interactive music composition system Mind Band, online singing voice synthesis system Sinsy, etc. In addition, Han et al. \cite{226} proposed an alarm sound recommendation system based on music generation. Inspired by MuseGAN \cite{32}, its core employed a music generation algorithm based on GAN. Before the generator, there is an encoding model that encodes result of emotion prediction and controls the input of the generator so as to determine the emotion and style of the generated music. It is the first application recommending real-time generated music rather than existing music, but the generated music is still far away from the existing music and the model is not very stable. Some people prefer to listen to sleepy music to help aid sleep. In response to this demand, Ouyang et al. \cite{227} proposed a MIDI-VAE-based style transfer method \cite{48} to generate sleepy music from existing non-sleepy music. Moreover, Lidiya et al. \cite{324} proposed a real-time music generator application named MuseBeat that can generate music synchronized with the user's heart rate, so as to establish more emotional ties with the body, such as helping people to concentrate during reading and calm down at the beginning of meditation.

Google's Magenta project \cite{281} and Sony CSL's Flow Machine project \cite{278} are two excellent research projects focusing on music generation. They have developed multiple AI-based tools to assist musicians to be more creative \cite{238}. The model MusicVAE \cite{23} proposed by Google Magenta is used to create many different tools: Melody Mixer and Beat Blender allow to generate interpolation between two melodies or drumbeats; Latent Loops creates a two-dimensional grid between four melodies and allows the users to draw a path in the space to create complex melodies. The Flow Machines project \cite{278} is led by Francois Pachet at Sony Computer Science Laboratories (Sony CSL Paris) and Pierre and Marie Curie University (UPMC). They created Flow Composer, a tool that helps composers generate new melodies or complete existing melodies. Users can choose the parts they like and the parts they want to change, and how AI fills in the blanks. Flow harmonizer can transfer styles on existing works, such as Ode to joy in the styles of Beatles, Bossa Nova, Bach, etc. Reflexive Looper \cite{334} enhances the functions of classic looper through AI, which can identify the style and instrument of recorded music. It can play these loops accordingly and fill in the missing parts of the performance, such as bass lines, guitar accompaniment and choirs. Over the past few years, there have also been several startups that use AI to create music. Jukedeck\footnotemark[1] and Amber\footnotemark[6] music focus on creating royalty-free soundtracks for content creators such as video producers. Hexachords' Orb Composer\footnotemark[7] provides a more complex environment for creating music, not just for short videos, it is more suitable for music professionals. Aiva Technologies designed and developed AIVA\footnotemark[8] that creates music for movies, businesses, games and TV shows. These are just a few of the many startups venturing into this uncharted territory, but what all of them have in common is to encourage cooperation between human and machines, to make non-musicians creative and make musicians more effective.
\section{Datasets}
\label{sec:5}
Since the development of automatic music generation, a variety of music datasets have been born, covering music genres such as folk, pop, jazz, etc. The storage forms include MIDI, pianoroll, audio and so on. The selection of the dataset is closely related to the task of music generation. For example, JSB Chorales dataset \cite{85} is often chosen in the study of polyphonic music. Whether good music can be generated well is also closely connected with the selection of datasets. For deep learning algorithms supported by a great quantity of data, the amount of data is necessary, followed by whether the data is homogeneous. As Hadjeres pointed out in \cite{309}, the trade-off between the size of the dataset and its consistency is a major issue when constructing deep generative models. If the dataset is pretty heterogeneous, a good generative model should be able to distinguish different subcategories and manage to generalize well. On the contrary, if there are only subtle differences between subcategories, it is important to know if the ``average model" can produce musically interesting results. This paper summarizes the common datasets in previous studies from the perspective of music storage forms, hoping to bring some convenience to future researchers. The classification and summary of datasets are shown in Table \ref{tab:222}.
\subsection{MIDI}
As introduced in Section 3, MIDI is a descriptive ``music language", which describes the music information to be performed in bytes, such as what instrument to use, what note to start with, and what note to end at a certain time. MIDI can be employed to listen or input into the analysis program that only requires the basic music description of music score. The MIDI file itself does not contain waveform data, so the file is very small. The pretty\_midi Python toolkit contains practical functions/classes for parsing, modifying and processing MIDI data, through which users can easily read various note information contained in MIDI.

Music21 \cite{85} is an object-oriented toolkit for analyzing, searching and converting music in symbolic form. J. S. Bach four-part chorus dataset can be directly obtained from music21 Python package, which contains 402 choruses. The four parts in the dataset are soprano, alto, tenor and bass. However, this data set is very small and lacks expressive information.

Ferreira et al. \cite{42} created a new music dataset VGMIDI with sentiment notation in symbolic format, which contains 95 MIDI labelled piano pieces (966 phrases of 4 bars) from video game soundtracks and 728 non-labelled pieces, all of them vary in length from 26 seconds to 3 minutes. MIDI labelled music pieces is annotated by 30 human subjects according to a valence-arousal (dimensional) model of emotion. The sentiment of each piece is then extracted by summarizing the 30 annotations and mapping the valence axis to sentiment. For the concrete operation of emotion annotation extraction, please refer to literature \cite{42}.

The Lakh MIDI Dataset (LMD) is the largest symbolic music corpus to date, including 176,581 unique MIDI files created by Colin Raffel, of which 45,129 files have been matched and aligned with the items in the Million Song Dataset (MSD) \cite{87}. However, the dataset has unlimited polyphonic, inconsistent expressive characteristics and contains various genres, instruments and time periods. LMD includes the following formats: 1) 176,581 MIDI files with duplicate data removed, and each file is named according to its MD5 checksum (called ``LMD full"); 2) subset of 45,129 files (called ``LMD matched") that match items in the MSD; 3) All LMD-matched files are aligned with the 7Digital preview MP3s in the MSD (called ``LMD aligned").

{
\normalsize
\setlength{\abovecaptionskip}{5pt}
\begin{center}
\begin{landscape}
\begin{longtable}{|m{50pt}|m{80pt}|m{27pt}|m{30pt}|m{27pt}|m{55pt}|m{120pt}|m{110pt}|}
\caption{Dataset summary}
\label{tab:222} \\

\hline
\multirow{2}{*}{\textbf{Format}}  & \multirow{2}{*}{\textbf{Name}}  & \multicolumn{3}{c|}{\textbf{Modality}} &\multirow{2}{*}{\textbf{\begin{tabular}[c]{@{}l@{}}Applicable\\task\end{tabular}}} &\multirow{2}{*}{\textbf{Size}} &\multirow{2}{*}{\textbf{Access}}   \\ \cline{3-5}
				& & \textbf{Score} & \textbf{\begin{tabular}[c]{@{}l@{}}Perfor-\\mance\end{tabular}} & \textbf{Audio} & & &   \\ \hline
\endfirsthead

% Appear the table header at the top of every page
\multicolumn{8}{r}{Continued table 2} \\ \hline
\multirow{2}{*}{\textbf{Format}}  & \multirow{2}{*}{\textbf{Name}}  & \multicolumn{3}{c|}{\textbf{Modality}} &\multirow{2}{*}{\textbf{\begin{tabular}[c]{@{}l@{}}Applicable\\task\end{tabular}}} &\multirow{2}{*}{\textbf{Size}} &\multirow{2}{*}{\textbf{Access}}   \\ \cline{3-5}
& & \textbf{Score} & \textbf{\begin{tabular}[c]{@{}l@{}}Perfor-\\mance\end{tabular}} & \textbf{Audio} & & &   \\ \hline
\endhead

% Appear \hline at the bottom of every page
\hline
\endfoot
% data begins here
\multirow{10}{*}{MIDI}  & JSB Chorus &\Checkmark  &   &   & Polyphonic  & 402 Bach four parts chorus   & Music21toolkit\cite{85} \\ \cline{2-8} 
                        & VGMIDI  & \Checkmark  & \Checkmark & & Polyphonic with sentiment & 823 piano video game soundtracks & Derived from \cite{42} \\ \cline{2-8} 
                        & Lakh MIDI Dataset & \Checkmark & \Checkmark  &  & Multi-instrumental & 176,581MIDI files  & http://colinraffel.com/pro-jects/lmd/ \\ \cline{2-8} 
				        & Projective Orchestral Database  & \Checkmark & & & Orchestral & 392 MIDI files grouped in pairs containing a piano score and its orchestral version  & https://qsdfo.github.io/LOP-/database \\ \cline{2-8} 
				        & e-Piano Competition Dataset  & \Checkmark & \Checkmark  & & Polyphonic \& Performance & $\sim$1400 MIDI files of piano performance & http://www.piano-e-competition.com \\ \cline{2-8} 
			        	& BitMidi & \Checkmark & & & Polyphonic & 113,244 MIDI files curated by volunteers around the world  & https://bitmidi.com/ \\ \cline{2-8} 
				        & Classical Archives & \Checkmark &  & & Polyphonic & Maximum number of MIDI files of free classical music  & https://www.classical-archives.com/  \\ \cline{2-8} 
				        & The largest MIDI dataset on the Internet  & \Checkmark  &  &  & Polyphonic \& Style & About 130,000 pieces of music from 8 distinct genres (classical, metal, folk, etc.)  & http://stoneyroads.com/20-15/06/behold-the-worlds-biggest-midicollection-on-the-internet/ \\ \cline{2-8} 
				        & ADL Piano MIDI & \Checkmark & \Checkmark & & Polyphonic  & 11,086 unique piano MIDI files  & https://github.com/lucasnfe/-adl-piano-midi \\ \cline{2-8} 
				       & GiantMIDI-Piano & \Checkmark & \Checkmark &  & Polyphonic & 10,854 MIDI files of classical piano, 1,237 hours in total & https://github.com/byte-dance/GiantMIDI-Piano \\ \hline
\multirow{4}{*}{MusicXML} & TheoryTab Database  & \Checkmark  &  & & Polyphonic  & 16K lead sheet segments  & https://www.hooktheory.-com/theorytab \\ \cline{2-8} 
				          & Hooktheory Lead Sheet dataset & \Checkmark &  & & Polyphonic & 11,329 lead sheet segments & Derived from \cite{74} \\ \cline{2-8} 
				          & Wikifonia & \Checkmark &  &  & Polyphonic & 2,252 western music lead sheets & http://marg.snu.ac.kr/chord\_-generation/(CSV format) \\ \cline{2-8} 
				          & MuseScore lead sheet dataset  & \Checkmark  & \Checkmark & & Performance  & lead sheet corresponding to Yamaha e-Competitions MIDI dataset  & https://musescore.com  \\ \hline
Pianoroll & Lakh Pianoroll Dataset & \Checkmark & \Checkmark  & & Multi-instrumental & Approximately equal to the size of LMD & https://salu133445.github.-io/musegan/  \\ \hline
\multirow{4}{*}{Text} & Nottingham Music Dataset  & \Checkmark  &  &  & Monophonic  & About 1,000 folk songs & abc.sourceforge.net/NMD/  \\ \cline{2-8} 
				      & ABC tune book of Henrik Norbeck & \Checkmark &  &  & Monophonic & More than 2,800 scores and lyrics in ABC format, mainly Irish and Swiss traditional music & http://www.norbeck.nu/abc/ \\ \cline{2-8} 
				      & ABC version of FolkDB  & \Checkmark  &  & & Monophonic  & Unknown & https://thesession.org/ \\ \cline{2-8} 
				      & KernScores  & \Checkmark &  &  & Polyphonic & Over 700 million notes in 108,703 files  & http://kern.humdrum.org \\ \hline
\multirow{6}{*}{Audio} & NSynth Dataset &  &  & \Checkmark & Music audio  & 306,043 notes & https://magenta.tensorflow.-org/datasets/nsynth \\ \cline{2-8} 
				       & FMA dataset &  &  & \Checkmark & Music audio  & 106,574 tracks of 917GiB  & Derived from cite{212} \\ \cline{2-8} 
				       & Minist musical sound dataset & & & \Checkmark & Music audio & 50,912 notes & https://github.com/ejhum-phrey/minst-dataset/ \\ \cline{2-8} 
				       & GTZAN Dataset & &  & \Checkmark & Music audio & 1,000 30s music audios & http://marsyas.info/down-load/data\_sets  \\ \cline{2-8} 
				       & Studio On-Line (SOL) & & & \Checkmark & Music audio  & 120,000 sounds & Derived from \cite{215} \\ \cline{2-8} 
				       & NUS Sung and Spoken Lyrics(NUS-48E) Corpus & & & \Checkmark & Sing Voice & 169 minutes recordings of 48 English songs & Derived from \cite{214} \\ \hline
\multirow{9}{*}{\begin{tabular}[c]{@{}l@{}}Multi-\\modality\end{tabular}} & MusicNet Dataset & \Checkmark & & \Checkmark & Fusion  & 330 recordings of classical music & https://homes.cs.washing-ton.edu/$\sim$thickstn/musicnet-.html  \\ \cline{2-8} 
				          	   & MAESTRO Dataset & \Checkmark & \Checkmark & \Checkmark & Fusion &172 hours of virtuosic piano performances & https://g.co/magenta/-maestrodataset \\ \cline{2-8} 
				          	   & NES Music Database & \Checkmark & & \Checkmark & Multi-instrumental & thousands of & Derived from \cite{115} \\ \cline{2-8} 
				          	   & Piano-Midi & \Checkmark & \Checkmark & \Checkmark & Polyphonic \& performance & 332 classical piano pieces & www.piano-midi.de/ \\ \cline{2-8} 
				               & Groove MIDI Dataset & \Checkmark & \Checkmark & \Checkmark & Drum & 13.6 hours recordings, 1,150 MIDI files and over 22,000 measures of tempo-aligned expressive drumming & https://magenta.tensorflow-.org/datasets/groove \\ \cline{2-8} 
				          	   & POP909 & \Checkmark & \Checkmark & \Checkmark & Polyphonic & multiple versions of the piano arrangements of 909 popular songs & https://github.com/music-x-lab/POP909-Dataset \\ \cline{2-8} 
				          	   & ASAP & \Checkmark & \Checkmark  & \Checkmark  & Polyphonic Performance\& Fusion & 222 digital musical scores aligned with 1,068 performances  & https://github.com/fosfrance-sco/asap-dataset                                             \\ \cline{2-8} 
				          	   & Aligned lyrics-melody music dataset & \Checkmark &  &  & Fusion & 13,937 20-note sequences with 278,740 syllable-note pairs  & https://github.com/yy1lab/-Lyrics-Conditioned-Neural-Melody-Generation \\ \cline{2-8} 
				          	   & MTM Dataset & \Checkmark &  & \Checkmark & Fusion & Unknown & https://github.com/Morning-Books/MTM-Dataset \\
% more data here
\hline
	
\end{longtable}
\end{landscape}
\end{center}
}
The Projective Orchestral Database (POD) is devoted to the study of the relationship between piano scores and corresponding orchestral arrangements. It contains 392 MIDI files, which are grouped in pairs containing a piano score and its orchestral version. In order to facilitate the research work, crestel et al. \cite{114} provided a pre-computed pianoroll representation. In addition, they also proposed a method to automatically align piano scores and their corresponding orchestral arrangements, resulting in a new version of MIDI database. They provide all MIDI files as well as preprocessed pianoroll representations of alignment and misalignment for free on the following website https://qsdfo.github.io/LOP/index.html.

The e-piano junior competition is an international classical piano competition. The e-piano junior competition dataset is a collection of professional pianists' solo piano performances. It is the largest public dataset that provides a substantial amount of expressive performance MIDI (~1400) of professional pianists. Most of them are late romantic works, such as Chopin and Liszt, as well as some Mozart sonatas. Since this dataset provides high-quality piano performance data in MIDI, including the fine control of timing and dynamics by different performers, the dataset is widely used in the research of performance generation, but it does not contain the corresponding music score of the pieces \cite{133}.

The ADL piano MIDI dataset is based on LMD. In LMD, there are many versions of the same song, and only one version is reserved for each song in the ADL dataset. Later, Ferreira et al. \cite{290} extracted from the LMD only the tracks with instruments from the ``piano family" (MIDI program number 1-8). This process generated a total of 9,021 unique piano MIDI files. These files are mainly rock and classical music, so in order to increase the genres diversity (as jazz, etc.) of the dataset, they have added another 2,065 files obtained from public resources on the Internet\footnote[11]{https://bushgrafts.com/midi/ and http://midkar.com/jazz/}. All the files in the collection are de-duped according to MD5 checksums, and the final dataset has 11,086 pieces.

Recently, ByteDance released GiantMIDI-Piano\cite{358}, the world's largest classical piano dataset, including MIDI files from 10,854 music works of 2,784 composers, with a total duration of 1,237 hours. In terms of data scale, the total duration of different music pieces in the dataset is 14 times that of Google’s MAESTRO dataset. In order to construct the dataset, researchers have developed and open-sourced a high-resolution piano transcription system\cite{359}, which is used to convert all audio into MIDI files. MIDI files include the onset, dynamics and pedal information of notes.

In addition, BitMidi\footnote[12]{https://bitmidi.com/} provides 113,244 MIDI files curated by volunteers around the world; Classical Archives\footnote[13]{https://www.classicalarchives.com/} is the largest classical music website, including the largest collection of free classical music MIDI files; the largest MIDI dataset\footnote[14]{http://stoneyroads.com/2015/06/behold-the-worlds-biggest-midicollection-on-the-internet/} on the Internet contains about 130,000 music from eight different genres (classical, metal, folk, etc.); FreeMidi\footnote[15]{https://freemidi.org/} comprises more than 25,860 MIDI files of assorted genres.
\subsection{Pianoroll}
\label{sec:5.2}
As described in Section 3, pianoroll represents each music piece as a 2D matrix. Through pypianoroll Python package, it is convenient to parse MIDI files into pianorolls or write multi-track pianorolls into MIDI files. Therefore, there are not many datasets stored in pianoroll format. Here we only introduce the Lakh Pianoroll Dataset and its subsets.

The Lakh Piano Dataset is derived from LMD. Dong et al. \cite{32} transformed MIDI files into multi-track pianorolls. For each mearsure, they set the height to 128 and the width (time resolution) to 96 to model common time patterns such as triplets and 16th notes, and used pretty\_midi Python library to parse and process MIDI files. Finally, they named the resulting dataset Lakh Pianoroll Dataset (LPD). They also put forward LPD-matched subset, which is obtained from LMD-matched subset. Since some tracks play only a few notes in the whole song, this will increase the sparsity of data and hinder the learning process. Therefore, Dong et al. \cite{32} deal with such a data imbalance problem by merging the soundtracks of similar instruments in LPD-matched (by summing their pianorolls). Each multi-track pianoroll is compressed into five tracks: bass, drum, guitar, piano and string. After this step, they got LPD-5-matched, which is composed of 30,887 multi-track pianorolls.
\subsection{MusicXML}
\label{sec:5.3}
MusicXML aims to completely represent western music symbols. Therefore, MusicXML can contain a variety of types of music symbols, such as rest, slur, beam, barine, key and time signature, articulation, organization marks, etc., which are not included in MIDI format. MusicXML is usually employed to store lead sheets. Lead sheet is a form of musical symbol utilized to specify the basic elements of a song: melody, lyrics, harmony, repetition marks, etc. The melody is written with modern western music symbols, the lyrics are the text below the score, and the harmony is specified by the chord symbol above the score. Lead sheets rarely involve information about instruments or accompaniment. This kind of music sheet is not arranged and is the most basic form of music or song.

Online music theory forum TheoryTab is hosted by Hooktheory, a company that produces pedagogical music software and books. For copyright reasons, TheoryTab forbids uploading full-length songs, so users upload snippets of a song (here referred to as lead sheet samples), which they voluntarily annotate with structural labels (e.g. Intro, Verse, and Chorus) and genre labels. A music piece can be associated with multiple genres \cite{74}. Theorytab Dataset (TTD) contains 16K lead sheet fragments stored in XML format. Yeh et al. \cite{74} collected a new dataset called the Hooktheory Lead Sheet Dataset (HLSD) from TheoryTab. Each lead sheet sample contains high-quality, human-transcribed melodies and their corresponding chord progressions, which are specified by both literal chord symbols (e.g., Gmaj7) and chord functions (e.g., VI7) relative to the provided keys. HLSD contains 11,329 lead sheet samples, all in 4/4 time signature, covering 704 different chord categories.

Wikifonia.org offers a public lead sheet repository. Unfortunately, the website ceased service in 2013, but Lim et al. \cite{28} obtained some data (including 5,533 western music lead sheets in MusicXML format), including music genres such as rock, pop c, jazz and so on. They collected 2,252 lead sheets from the collected dataset. If a bar is composed of two or more chords, only the first chord is selected. Finally, the music features are extracted and converted into CSV format.

What’s more, Jeong et al. \cite{133} collected scores corresponding to Yamaha e-piano junior competition MIDI dataset from MuseScore\footnote[16]{https://musescore.com}, a community-based web platform of music score. These scores are transcribed voluntarily by the community users and can be exported in MusicXML format.
\subsection{Text}
\label{sec:5.4}
ABC, developed by Chris Walshaw, is a format for recording music in plain text. It was originally designed for folk music originated in Western Europe and was later extended to support a complete representation of classical music scores. The abc2midi program in abcMIDI package developed by James Allwright can create MIDI files directly from ABC files. The Nottingham Music Dataset (NMD) is composed of simple melodies on chords, containing 1200 British and American folk songs in special text format created by Eric Foxley and published in his music database. Then, using a program written by Jay Glanville, NMD2ABC and some Perl scripts, most of this dataset can be converted into ABC notation. Recently, Seymour Shlien edited these files to correct missing beats during repetition and conversion. The latest modified dataset can be downloaded from following website https://ifdo.ca/~seymour/nottingham/not tingham.html. In addition, the artificial intelligence start-up company Jukedeck also preprocessed the NMD to generate a clean version of the dataset, in which each song has a separate chord and melody part. In addition, Henrik Norbeck has collected more than 2,800 music scores and lyrics in ABC format, mainly including traditional music from Ireland and Sweden, which are available for free on http://www.norbeck.nu/abc/. The ABC version of traditional Scottish and Irish music is also available on https://thesession.org/.

The Humdrum file format is text-based as well, allowing for easy viewing and editing in any standard text editor. Each part in a music piece forms a column with musical time progressing by successive rows, and with all elements on the same row occurring at the same performance time. In addition to musical data, other forms of data can also be included in its own column (called spines in Humdrum terminology) to facilitate the analytic mark-up of the musical data \cite{331}. The KernScores website\footnote[17]{http://kern.humdrum.org} is developed to organize music scores from various sound sources in the humdrum **kern data format. Therefore, the native storage format of music scores in KernScores is Humdrum file format. Through this website, humdrum data can be directly translated into MIDI files. If the music is scanned from music scores outside copyright, the scanned music is usually provided as a reference in PDF format. The KernScores library contains more than 7 million notes in 108,703 files. Moreover, Cherla et al. \cite{245} employed Essen Folk Song Collection from KernScores library, consisting of 7 distinct traditional folk songs and chorus melodies.
\subsection{Audio}
\label{sec:5.5}
Audio files are usually divided into two categories: sound files and MIDI files. Sound files are the original sounds recorded by sound recording devices, which directly record the binary sampling data of real sounds; MIDI files are sequences of musical performance instructions that can be performed by drawing support from sound output devices or electronic instruments connected to the computers. The datasets described in this section belong to the first category.

Nsynth is a four-second monophonic music dataset \cite{152} with a sampling rate of 16KHz. It contains 30,6043 notes, each with a unique pitch, timbre and envelope. For 1,006 instruments from the commercial sample library, four-second audio snippets are generated by ranging over every pitch of a standard MIDI piano (21-108) as well as five different velocities (25, 50, 75, 100, 127). The note was held for the first three seconds and allowed to decay for the final second. Based on a combination of human evaluation and heuristic algorithms, three additional pieces of information are added to each note: Source: the sound production method of the note instrument; family: the advanced family of the note instrument; and quality: the sound quality of the note. Nsynth is an order of magnitude larger than the comparable public dataset Minst \cite{310}. The Minst dataset here is not the handwritten digit recognition dataset in the image field, but a large, standardized and easy-to-use dataset created by Humphrey via integrating four different solo instrument datasets into one. This dataset can be regarded as the Minst dataset in the field of music audio processing. The four solo instrument datasets are: University of Iowa – MIS; Philharmonic; RWC – Instruments; Good Sounds. The final Minst music audio dataset contains 12 instruments with 50,912 notes.

The FMA dataset \cite{212} is a free and open library guided by WFMU (the longest-running freeform radio station in the United States), which provides 917GiB, 343 days of Creative Commons-licensed audio from 106,574 tracks of 16,341 artists and 14,854 albums, arranged in a hierarchical taxonomy of 161 genres. It provides full-length and high-quality audio, pre-calculated features, together with track- and user-level metadata, tags, and free-form text (as biographies). FMA also has fine genre information annotated by the artists themselves, i.e. multiple (sub-) genres associated to individual tracks, has a built-in genre hierarchy. The audio for each track is stored in a file which name is the track id, and all tracks are mp3-encoded. To make the dataset available as a development set or useful for people with low computing resources, Defferrardy et al. \cite{212} proposed the following subsets: (1) Full: the complete dataset; (2) Large: the complete dataset with audio limited to 30 second clips extracted from the middle of the tracks, reducing the size of the data by 10 times; (3) Medium: select tracks with only one top genre and sampled the clips according to the completeness of their metadata and their popularity resulting in 25,000 30s clips; (4) Small: select the top 1000 clips from the 8 most popular genres of the medium set, resulting in 8000 30s clips.

The GTZAN dataset\cite{356} is composed of 1000 30s music audio excerpts, including a total of 10 music genres, namely blues, classical, country, disco, hip hop, jazz, metal, pop, reggae, and rock. Each genre contains 100 clips. However, there are several faults in the dataset, such as duplication, mislabeling and distortion.

The NUS Sung and spoke lyrics (NUS-48E) corpus \cite{214} is a 169-minute audio recording set containing 48 recordings of English songs, including 115 minutes of singing data, 54 minutes of voice data, in which the lyrics were sung and read by 12 subjects composed of 6 males and 6 females, covering all sound types and accent types. There were 20 unique songs in the 48 English songs. Each subject was asked to annotate four different songs and each song was sung or read by at least one male and one female. 39-phoneme set used in CMU Dictionary was used for phonetic annotation \cite{17}. All singing recordings have been phonetically transcribed with duration boundaries, and the total number of annotated phones is 25,474. NUS-48E corpus is the first phonetically annotated dataset for singing voice research, and the largest paired dataset for Speech-to-Singing (STS) tasks \cite{186}. Plenty of SVC studies have also employed this dataset \cite{187,188}.

Studio Online \cite{215} is an application that provides high-quality instrument sounds “online” for music researchers, composers and professional audio studios, with a total of 120,000 sounds. Studio Online allows an instantaneous, around-the-clock access to IRCAM sound archiving and processing power, a remotely configurable and controlled personal music studio for music professionals and laymen. Users can not only download sound from the IRCAM database but also upload their own sound files to the private user space and process them through the IRCAM processing tools. IRCAM is a world-renowned institution with specific capabilities in acoustic and psychoacoustics of instrumental sounds, sound analysis and conversion software, and computer aided composition.

In addition, a collection of all 32 Beethoven piano sonatas is publicly available on the website https://archive.org/, equivalent to 10 hours of non-human voice audio. The audio dataset employed in \cite{164} was crawled from the MuseScore web forum\footnotemark[16], and a total of about 350,000 unique MIDI files and corresponding MP3 files were obtained. Most MP3 files are synthesized from MIDI by uploaders using the MuseScore synthesizer. The NIT-SONG070-F001 singing voice dataset used in \cite{201} consists of studio-quality recordings of a female singer singing Japanese children songs. The original dataset contains 70 songs, but the public version only contains a subset of 31 songs which is available on http://hts.sp.nitech.ac.jp/archives/2.\\3/HTS-demo\_NIT-SONG070-F001.tar.bz2.

There are also multiple singing voice separation datasets that are applied to singing style transfer. The MIR-1K dataset \cite{311} consists of 1000 segments of 110 unique Chinese songs sung by 19 amateur singers with a total length of 133 minutes. Since this dataset is used for singing voice separation, annotations include pitch, lyrics, unvoiced frame types, and vocal/non-vocal segments, but do not contain segmentation on the word level or below. Later, a publicly available dataset iKALA dataset \cite{213} with longer duration and higher quality than MIR-1K dataset was proposed. The iKALA dataset contains 352 30-second clips of Chinese popular songs in CD quality. The singers and musicians are professionals. The dataset is human-labeled for pitch contours and timestamped lyrics.

\subsection{Multimodality}
\label{sec:5.6}
Another kind of dataset contains multi-modal information, such as scores, lyrics and music audio. The more common type is dataset that contains MIDI files and aligned audio files. Such multi-modal data sets are often used in fusion generation tasks.

MusicNet dataset \cite{332} is a collection of 330 freely licensed classical music recordings, including 10 composers, 11 instruments, plus instrument/note annotations, resulting in more than 1 million temporal labels on 34 hours of chamber music performance. The labels indicate the precise time of each note in each recording, the instrument playing each note, and the position of the note in the rhythm structure of the music. The average recording length is 6 minutes. The shortest recording time is 55 seconds and the longest is nearly 18 minutes. MusicNet labels come from 513 label classes using the naivest definition of a class: distinct instrument/note combinations. Labels are obtained from music scores aligned with recordings through dynamic time warping. Beethoven’s composition occupies most of the dataset, and the dataset is inclined to be piano solos. However, as discussed by Hawthorne et al. \cite{116}, the alignment between audio and score is not completely accurate.

The MAESTRO dataset \cite{116} contains more than 172 hours of virtuosic piano performance, which is derived from 1282 real performances of approximately 430 different music pieces in the nine-year International Piano e-Competition, in which the note labels and audio waveforms are precisely aligned ($\approx$3ms). The MAESTRO is about an order of magnitude larger than the previous audio and MIDI paired datasets. The MIDI data contains key strike velocities and sustain pedal positions. The audio files and MIDI files are divided into individual music pieces and annotated with composer, title and year of performance. Repertoire is mainly classical, including composers from the 17th century to early 20th century.

The Nintendo Entertainment System Music Database (NES-MDB) \cite{115} contains thousands of multi-instrument songs, which are synthesized by the Nintendo Entertainment System (NES) audio synthesizer, forming a total of about 46 hours of chiptunes. All of the music was created in a limited time period, so it is more cohesive in style than other large multi-instrument music datasets. For each song, the dataset contains a musical score for four instrument voices as well as expressive attributes for the dynamics and timbre of each voice. The four monophonic instruments are: two pulse wave generators (P1/P2), a triangle wave generator (TR) and a noise generator (NO). Unlike datasets composed of general MIDI files, NES-MDB contains all the information needed to render exact acoustic performances of the original compositions. Donahue et al. \cite{39} would like to fine-tune the pre-trained model on the LMD to the NES-MDB dataset, so they stored each chiptune in the NES-MDB as a MIDI file. Donahue et al. \cite{115} also released NES-MDB in MIDI format with no down-sampling.

Piano-Midi is a MIDI database of classical pianos, which contains the music pieces of 25 composers from three major classical periods (baroque, classical and romantic). The dataset is rich in timing and dynamic information, but possesses different time periods. The MIDI files of the dataset and their corresponding mp3 and ogg audio files are available on www.piano-midi.de/. Some music pieces have corresponding music score in PDF format as well.

The Groove MIDI Dataset (GMD) is a dataset created by Gillick et al. \cite{145,146} for the research of drum performance. It contains 13.6 hours, 1150 MIDI files and 22,000 bars of tempo-aligned expressive drumming performed by 10 drummers. The dataset was record on the Roland td-11\footnote[18]{https://www.roland.com/us/products/td-11/} electronic drum kit equipped with sensors to capture precise performance features in MIDI format. The drummer plays each performance with a metronome at a specific tempo. Since the metronome provides a standard measurement of where the musical beats and subdivisions lie in time, it can accurately quantize all the notes to the nearest musical division, yielding a musical score. MIDI notes are related to an instrument, a time and a velocity, micro-timings describe how note timings stray from a fixed grid, and velocities indicate how hard notes are struck. Each performance in the dataset also contains relevant metadata (such as music style annotations and tempo) and corresponding synthesized audio outputs.

Wang et al. \cite{293} proposed a novel dataset named POP909 for music arrangement generation, which contains multiple versions of piano arrangements of 909 popular songs created by professional musicians. The total length of POP909 is about 60 hours, including 462 artists. The release of all songs spans around 60 years. The main body of the dataset contains vocal (main) melody, secondary melody or main instrument melody (bridge) in MIDI format and piano accompaniment of each song, which are aligned with the original audio file. In addition, beat, chord, and key annotations are provided, and each note event contains expressive dynamics.

Foscarin et al. \cite{294} proposed a new dataset, aligned scores and performances (ASAP), which consists of 222 digital scores and 1068 aligned performances of western classical piano music from 15 composers. 548 of the recordings are available as MIDI only, and all the others (520) are provided as MIDI and audio recordings aligned with approximately 3ms precision. Music scores are provided in pairs of MusicXML files and quantified MIDI files, while performance is provided in the form of paired MIDI files and partial audio recordings. For each MIDI score and performance, ASAP offers the positions of all beats, downbeats, time signature changes and key signature changes. Each score and performance in ASAP are labeled with metadata including the composer and title of the piece. This dataset is the largest dataset known containing fine-grained alignment between music score, MIDI and audio performance data. It can help to achieve audio-to-score transcription task and other tasks based on metrical structure, e.g., key signature estimation and beat or downbeat tracking.

Due to the lack of publicly available aligned lyrics-melody datasets, Yu et al. \cite{329} acquire lyrics-melody paired data from two sources based on considering melodies with enough English lyrics, where 7,998 MIDI files come from the ``LMD full" MIDI dataset and 4,199 MIDI files come from the reddit MIDI dataset. The processed dataset contains 13,937 sequences each with 20 notes and 27,8740 syllable-note pairs. Zeng et al. \cite{360} also introduced a Music Ternary Modalities Dataset (MTM Dataset) containing sheet music, lyrics and music audio. In detail, the syllable of lyrics, and audio snippet, and sheet music segment generated from music notes are aligned.
\subsection{Others} 
\label{sec:5.7}
\quad\,
1) dMELODIES for Disentanglement Learning

There is no standard dataset in music for disentanglement learning, so it is impossible to compare different disentanglement methods and technologies directly, which hinders the systematic research on understanding disentanglement in music background. Therefore, Pati et al. \cite{250} used music21 Python package to create a symbol dataset called dMelodies composed of two-bar monophonic melodies (8th note is the minimum note duration) for music representation learning, where each melody is the result of a unique combination of nine latent factors that span ordinal, categorical, and binary types. The entire dataset contains about 1.3 million data points.

2) Bach Doodle Dataset

Each user interacting with Bach Doodle can add their pieces to a dataset called Bach Doodle \cite{43}. Through anonymous interaction with Bach Doodle system, 8.5 million users generated 21.6 million micro-compositions. Each interaction generated a data point, including the following contents: the melody input by the user, the four-part accompaniment returned by the Coconet, and other metadata (source country, tune/beat of the piece, composition duration, and the number of times it was listened to, etc.). The dataset can be downloaded from https://g:co/magenta /bach-doodle-dataset.
\section{Evaluation}
\label{sec:6}
There is no unified evaluation criterion for the results of generative algorithms that work in the area of arts. Indeed, since music, literature, cinema etc. are intrinsically subjective it is rather hard to approach them with a certain rigorous metric \cite{60}. Different from the evaluation of tasks such as classification and prediction, in which only quantitative indicators such as accuracy are required, the quality of generated content often needs subjective evaluation as the final judge of creative output is human. In addition, the results of some objective evaluation metrics also have certain reference value for the generation quality. Therefore, this paper summarizes the music evaluation methods from both objective and subjective aspects. Although subjective evaluation is generally preferable for evaluating generative modeling, it might require significant resources and face challenges in terms of reliability, validity and reproducibility of the results. Objective methods, on the other hand, can be easily executed yet often lack musical relevance as they are often not based on musical rule systems or heuristics. \cite{117}. Whether it is objective or subjective evaluation, the results generated by multiple models are usually compared together. However, comparing models trained on various datasets is problematic as it is tough to have a standardized definition of improvement and quality, so it is better to retrain the models that need to be compared on a unified training dataset. Effective evaluation is able to find out the shortcomings of the generated music, so as to improve continuously to generate music with better performance.
\subsection{Objective}
\label{sec:6.1}
The objective evaluation mainly includes the quantitative evaluation of music generative model and generated music. The former often employs metrics such as loss and accuracy to evaluate models. These metrics only reflect the ability of the model to process data, but cannot truly reflect the generation effect, especially for music that requires a highly innovative form of artistic expression. For the latter, the selection of music metrics is mainly affected by the music generation task. To define these music metrics, researchers need to master certain knowledge of music theory. The definition and selection of music metrics will be elaborated in section 6.1.2. There are also several methods to detect certain patterns of generated music, so as to judge whether the music achieves the desired effect. For example, the VMO mostly detects repetitive patterns in music, and the Keyscapes is conductive to analyze the tone of generated music, etc. Yang et al. \cite{117} divided the objective evaluation of music generation into three categories: the first is the probability measures (as likelihood) without music domain knowledge, the second is the metrics with the help of music theory, such as tonal distance, rhythmic patterns, etc., and the third is the metrics for specific tasks/models. The classification of objective evaluation methods in this paper is in common with that of Yang et al. \cite{117} to some extent.
\subsubsection{Model Metrics}~{}\\
\label{sec:6.1.1}
Model metrics refer to the evaluation metrics that does not contain music domain knowledge. Usually, the pros and cons of the model are evaluated by comparing the data differences between the generated samples and the original samples, analyzing the distribution statistical characteristics of the generated samples, or calculating the classification accuracy.

The most commonly used model metrics for score generation are Loss, PPL (Perplexity), BLEU score, precision, recall, F1, etc. The lower perplexity on the test set indicates that the model is suitable for invisible data, that is, the model has better generalization performance and has more advantages in generating novel music. Loss is employed to estimate the degree of inconsistency between the predicted result of the model and the ground truth and is utilized to select the hyperparameters of the network. Loss reduction can only mean that the model can understand the problem numerically, but doesn’t represent the generated music must be brilliant when loss is small enough, and the model whose loss cannot be converge is more difficult to generate satisfactory music. The BLEU score is exploited to measure the similarity between the validation set and the generated samples. There are also other metrics for evaluating model performance for specific tasks, such as reconstruction accuracy mostly appeared in VAE models \cite{23}, chord prediction accuracy \cite{28}, log likelihood on test sets \cite{25}, style likelihood \cite{48}, style classification accuracy \cite{65}, and so on. It should be mentioned that the ultimate goal of models is to generate music, not to make predictions. Therefore, the above model metrics can only be exploited as references, not a decisive metric to evaluate the quality of generated music.
{
%\normalsize
\begin{table}[bp]
	\centering
	%	\scriptsize
	\footnotesize
	\setlength{\abovecaptionskip}{5pt}
	\caption{Audio evaluation metrics}
	\label{tab:333}
	\resizebox{\textwidth}{!}{%
		\begin{tabular}{|m{60pt}|m{220pt}|}
			\hline
			\textbf{Metrics}                             & \textbf{Definition}                                                                                                                                                                                                                                                                                                                                                                                                                         \\ \hline
			Inception Score (IS) \cite{170,174} & Generated examples are run through a pretrained Inception classifier and the Inception Score is defined as the mean KL divergence between the image conditional output class probabilities and the marginal distribution of the same. IS penalizes models whose examples aren’t each easily classified into a single class, as well as models whose examples collectively belong to only a few of the possible classes.                     \\ \hline
			Frechet Inception Distance (FID) \cite{174}   & A GAN evaluation metric based on the 2-Wasserstein (or Frechet) distance between multivariate Gaussians fit to features extracted from a pretrained Inception classifier and show that this metric correlates with perceptual quality and diversity on synthetic distributions.                                                                                                                                                             \\ \hline
			Pitch Accuracy (PA) and Pitch Entropy (PE)   & The accuracy of the same pretrained pitch classifier on generated examples (PA) and the entropy of its output distribution (PE).  \\ \hline
			Number of Statistically-Different Bins (NDB) & Measure the diversity of generated examples: the training examples are clustered into k = 50 Voronoi cells by k-means in log-spectrogram space, the generated examples are also mapped into the same space and are assigned to the nearest cell. NDB is reported as the number of cells where the number of training examples is statistically significantly different from the number of generated examples by a two-sample Binomial test. \\ \hline
		\end{tabular}%
	}
\end{table}
}
In addition to the general model metrics mentioned above (as loss and perplexity, etc.), the commonly used model indicators in performance modeling include the mean square error (MSE) between the characteristics of human performance and generated performance, and the correlation between generated performance and human performance \cite{135}. Gillick et al. \cite{146} proposed Timing MAE, timing MSE, Velocity KL, Timing KL and other metrics to evaluate the performance of drum performance generation. KL divergence is an asymmetric measure of the difference between two probability distributions, which is usually used to evaluate the distance between the training dataset and the generated data distribution.

Common evaluation metrics for audio generation include performance reconstruction metrics, such as Root Mean Square Error (RMSE), Logarithmic Spectral Distortion (LSD), Mel Cepstrum Distortion (MCD), mean difference between original and reconstructed audio, and transfer quality metrics such as Maximum Mean Deviation (MMD), non-differentiable K-Nearest Neighbor (KNN) test, etc. In \cite{158}, the Kullback Leibler (KL) divergence of the Inter-Onset Interval Length Distributions was calculated to evaluate the rhythm style similarity between texture and its source audio, and the maximum cross-correlation value between the two signals was calculated to determine whether the texture is simply a distorted duplicate of the source. \cite{191} put forward the audio evaluation metrics in Table \ref{tab:333}, among which IS and FID have become a de-facto standard for measuring the performance of any GAN-based model \cite{199}. In particular, the commonly used system evaluation metrics for singing voice synthesis include Root Mean Square Error and Correlation of log F0 (F0-RMSE, F0-CORR) and Duration (Dur RMSE, Dur COR), Band Aperiodic Distortion (BAPD), Voiced/Unvoiced error rate (V/UV) (or FPR and FNR), etc. Hono et al. \cite{176} employed root mean squared error in phoneme boundary (Boundary-RMSE) to evaluate the accuracy of prediction time-lag and duration. Blaauw et al. \cite{201} proposed the metrics of singing voice synthesis in Table \ref{tab:444}.
\begin{table}[H]
	\centering
	\normalsize
	\setlength{\abovecaptionskip}{5pt}
	\caption{Singing voice synthesis metrics}
	\label{tab:444}
		\begin{tabular}{|m{63pt}|m{250pt}|}
			\hline
			\textbf{Type}                    & \textbf{Features}                                                                                                                                                                                                                           \\ \hline
			\makecell*[l]{Timbre metrics}  & \begin{tabular}[c]{@{}l@{}}Mel-Cepstral Distortion (MCD)\\ Band Aperiodicity Distortion (BAPD)   \\ Modulation Spectrum Log Spectral Distortion (MS-LSD)  \\ Modulation Spectrum (MS) for Mel-Generalized Coefficients\\ (MGC)\end{tabular} \\ \hline
			\makecell*[l]{Voiced/unvoiced\\ decision metrics} & \begin{tabular}[c]{@{}l@{}}FPR  \\ FNR\end{tabular}  \\ \hline
			\makecell*[l]{Timing metrics} & \begin{tabular}[c]{@{}l@{}}Mean Absolute Error (MAE)  \\ Root Mean Squared Error (RMSE)   \\ Pearson correlation coefficient r between onsets or durations\end{tabular} \\ \hline
			\makecell*[l]{F0 metrics} & \begin{tabular}[c]{@{}l@{}}F0 RMSE  \\ F0 Modulation Spectrum Log Spectral Distortion (MS-LSD) \\ Modulation Spectrum (MS) for log F0 \\ Pearson correlation coefficient r\end{tabular}                                                     \\ \hline
		\end{tabular}%
	
\end{table}
\subsubsection{Music Metrics/Descriptive Statistics}~{}\\
\label{sec:6.1.2}
Music metrics are derived from some music concepts, also known as statistical descriptors of music. Those who make music metrics need to master certain music background knowledge. Ideally, with the training process going on, the distribution of the generated music and the real music should be closer and closer. Therefore, many methods judge the quality of the music generated by the model via comparing the descriptive statistics of the generated music with the real music or music generated by other models. Researchers can develop different music evaluation metrics according to their research tasks, and the definition of metrics needs to be explained in detail, especially the corresponding relationship between the value of metrics and the quality of music. Music metrics are widely used in score generation evaluation, because score generation can explicitly represent various characteristics of notes, such as pitch, duration, but cannot represent timbre. We divide these metrics into four categories: pitch-related, rhythm-related, chord/harmony-related, and style transfer-related. While audio generation usually compares different samples at the level of acoustic features or sound signals. These comparisons are also applicable in the fields of speech synthesis and TTS, and are not specific to musical characteristics. This section mainly introduces the common or distinctive music metrics under three levels of music generation, which does not cover all the studies.
\begin{table}[]
	\normalsize
	\setlength{\abovecaptionskip}{5pt}
	\caption{Signature vectors}
	\label{tab:555}
	\scalebox{1.0}{%
		\begin{tabular}{|m{3cm}|m{8.2cm}|}
			\hline
			\textbf{Signature vectors} & \textbf{Definition}                                                                                                                                                                                        \\ \hline
			Number of notes            & Number of notes in the piece divided by the length of the piece.                                                                                                                                           \\ \hline
			Occupation rate            & The ratio between the number of non-null values in the pianoroll representation and the length of the piece.                                                                                               \\ \hline
			Polyphonic rate            & The number of time steps where two or more notes were played simultaneously, divided by the total number of notes in the piece.                                                                            \\ \hline
			Pitch range descriptors    & The maximum, minimum, mean and standard deviation of the non-null pitches in the piece, all values were divided by 127 in order to force these descriptors to be bounded between 0 and 1.                  \\ \hline
			Pitch interval range       & An interval is a difference in pitch between two consecutive notes. All intervals were scaled between 0 and 1 (i.e., divided by 127) and the maximum, minimum, mean and standard deviation were computed.  \\ \hline
			Duration range             & The duration is the number of time steps during which a note is held. As before, the maximum, minimum, mean and standard deviation of all durations in the piece were computed (no scaling was performed). \\ \hline
		\end{tabular}%
	}
\end{table}

Since the development of music score generation research, numerous research teams have defined a large number of music metrics for various generation tasks. It is unrealistic to introduce all of them. Here we only point out some commonly used and representative metrics. See Table \ref{tab:666} for details. In addition, Sabath et al. \cite{119} proposed a novel objective evaluation metrics to quantify the music similarity between music distributions. They first used the signature vectors (see Table \ref{tab:555}) composed of high-level symbolic music descriptors to describe the given music. Then, the music similarity between music distributions is quantified by calculating Mahalanobis distance \cite{312} given the average feature vector $\mu$ and covariance matrix $\Sigma$ of music $x$ and corpus $\mathcal{D}$. The calculation formula is as follows:
\begin{equation}
D(x,\mathcal{D})=\sqrt{(x-\mu)^T\Sigma^{-1}(x-\mu)} 
\end{equation}

Yang et al. \cite{117} also put forward a set of simple musically informed objective metrics enabling an objective and reproducible way of evaluating and comparing the output of music generative systems. Their evaluation strategy includes absolute metrics and relative metrics. Absolute metrics are used to give insights into the properties and characteristics of a generated or collected set of data, and relative metrics are used to compare two sets of data, e.g., training and generated. First, two datasets are gathered as input datasets of the evaluation system. After that, a set of custom-designed features are extracted, including pitch-based and rhythm-based features. Finally, two objective evaluation metrics of generative systems from the training dataset’s intra-set distance PDF (target distribution) and the inter-set distance PDF between the training and generated datasets: (1) the Overlapping Area (OA) and (2) the Kullback–Leibler Divergence (KLD).  For the design of features and the details of calculation and implementation, please refer to article \cite{117}. The evaluation framework has been released as an open-source toolbox, which the demonstrated evaluation and analysis methods along with visualization tools. The evaluation system has also been recognized by many researchers, e.g. \cite{73} and \cite{295} used OA evaluation metrics; \cite{100} used several features proposed in \cite{117}, such as pitch count per bar (PC/bar), average pitch interval (PI), etc.; and \cite{31} directly used the toolbox MGEval to evaluate their music generation system.

{
\normalsize
\setlength{\abovecaptionskip}{5pt}
\begin{center}
\begin{landscape}
\begin{longtable}{|m{3cm}|m{3.7cm}|m{10cm}|}
\caption{Music metrics}
\label{tab:666}\\
\hline
\multirow{2}{*}{\textbf{Types}} & \multicolumn{2}{c|}{\textbf{Metrics}} \\ \cline{2-3} & \textbf{Name} & \textbf{Definition} \\ \hline
\endfirsthead
\multicolumn{3}{r}{Continued table 6} \\ \hline
\multirow{2}{*}{\textbf{Types}} & \multicolumn{2}{c|}{\textbf{Metrics}} \\ \cline{2-3} & \textbf{Name} & \textbf{Definition} \\ \hline
\endhead	
% Appear \hline at the bottom of every page
\hline
\endfoot
% data begins here
\multirow{14}{*}{\begin{tabular}[c]{@{}l@{}}Pitch-related\\ \cite{24,32,104,219}\end{tabular}}  
& EB & ratio of empty bars (in \%) \\ \cline{2-3} 
& UPC & number of used pitch classes per bar (from 0 to 12) \\ \cline{2-3} 
& QN & ratio of ``qualified” notes (in \%).\cite{32} consider a note no shorter than three timesteps (i.e. a 32th note) as a qualified note. QN shows if the music is overly fragmented. \\ \cline{2-3} 
& PP/PR & or polyphonicity, ratio of the number of time steps where more than two pitches are played to the total number of time steps. \\ \cline{2-3} 
& Scale consistency & computed by counting the fraction of tones that were part of a standard scale, and reporting the number for the best matching such scale. \\ \cline{2-3} 
& NS & or note in scale, ratio of the white key to the sum of the black and white keys in the piano-rolls format data. It can be used to roughly judge whether the generated music has the same style as that of training data. \\ \cline{2-3} 
& Repetitions & repetitions of short subsequences were counted, giving a score on how much recurrence there is in a sample. This metric takes only the tones and their order into account, not their timing. \\ \cline{2-3} 
& Tone Span & number of half-tone steps between the lowest and the highest tone in a sample. \\ \cline{2-3} 
& Number of Unique Note Pitches/Duration & count of how many different pitches/time durations are used in a music piece. \\ \cline{2-3} 
& 2/3/4-Note Repetitions & number of repetitions of 2/3/4-gram notes occurred throughout the sequence. \\ \cline{2-3} 
& Consecutive Pitch Repetitions (CPR) & For a specified length $l$, CPR measures the frequency of occurrences of $l$ consecutive pitch repetitions \\ \cline{2-3} 
& Tone Spans (TS) & For a specified tone distance d, TS measures the frequency of pitch changes that span more than d half-steps. \\ \cline{2-3} 
& Rote Memorization frequencies (RM) & Given a specified length $l$, RM measures how frequently the model copies note sequences of length $l$ from the corpus. \\ \cline{2-3} 
& Pitch Variations (PV) & PV measures how many distinct pitches the model plays within a sequence. \\ \hline
\multirow{6}{*}{\begin{tabular}[c]{@{}l@{}}Rhythm-related\\ \cite{24,104}\end{tabular}} 
& Qualified Rhythm frequency (QR) & QR measures the frequency of note durations within valid beat ratios of \{1,1/2,1/4,1/8,1/16\}, their dotted and triplet counterparts, and any tied combination of two valid ratios. \\ \cline{2-3} 
& DP & or drum pattern, ratio of notes in 8- or 16-beat patterns, common ones for Rock songs in 4/4 time (in \%). \\ \cline{2-3} 
& Number of Concurrent Three/Four Notes & means that the consecutive three/four notes have the same start and end time. \\ \cline{2-3} 
& Off-beat Recovery frequency (OR) & Given an offset d, OR measures how frequently the model can recover back onto the beat after being forced to be off by d timesteps. \\ \cline{2-3} 
& Rhythm Variations (RV) & RV measures how many distinct note durations the model plays within a sequence. \\ \cline{2-3} 
& Durations of Pitch Repetitions (DPR) & For a specified duration d, measures the frequency of pitch repetitions that last at least d long in total. \\ \hline
\multirow{8}{*}{\begin{tabular}[c]{@{}l@{}}Chord/Harmony\\-related\\ \cite{24,74,104}\end{tabular}} 
& Harmonic Consistency (HC) & The harmonic consistency metric is based on the Impro-Visor \cite{333} note categorization, represented visually by coloration, which measures the frequency of black, green, blue, and red notes. \\ \cline{2-3} 
& TD & or tonal distance, It measures the harmonicity between a pair of tracks. Larger TD implies weaker inter-track harmonic relations. \\ \cline{2-3} 
& Chord histogram entropy (CHE) & Given a chord sequence, create a histogram of chord occurrences with $|C|$ bins. Then, normalize the counts to sum to 1, and calculate its entropy: $H=-\sum_{i=1}^{|C|}{p_ilogp_i}$, where $p_i$ is the relative probability of the i-th bin. The entropy is greatest when the histogram follows a uniform distribution, and lowest when the chord sequence uses only one chord throughout.\\ \cline{2-3} 
& Chord coverage (CC) & The number of chord labels with non-zero counts in the chord histogram in a chord sequence. \\ \cline{2-3} 
& Chord tonal distance (CTD) & CTD is the average value of the tonal distance computed between every pair of adjacent chords in a given chord sequence. The CTD is highest when there are abrupt changes in the chord progression (e.g., from C chord to B chord). \\ \cline{2-3} 
& Chord tone to non-chord tone ratio (CTnCTR) & define CTnCTR as $\frac{n_c+n_p}{n_c+n_n}$, $n_c$ is the number of the chord tones, $n_n$ is the number of the non-chord tones, $n_p$ is the number of a subset of non-chord tones ($n_p$) that are two semitones within the notes which are right after them, where subscript $p$ denotes a “proper” non-chord tone. CTnCTR equals one when there are no non-chord tones at all, or when $n_p$ = $n_n$. \\ \cline{2-3} 
& Pitch consonance score (PCS) & The consonance score is set to 1 for consonance intervals including unison, major/minor 3rd, perfect 5th, major/minor 6th, set to 0 for a perfect 4th, and set to -1 for other intervals, which are considered dissonant. PCS for a pair of melody and chord sequences is computed by averaging these consonance scores across a 16th-note windows, excluding rest periods. \\ \cline{2-3} 
& Melody-chord tonal distance (MCTD) & MCTD is the average of the tonal distance between every melody note and corresponding the chord label calculated across a melody sequence, with each distance weighted by the duration of the corresponding melody note. \\ \hline
\multirow{3}{*}{\begin{tabular}[c]{@{}l@{}}Style Transfer\\ \cite{40,65}\end{tabular}} 
& Style fit & how well the transformed music fits the desired style. Computed by computing a particular set of outputs’ style profile and measure its cosine similarity to a reference profile. For the specific calculation of style profile, please refer to reference \cite{40}. \\ \cline{2-3} 
& Content preservation & how much content it retains from the original. Computed by correlating the chroma representation of the generated segment with that of the corresponding segment in the source style. \\ \cline{2-3} 
& Strength of transfer & When considering a transfer from $A$ to $B$, $C_{A,B}$ reports the probability $P_A(x)$ if the source genre is $A$, and $P_B(x)$ if the source genre is $B$. Considering a domain transfer from $A$ to $B$ is successful if $P_A(x_A)=C_{A,B}(x_A)>0.5$ AND $P_A(\widehat{x}_B)=C_{A,B}(\widehat{x}_B)<0.5$. Define the strength of the domain transfer in one direction $(A\rightarrow B\rightarrow A)$ as: $S_{A\rightarrow B}^D=\frac{P(A|x_A)-P(A|\widehat{x}_B)+P(A|\widetilde{x}_A)-P(A|\widehat{x}_B)}{2}$. The final domain transfer strength of a particular model is defined as the average of the strengths in both directions: $S_tot^D=\frac{1}{2}(S_{B\rightarrow A}^D+S_{A\rightarrow B}^D)$.\\ \hline				
% more data here
\end{longtable}
\end{landscape}
\end{center}
}

Music performance can be regarded as an interpretation of a given score by performers. Most music metrics of score generation can also be employed for the evaluate of performance generation. Moreover, some additional evaluation metrics have been proposed for performance characteristics (as velocity and timing), such as Mean Velocity (MV), Variation of Velocity (VV), Mean Duration (MD), and Variation of Duration (VD) proposed by Choi et al. \cite{217}.
\subsubsection{Other Methods}~{}\\
Apart from model metrics and music metrics, there are also some methods that use other theories or algorithms to evaluate specific aspects of music, such as the structure, style, and tonality. This section briefly introduces these methods, and the specific implementation of the methods needs further reading of relevant literature.\\

1) VMO(structure) \\
\begin{figure}[H]
	\setlength{\abovecaptionskip}{0pt}
	\setlength{\belowcaptionskip}{0pt}
	\includegraphics[width=0.5\textwidth]{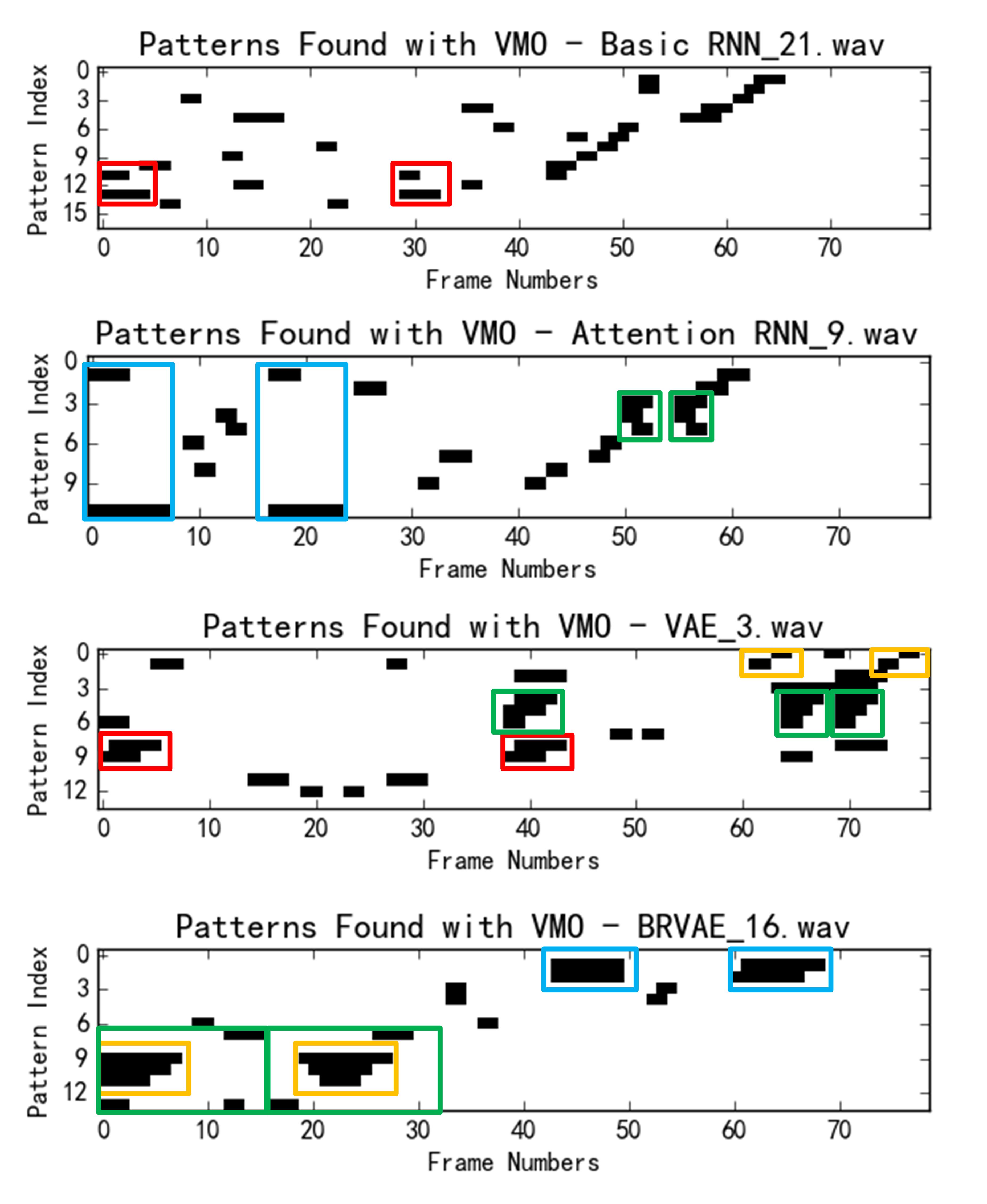}
	\caption{Pattern discovered in each sample by VMO}
	\label{fig:2121}       
\end{figure}
Variable Markov Oracle (VMO) is an information dynamics method developed by Wang et al \cite{313}. It can be exploited to detect repeated music patterns in a time series. Its application in music evaluation is mainly to visualize the music patterns generated by different models, so as to analyze the music structure. Before using this method, it is necessary to convert the music data in symbol domain (. mid) to the audio domain (. wav). VMO uses a string-matching algorithm called Factor Oracle (FO) to search for repeated segments (suffixes) at every time instance in the signal \cite{239}. A key step of VMO is to find a threshold $\theta$ to establish similarity between features. Under different $\theta$, VMO constructs different symbol sequences and suffix structures from signals. In order to select a sequence with the most abundant information patterns, the optimal threshold  $\theta$ is usually selected by calculating the Information Rate (IR). When using VMO for pattern discovery, VMO with higher IRs captures more repeat subsequences than VMO with low IRs. In addition, the IR itself can reflect the self-similarity structure in the sequence \cite{314}, and the higher IRs means that there are more self-similarity structures in the sequence. Music is usually self-similar except for some avant-garde pieces, and structure and repetition are common features of almost all music. Especially in pop music, continuous bars are often repeated on a shorter time scale. Chen et al. \cite{30} reflected the self-similarity structure in music by comparing the IRs of the generated music of different models, and generated the pattern plots of music generated by distinct models using VMO, then detected the repeated patterns in the plots, as shown in Figure \ref{fig:2121}; \cite{216} and \cite{252} used IR as the evaluation metrics as well, which is the mutual information between current and past observations. Intuitively, larger values of IR occur when repetition and variation are in balance, and smaller values of IR occur when sequences are either random or very repetitive.\\

2) Originality\\

In order to evaluate the creativity and originality of the generated music and judge whether the generated music ``plagiarizes" the music in the original corpus, Hadjeres et al. \cite{26} proposed a plagiarism analysis method: given a music, find the length of the longest subsequence which can be found identically in the training dataset, and draw histograms of this quantity. The shorter the corresponding length of the histogram peak, the more innovative the generated music. Chu et al. \cite{58} also conducted a creativity evaluation of the generated music. Same as \cite{26}, they calculated the length of the longest subsequence matched with the training data to evaluate the extent to which the model copied the training data. In addition, they divided each generated melody into segments of 2-bars, and then compared each segment to all segments in the rest of the 100 generated songs, and record the repeat time. This evaluates how much the model repeats itself. Hakimi et al. \cite{353} evaluated plagiarism by measuring the percentage of n-grams that appear in that solo that also appear in any other solo in the source.\\

3) Style\\

Tasks involved in style include generating music with a specific style and style transfer. Researchers usually train a CNN-based style classifier to judge whether the generated music style meets the expectation. For example, Jin et al. \cite{221} used the Minimum Distance Classifier (MDC) algorithm to determine whether the generated music is classical music. In order to evaluate the effect of style transfer, Brunner et al. \cite{65} constructed a binary style classifier using real data to output the probability distribution of distinct styles. In addition, several evaluation metrics related to style transfer are also listed in Table \ref{tab:666}. Commonly used evaluation metrics for singing style transfer include Mel Cepstrum Distortion (MCD), Signal-to-Noise Ratio (SNR) between target waveform and converted waveform, log spectral distance (LSD), F0 raw chroma accuracy (RCA), singer identity (SD). SD needs to train a classifier to represent the probability that a segment belongs to the target singer; LSD is a metrics of phoneme clarity, lower value means better performance; RCA is employed to evaluate melody transfer, the higher the value, the better \cite{186}.\\

4) Dimensionality reduction\\

High-dimensional feature space is nonintuitive and difficult to understand for human beings, thereby is not conducive to evaluation and analysis. The solution to solve this problem is usually to map high-dimensional features to a low-dimensional feature space, such as a two-dimensional space, and then carry out further operation and analysis. The common methods of feature mapping include t-SNE, PCA, etc. In \cite{292}, PCA is exploited to reduce the dimensionality of note embedding to display the latent note space, and reduce the dimensionality of simu\_note embedding to visualize the latent chord space. The application of t-SNE is more extensive. \cite{62} employed this technology to map the features of each track into a 2D space, and then used K-means clustering to cluster all the features into five clusters. The assessment of the characteristics of each cluster is performed by counting the occurrences of different training “keywords” in each cluster. In addition, when it comes to the generation of music of different styles, t-SNE technology is often utilized to map the style embedding space, so as to analyze the ability of model learning style. E.g. in \cite{48}, t-SNE is used to visualize the latent vectors of 30 jazz songs and 20 classical songs\iffalse, as shown in Figure \ref{fig:2222}\fi; \cite{164} used t-SNE technology to project the timbre code from the k-dimensional space to a 2D space, thereby evaluating the performance of timbre and pitch disentanglement.\\
\begin{comment}
\begin{figure}[H]
\includegraphics[width=0.4\textwidth]{img//22.png}
\caption{T-SNE plot of latent vectors for bars from Jazz
and Classic songs
}
\label{fig:2222}       
\end{figure}
\end{comment}

5) Keyscapes\\

Keyscapes are used to illustrate the tonality of musical works, with different color for each key. George et al. And Lattner et al. \cite{216,252} utilized the humdrum keyscape tool by David Huron \cite{315} to parse the tonality of a given MIDI file using the Krumhansl-Schmuckler key-finding algorithm \cite{316}, and converted the result to a PNG image file from the humdrum native data format. The peak of the pyramid shows the key estimates of the whole piece, M and moving down toward the base is a recursive set of decreasing window sizes which give an estimation of the key estimates within that context window. Intuitively, the highly-structured keyscape indicates that there are more self-references existing in the segment, with tonal centers that recur often in fractal patterns.\\
\begin{comment}
\begin{figure}[H]
\includegraphics[width=0.4\textwidth]{img//23.png}
\caption{Toy example demonstrating tonal clustering within a keyscape}
\label{fig:2323}       
\end{figure}
\end{comment}

6) Fitness scape plot\\

Fitness scale plot algorithm \cite{317,318} and related SM toolbox \cite{319} offer an aesthetic method for detecting and visualizing repetitive structures in music. The fitness scape is a matrix $S_{(N\times N)}$, where $s_{ij}\in[0,1]$ is the fitness, that is, the repeatability in the music segment derived from the self-similarity matrix of the segment specified by $(i,j)$. As an example, Wu et al. \cite{256} compared the fitness scale plots of the generated work and the real work\iffalse, as shown in Figure \ref{fig:2424}\fi.
\begin{comment}
\begin{figure}[H]
\includegraphics[width=0.6\textwidth]{img//24.png}
\caption{Fitness scape plots}
\label{fig:2424}       
\end{figure}
\end{comment}
\subsection{Subjective}
\label{sec:6.2}
Although many objective evaluation metrics have been proposed, there is still a lack of correlation between the quantitative evaluation of music quality and human judgement, so subjective evaluation is indispensable. Subjective evaluation is human assessment, and human feedback may be the most reasonable and persuasive post-hoc evaluation method. Human evaluation unconsciously takes into account local music statistics and establishes expectations for tracking longer musical structures, such as identifying salient themes and styles \cite{320}. The most common subjective evaluation method is listening test. The design of listening test is a complicated process, involving many variables \cite{117}, such as the selection and presentation of audio samples, listening environment, selection of subjects, expression of questions, etc. In addition, there are some subjective evaluations that do not involve listening, mostly conducted by music experts to analyze the generated score, lead sheet, etc. The number of subjects involved in subjective evaluation may vary greatly, and it faces the following two problems: one is that it takes more time compared with machine evaluation; the other is the evaluation error caused by auditory fatigue. Continuous listening will make the listeners produce relatively unreliable judgments. Nevertheless, human subjective evaluation is still significant and indispensable.
\subsubsection{Listening test}~{}\\
\label{sec:6.2.1}
Listening test is the most basic, most common and most convincing evaluation method in music generation tasks. No matter what level of music generation is, listening test is inseparable. The difference is that score generation needs to render performance characteristics to the generated results to obtain the same results as performance generation, and then synthesize audio files for human testing. Audio generation does not need these processes and can directly generate auditory results. Generally speaking, subjective listening test requires \cite{321}: 1) there are enough listening subjects with sufficient diversity to offer statistically significant results; 2) the music knowledge level of subjects is evenly distributed, including both music amateurs who are lack of music knowledge and experts in the field of music composition; 3) the experiment was conducted in a controlled environment with specific acoustic characteristics and equipment; 4) each subject received the same instructions and stimuli. The simplest listening test is Turing test. Beyond that, the generated music can also be compared with original corpus or music generated by other models. In order to prove that the evaluation results have statistically significant differences, hypothesis tests are usually conducted, such as Wilcoxon signed-rank test \cite{23}, t-test \cite{39}, Kruskal Wallis H-Test \cite{53}, etc.

Here are five common listening tests in various studies:
\begin{itemize}
	\item Turing test, which requires the subjects to judge whether the music is machine-generated or human-created \cite{26,40}.
	\item Carry out two/multiple comparison and selection of music. The choice can be preference selection, that is, to choose the most favorite one from a group of music \cite{34,40}, or based on some criteria, such as selecting sample generated by the model which is closest to the original corpus style \cite{145,274};
	\item Side by side rating. For each pair of prediction and real data, the subjects are required to give scores such as - 1 (generated music is worse than the real) to + 1 (generated music is better than the real) \cite{171}, which can be regarded as the further quantification of binary selection;
	\item According to several evaluation metrics, rank a set of music samples generated by different models or different configurations \cite{61,173}.
	\item Score the generated music based on several evaluation metrics, or by answering several subjective questions, such as how pleasant the melody is and how natural the music is, the common scoring rule is 5-point Likert \cite{21,29}, and there are other researcher-defined scoring rules, such as the 4-point scale adopted by \cite{47}, the hundred percentage system adopted in MUSHRA test \cite{207}. The 5-point Mean Opinion Scores (MOS) evaluation method is widely used in the field of audio generation (as singing voice synthesis \cite{177}, singing voice conversion \cite{188}, etc.), and CrowdMOS toolkit \cite{321} can be employed to collect MOS.
\end{itemize}

Apart from the above listening tests, there are some other listening tests, e.g. in \cite{27}, so as to evaluate the extent to which the model can create music with clear style, the subjects are required to identify the type of generated music from given types, and judge whether the music generated by models has a clear style by calculating the classification accuracy rate of the type; similarly, in \cite{67}, to evaluate whether the music generated by models has a clear emotion, the subject is invited to identify the emotion category of each piece of music; another listening test only invites professional composers to score the music. The scoring criteria here are not subjective questions, but professional music evaluation metrics, which requires subjects to possess a strong music background. E.g. Yang et al. \cite{38} evaluated the notes and rhythms in the generated music according to the music theory consistent with the Beatles’ music; Wei et al. \cite{98} invited professional drummers and composers to give descriptive feedback and detailed comments on the structural compatibility, stability and variability of the generated drum patterns.

Listening tests can be conducted on online platforms as well, such as Amazon Mechanical Turk \cite{274} and some online websites containing questionnaires \cite{210}. Although online testing saves manpower, it cannot ensure the authenticity and validity of the collected results, for it does not rule out the situation that someone hastily completes the task so as to get rewards. However, online evaluation can indeed invite more participants. Liang et al. \cite{51} conducted a public music discrimination test on bachbot.com, and a total of 2,336 participants involved in the test, which is the largest human evaluation so far.
\subsubsection{Visual analysis}~{}\\
\label{sec:6.2.2}
Visual analysis means that no auditory perception, people evaluate the quality of generated music subjectively only through observing the music representation forms which are visual, such as music score, pianoroll, waveform, spectrogram, etc.\\

1) Score analysis\\

Score analysis is to discuss the music information implied in the score, which is usually carried out by experts in music field. The analysis of music score focuses on different aspects according to different generation tasks, such as pitch change, rhythm, repetition pattern, transition between bars, etc. Dong et al. \cite{32} analyzed the stability, smoothness, musicality of the generated melody given chord and rhythm pattern in music, Yan et al. \cite{47} recorded the generated accompaniment score in .pdf format and recruited three PhD students of music theory as raters, then asked them to score accompaniment according to certain scoring standards; Pati et al. \cite{70} compared and analyzed the inpainting scores generated by different models. Although the analysis of music scores relies on the relevant evaluation criteria, different judges may have different opinions on the same score, so we summarize it into the scope of subjective evaluation.\\

2) Waveform/Spectrogram analysis\\

The evaluation of the quality of generated audios usually displays the waveform, spectrogram, rainbowgram of the audio samples. Waveform or spectrogram analysis is to observe the audio information contained in the generated waveform or spectrogram and make corresponding analysis. For example, Engel et al. \cite{152} showed each note as a ``Rainbowgram",a CQT spectrogram with intensity of lines proportional to the log magnitude of the power spectrum and color given by the derivative of the phase. Time is on the horizontal axis and frequency on the vertical axis. Then they observed the rainbowgram of the reconstructed notes of different instruments, and compared the rainbowgram of audio produced by interpolation in the latent space and the original audio space. Andreux et al. \cite{163} compared the waveform and log-spectrogram of the original and reconstructed audio. Also common is the comparison of mel-spectrograms and F0 contour maps. For instance, \cite{166} compared the mel-spectrograms of the input and output of different models, \cite{178} compares the F0 contours generated by distinct models, and so on.
\section{Challenges and Future Directions}
\label{sec:7}
According to the above research, we can see that automatic music generation has made great progress in the past decade. Compared with traditional methods, deep learning technology shows its powerful capabilities. However, there are still many challenges in using deep learning to generate music. Moreover, deep learning itself has the nature of unexplainable. This section will mainly discuss the difficulties and challenges faced by different levels of music generation, and point out several future research directions.
\subsection{Challenges}
\label{sec:7.1}
\subsubsection{Top-level: Score Generation}~{}\\
\quad\,\ \textbf{Structure.} The generated music lacks long-term structure, i.e., recurring themes, motivations and patterns. Especially when using recurrent sequential model for music generation, the music generated by these models gradually becomes boring because there is no new external inducement during the generating process. The most extreme case is that the generated notes only contain one pitch value. When the length of music increases, it is hard to model the local structure and global structure at the same time. Global patterns are about long structures that span multiple paragraphs. Local patterns refer to elements from the past or even from distant past, which are repeated and further developed to create contrast and surprise \cite{219}. Although a lot of work has been done to generate longer music pieces, it is still far from reaching the goal of generating a normal song length. There are relatively few studies focusing on self-repetitive structures \cite{236}. The extraction of theme \cite{27} is still an unexplored area. Additionally, music also contains a whole structure of its own, such as AABA or AAB form in jazz, which inevitably requires different modeling of various components of music, and the change of structure makes automatic music generation more complicated. At present, only the template-based method proposed by Zhou et al. \cite{37} can generate a specific overall structure of music.

\textbf{Closure.} Some of the current models can only generate fixed-length music segments, while others can generate music of arbitrary length. However, there is no clear control over when the music will end, resulting in the sudden end of the generated music. The quality of music closure is one of the important aspects to distinguish real music from machine-generated music.

\textbf{Creativity.} The distribution of the music generated by the data-driven deep learning algorithm is close to that of training dataset. Creativity is the extrapolation of outlier patterns beyond the observed distribution. Present machine learning regimes, however, are mainly capable of handling interpolation tasks and not extrapolation \cite{353}, resulting in that the generated music lacks certain innovation. Moreover, a significant percentage of generated sequences, despite their statistical similarity to the training data, are flagged as wrong, boring or inconsistent, when reviewed by human peers \cite{60}. The boundary between music creativity and plain clumsiness is highly uncertain and difficult to quantify. There are relatively few researches focusing on the creativity of generative music for now, such as \cite{218}.

\textbf{Emotion.} As German philosopher Hagel has said \cite{67}: ``Music is the art of mood. It is straightly directed against mood." The research on emotional music is not much, and the classification of emotion is relatively simple. It still needs lot of effort to understand the relationship between music and emotion and integrate this important relationship into the music generation system.

\textbf{Style.} Currently, the music style characteristics generated are often consistent with the music styles of training datasets. DeepJ \cite{27} is able to generate music with a specific style, but it is limited to music styles in the classical period, and the generated music style is also included in the classical music dataset exploited during training. For style transfer, learning the disentanglement of music representations with respect to different features automatically is still an effortful task. Different music styles require different model settings, hence a universal framework for all music styles is desired \cite{50}.

\textbf{Interaction.} For now, the interaction during music generation process is insufficient, and users are unable to add more complex constraint information, such as forcing two notes to be equal or introducing soft constraints, only can add location constraints \cite{22} or some unary constraints, such as rhythms, notes, parts, chords and cadences \cite{26}. Since the neural network has not been designed to be controllable, the output of deep learning model is uncertain for users, which makes it difficult for users to control AI and express their creative goals. Arbitrary control is a tough problem for deep learning architecture and technology.

\textbf{Impromptu.} It is laborious to realize automatic music improvisation, which involves a combination of skills that are difficult to reproduce algorithmically. Automated music composition, co-operative agency, responsive behavior and expressive performance are individually complicated for model, and particularly so in combination. Useful patterns in music creation may be from a large number of pre-composed data, but specific characteristics important to the development of collaborative, spontaneous composition seen in improvised performances are never exposed to the system \cite{102}.

\textbf{Representation.} Most existing studies only use pitch and duration to represent music, which are too simple to support the diversity of musical symbols, such as trios, vibrato and ornaments. In the future, it is expected that new forms representing the diversity of music symbols could be put forward. In addition, the representation of chords has been greatly simplified for now. Most researches only focus on the representation and generation of triads, ignoring other chord types.

\textbf{Evaluation.} Different research is not completely inconsistent in the evaluation methods for generated music. The common evaluation method is that researchers define some evaluation metrics related to music characteristics, and calculate the corresponding metrics values of a set of music, then compare them. However, the metrics defined by distinct researchers are often diverse, and the evaluation results of the same piece of music may be quite different under the metrics defined by other researchers. As Theis et al. \cite{118} observed, ``A good performance with respect to one criterion does not necessarily imply a good performance with respect to another criterion", which makes it difficult to compare the music generated in different research work. And more metrics are needed to sort out the discrepancies between more and more music models. There is also a lack of correlation between quantitative metrics and subjective evaluation. Music that meet good standards on objective metrics may perform poorly in subjective evaluation, and even extreme situations would happen that the two evaluation results are completely opposite. Music achieving good subjective evaluation lacks the explanation of quantitative metrics. Moreover, current systems are unable to evaluate the generated music automatically. Although great progress has been made in the automatic analysis and categorization of music, they are still not close to human-level performance, which makes assessing music very difficult and partly explains why music evaluation cannot be automated through computational models so far \cite{117}.

\textbf{Application.} The magic of algorithmic composition lies in that its goal is to reproduce the real intelligence of human beings, rather than complex imitation \cite{218}. Obviously, it is quite challenging to build the artistic creativity of machines. People have been committed to developing novel architectures and proving that the quality of music generated by these architectures is comparable to the music created by human beings. Although researchers have made a lot of efforts over the years, algorithmic composition still remains in laboratory research, while other artificial intelligence technologies, such as search agent and face recognition, have been widely used in the industrial field \cite{218}. The application of music generation system in the actual environment is limited for two reasons. First, most models place restrictions on the nature of the input. In most cases, there are limitations places on the number and type of tracks. Second, users cannot control the generation process in a fine-grained way, which is crucial for the system to be useful in the context of computational assisted composition \cite{289}.
\subsubsection{Middle-level: Performance Generation}~{}\\
\label{sec:7.1.2}
\quad\,\ \textbf{Pedal.} The note duration in piano performance is hard to model clearly because it involves modeling the influence of the sustain pedal, which will extend the duration of all notes under the pedal until the pedal is released \cite{137}. The modeling of piano pedal is one of the most challenging problems in performance generation.

\textbf{Dataset.} One of the reasons for the few researches in performance generation is the lack of available datasets. One is the alignment dataset of score and performance, and the other is the dataset of other instruments except piano.

\textbf{Evaluation.} The current evaluation methods for performance generation are too plain, almost all of them adopt MSE or listening test between real performance and generated performance, which lacks more meaningful evaluation metrics about music. It's really hard to judge whether the performance is good or not. After all, human performance competitions still require music experts as judges to score performances.

\textbf{Instrument.} Existing performance modeling methods may be more effective for instruments that use relatively simple mechanisms to produce sound (e.g. percussion instruments such as pianos and drums). It is hard to model the performance of instruments (e.g. stringed instruments and wind instruments) that produce sound in more complicated ways with pre-defined attributes such as dynamics. 
\subsubsection{Bottom-level: Audio Generation}~{}\\
\label{sec:7.1.3}
\quad\,\ \textbf{Model.} The neural generation of realistic audio remains a challenge, because of its complex structure, with dependencies on various temporal scales \cite{174}. The current models are still tough to learn and synthesize complex sounds, including multi-track and multi-instrument audio \cite{199}. Neural networks have achieved some success in modeling pre-extracted synthesis parameters, but these models lack end-to-end learning. The analysis parameters still have to be tuned manually and the gradient cannot flow through the synthesis procedure. As a result, small errors in the parameters can cause large errors in the audio that cannot propagate back to the network. The key point is that the authenticity of vocoders is limited by the expressivity of a given analysis/synthesis pair \cite{198}.

\textbf{Representation.} Fourier coefficient can general and can represent any waveform, but they are not free from bias. This is because they usually prefer to produce audio with aligned packets rather than oscillations. For example, stripe convolution models (such as SING \cite{274}) generates waveforms directly with overlapping frames. Since audio oscillates at many frequencies, all with different periods from the fixed frame hop size, the model must accurately align waveforms between different frames and learn filters to cover all possible phase variations. Fourier-based models, such as GANSynth \cite{191}, also have phase-alignment problems because STFT is a representation over windowed wave packets. Additionally, they must solve the problem of spectrum leakage, that is, when the Fourier fundamental frequency does not match the audio frequency exactly, it is necessary to combine sinusoids at multiple sinusoids and phases to represent a single sinusoid. Autoregressive waveform models, such as WaveNet \cite{149}, SampleRNN \cite{151}, are not restricted by the bias over generating wave packets and can express arbitrary waveforms. But they don't take advantage of oscillation bias, so they need larger and more data-hungry networks. Moreover, the use of teacher-forcing during the training process will lead to the aggravation of feedback errors, which also makes these models incompatible with perceptual losses such as spectral features, pretrained models and discriminators. Since the shape of the waveform does not perfectly conform to perception, the efficiency of these models is further reduced \cite{198}.

\textbf{Computation.} The best sound systems are still computationally expensive and difficult to migrate to low-cost devices \cite{199}. Training usually depends on many hours of data, which must capture not only the sound, but control factors that are difficult to obtain such as the air pressure inside a mouth for a wind instrument model. Even on the best GPU platform, the amount of data needed to create a good model may take days to train, which is both time-consuming and expensive.
\subsection{Future Directions}
\label{sec:7.2}
Through above description of the challenges faced by the three levels of music generation, we can see that though a lot of research has been done on score generation, there is still room for improvement; there is not much research on performance generation, which may be caused by the lack of performance datasets and the difficulty in evaluating the generation effect. The implementation methods of performance generation and score generation overlap to some extent, and the results of performance generation is further optimization of score. Plenty of research has been carried out on audio generation, maybe the next work is to improve the audio synthesis system in the light of the music characteristics.

We propose multiple future research directions of score generation next. Some of them may be in the process of research, but have not achieved the desired results. First of all, generate long-term music with clear themes and self-repetitive structure; at the same time, control the timing of music closure; generate the overall structure of music according to the task requirements, such as the verse-chorus-bridge-verse structure in pop music and the AABA structure in jazz music; next, generate innovative music, not just ensure that the generated music distribution is close to the training dataset distribution; generate music with specific style according to the users’ requirements. Generating music with a different style from the training dataset also involves style transfer, and current research can only achieve the conversion of specific style pairs, a general framework could be developed to achieve arbitrary conversion between styles in the future; furthermore, combine music creation process with emotion to generate music with specific emotion, and then use generated results to adjust human emotion. This may promote practical applications in real life, such as real-time music generation of games and automatic generation of background music of movies and videos; more importantly, strengthen the interaction with users during music creation, such as improvisation through real-time interaction with users. After all, the ultimate purpose of music generation is to serve human beings; in terms of music representation, improve the representation method to express the diversity of music, such as triplets, vibrato, etc. Meanwhile, the representation should not be too complicated to increase the burden of model learning; in the aspect of evaluation, explain the correlation between quantitative metrics and subjective evaluation, and gradually realize the automation of music evaluation; employ automatic features extraction methods to integrate complex music knowledge into the generation model, such as \cite{113}; consider using other methods (as reinforcement learning) to improve generation results. In recent years, reinforcement learning has shown promising results in many tasks, especially for tasks that require long-term learning. Currently, there are relatively few researches with the help of reinforcement learning for automatic music generation. 

In terms of performance generation, we can further explore the modeling of piano pedal, note closing speed, etc., create a large dataset aligned with music score and performance, and model the performance of any instrument, not only limited to piano and drum. There are not too many researches on performance generation using deep learning, and most of them utilize simple models based on LSTM, so there is still much room for further improvement. In addition, the expressive performance can express and convey the performer's understanding of the internal structure and emotional content of the music piece, and show the dramatic and emotional nature of the music in this way. In the most ideal situation, performance can emotionally attract and influence the audience, so researchers can explore the generation of performance with emotional expression in the future.

As for audio, the musicality of samples can be improved by providing high-level conditional information (as the composer of a piece) or incorporating prior knowledge about musical forms into the model \cite{168}. In particular, for the singing synthesis, we can consider the singing voice conversion under the background music, and create a virtual singer similar to Hatsune Miku in the future, who can sing more timbres. Moreover, current technology synthesizes words in lyrics without knowing the meaning of the lyrics. In the future, we hope that the meaning of the lyrics can be reflected in the singing expression like real human singers, so as to make the singing voice more emotional
\subsubsection{Emotion}~{}\\
\label{sec:7.2.1}
Music and language possess equally significant status, it can not only promote human communication, but also affect human emotions. Emotion is the main object of music creation, and music is able to express emotion and arouse emotion. Therefore, listening to music can not only be used as a way of entertainment, but as a method of emotional adjustment. For example, when you are in a bad mood, listening to cheerful music can appropriately alleviate negative emotions; when you feel tired, listening to impassioned and inspiring music can inject new strength into the exhausted body; listening to some soothing and quiet light music before going to sleep can achieve the effect of sleep-enhancing. By acquiring of human current emotions, and then adjusting them through music in a targeted manner, music can even be used as a method for depression treatment, psychotherapy and other aspects. At present, the researches on generating music with emotion are relatively few, and the types of involved emotion are relatively simple. The main reason is that emotion itself is very complicated, it is not easy to express emotion as a feature vector that can be understood by computer accurately. Secondly, the diversity of emotions makes our understanding of music emotion quite vague. Simply dividing emotions into the type of happy or excited through models such as the valence-arousal model may not be enough to express the emotional experience that music really brings to people; in addition, a major obstacle to the use of deep learning technology to generate emotional music is the lack of available emotional music datasets. Therefore, future work can focus on the following points: the representations of emotional characteristics, the classification of music emotion types and the creation of music emotion datasets.
\subsubsection{Interaction}~{}\\
\label{sec:7.2.2}
The fundamental purpose of various research directions under music information retrieval, whether it is music analysis, transcription, generation or recommendation, is to better serve human beings, especially for the more creative task of music generation. It is possible to completely use computers to generate innovative and fluent music end-to-end, but during this process humans lack a sense of participation, and the generated music may not meet human creative goals (as pitch contours, emotional trends). Recent studies have found that users would like to retain a certain level of creative freedom when composing music with AI, and that semantically meaningful controls can significantly increase human sense of agency and creative authorship when co-creating with AI \cite{355}. Therefore, interactive music generation methods/tools are becoming more and more important in the process of human-AI music collaborative creation. Although there have been some studies in this area, only a few interactive interfaces/tools are available for users. Moreover, some interactive systems only meet entertainment needs or are limited to studio use, such as Bach Doodle, FlowComposer, etc. Most of the music generated by these system does not reach the level of industrial applications. In the future, it is worth trying to develop an interactive generation system that can not only meet the specific creative needs/objectives of users but also generate satisfactory music so as to further promote it to industrial applications.
\subsubsection{Style}~{}\\
\label{sec:7.2.3}
During the decades of research and development of automatic music generation, it has always focused on music genres like classical, popular, folk. With the changes of times and the development and progress of human society, more and more niche music types have gradually entered the mainstream market, such as Chinese style music, rap, electronic music, etc., which has injected fresh blood into the field of automatic music generation. Once the corresponding datasets are available, the previous model architecture can be applied directly to these music genres, but the generation effect cannot be guaranteed, and the representations, model structures and hyperparameters may need to be adjusted. Future work can propose new representation methods or model architectures for these music genres. What’s more, some researchers have studied the automatic generation of music genres with more artistic characteristics, such as Peking Opera and opera. Compared with pop music, these may be tough to achieve wide application, but they play a good role in promoting the inheritance of national culture. Interested researchers can pay more attention to this aspect in the future.
\section{Conclusion}
In this paper, automatic music generation is divided into three levels in the light of three stages of music generation process, and then we provides the latest progress in applying deep learning techniques to achieve various music generation tasks under three levels. This paper focuses on different generation tasks, rather than algorithm techniques, so that we can have a comprehensive and profound understanding of what tasks are involved in the field of music generation, where it has developed, and what shortcomings or challenges exist, for the sake of laying the foundation for new breakthroughs in this field. Moreover, this classification method of focusing on generation tasks also brings convenience to researchers. They can quickly find out which methods can be used to realize the tasks they are interested, so as to clarify the implementation methods and improvement directions. We hope that this review can help to better understand the research field and developing trend of music generation based on deep learning. In addition, we also introduced various music data representations, music evaluation methods and some commonly used datasets that have appeared in the researches so far.

In the past decade, music generation using deep learning has made rapid progress, but its generation effect is still far from the expected results, existing plenty of problems such as the lack of structure and innovation in the generated music; the research on the modeling of music expressive performance is relatively less; it is difficult to capture the emotion in music; the interaction with users is limited, and there is no unified music evaluation standard, etc. In the future, we consider using new methods to improve the learning process and generate more satisfactory music, such as the reinforcement learning algorithm which is being explored; building a unified evaluation metrics for the generated music; combining emotion with music generation to explore the use of music to adjust human emotions; further strengthening the interaction between AI and human, so as to achieve the goal of music ultimately serving humans; applying automatic music generation technology to real life to promote social progress and development.

\bibliographystyle{unsrt}      % basic style, author-year citations
\bibliography{ref}   % name your BibTeX data base

\end{document}